# Flexo-Elastic Control Factors of Domain Morphology in Core-Shell Ferroelectric Nanoparticles: Soft and Rigid Shells

Eugene A. Eliseev[1], Anna N. Morozovska[2*], Riccardo Hertel[3†], Hanna V. Shevliakova[4], Yevhen M. Fomichov[5], Victor Yu. Reshetnyak[6‡], and Dean R. Evans[7§]

[1] Institute for Problems of Materials Science, National Academy of Sciences of Ukraine,
Krjijanovskogo 3, 03142 Kyiv, Ukraine

[2] Institute of Physics, National Academy of Sciences of Ukraine,
46, pr. Nauky, 03028 Kyiv, Ukraine

[3] Université de Strasbourg, CNRS, Institut de Physique et Chimie des Matériaux de Strasbourg, UMR 7504, 67000 Strasbourg, France

[4] Department of Microelectronics, National Technical University of Ukraine "Igor Sikorsky Kyiv Polytechnic Institute", Kyiv, Ukraine

[5] Charles University in Prague, Faculty of Mathematics and Physics,
V Holešovičkach 2, Prague 8, 180 00, Czech Republic

[6] Taras Shevchenko National University of Kyiv, Volodymyrska Street 64, Kyiv, 01601, Ukraine

[7] Air Force Research Laboratory, Materials and Manufacturing Directorate, Wright-Patterson Air Force Base, Ohio, 45433, USA

---

\*      Corresponding author 1: anna.n.morozovska@gmail.com

†      Corresponding author 2: riccardo.hertel@ipcms.unistra.fr

‡      Corresponding author 3: victor.reshetnyak@gmail.com

§      Corresponding author 4: dean.evans@afresearchlab.com

**Abstract**

Within the framework of the Landau-Ginzburg-Devonshire approach we explore the impact of elastic anisotropy, electrostriction, flexoelectric couplings, and mismatch strain on the domain structure morphology in ferroelectric core-shell nanoparticles of spherical shape. We perform finite element modelling (**FEM**) for multiaxial ferroelectric nanoparticle cores covered with an elastically-isotropic soft or elastically-anisotropic rigid paraelectric shell, with and without mismatch strains. The latter are induced by the difference of the core and shell lattice constants.

In the case of a core covered with a **soft shell**, the FEM results show that at room temperature a single polarization vortex with a dipolar kernel can be stable if the electrostriction coupling is relatively weak. With increasing anisotropic electrostriction coupling, the vortex disappears and is replaced by complex flux-closure structures, which are formed in the equatorial plane and transform into an elongated vortex with a central 180º domain wall near the core poles. This complex domain morphology develops in the core due to the anisotropic electrostriction, and the flexoelectric coupling leads to an additional curvature and twist of the polarization isosurfaces.

In contrast to this, FEM performed for a core covered with a **rigid shell** shows that, at room temperature, the anisotropic elastic properties of the shell can stabilize vortex-like structures with three flux-closure domains, which gradually "cross" in the equatorial plane of the core and transform into 120º-type domains near the core poles. The flexoelectric coupling leads to a noticeable curling of the flux-closure domain walls. A mismatch strain compensates the curling of the flux-closure domains in the core confined by the elastically-anisotropic rigid shell. Our analysis of the configuration of the polarization reveals different types of topological defects, namely Bloch point structures (**BPS**) and Ising lines, forming in a ferroelectric core covered with a soft or rigid shell.

Furthermore, we study the influence of the core radius on the temperature behavior of domain structure morphology, polarization value, and phase transition temperatures, and derive approximate analytical expressions to analyze the influence of the elastic properties of the shell as well as mismatch strain on the phase diagrams. The phase diagram for a core covered with an elastically-isotropic **soft shell** shows a relatively small but noticeable increase of the paraelectric-ferroelectric phase transition temperature induced by the flexoelectric coupling, whereas the phase diagram for a core covered with an elastically-anisotropic **rigid shell** reveals a relatively strong influence of mismatch strain.

The analytical results obtained from this study can be used for the optimization of core-shell ferroelectric nanoparticle sizes for advanced applications in nanoelectronics and nano-coolers. Specifically, the obtained analytical results allow for the selection of optimal parameters to reach high negative values of an electrocaloric response from an ensemble of non-interacting core-shell nanoparticles, which is important for energy convertors and cooling systems. Core-shell ferroelectric nanoparticles, whose polarization arranges in a vortex-like structure with different types of BPS and/or dipolar kernels, can be considered as promising candidates for nanosized field effect transistors and logic units.

# I. INTRODUCTION

Ferroelectrics are among the most interesting objects for fundamental and applied studies of spontaneous polarization dynamics, which is often characterized by a versatile morphology of multi-domain states with complex topology of electric dipoles [1, 2, 3, 4]. Special efforts are intended to answer the question on how complex topological states [5, 6, 7, 8, 9], such as flux-closure domains, polarization vortices, or skyrmions, which sometimes exist in nanosized ferroelectrics, can be controlled by elastic forces and/or electric fields (see e.g. Refs. [10, 11, 12, 13, 14] and citations therein).

Many recent works are devoted to the phase-field modeling of polarization vortices in nanosized ferroelectrics, such as nanodots and nanoparticles; their reaction to external stimuli, such as temperature changes [15, 16] electric fields [17]; elastic strains induced by the substrate, dislocations and local clamping forces [14]. This is typically done in the framework of a continuum phenomenological Landau-Ginzburg-Devonshire (**LGD**) approach combined with electrostatic equations and phase-field modeling (see Ref. [18] and refs. therein). Special attention is paid to the role of size and shape effects [17], for instance, Mangeri et al. [19] considered noninteracting spherical ferroelectric nanoparticles embedded in a dielectric matrix and showed that the vortex-like polarization morphology is strongly affected by the particle diameter. Mangeri et al. [20] then proposed different ways for the electromechanical control of polarization vortices in interacting ferroelectric-dielectric dimers. Pitike et al. [21] revealed that the critical sizes of ferroelectric nanoparticles with vortex-like polarization textures are strongly dependent on the particle shape. Chen and Fang [22] studied the electrocaloric effect (**ECE**) in barium titanate nanoparticles with vortex polarization using a core–shell model.

Recently, we predicted that it is possible to control the domain structure of core-shell ferroelectric nanoparticles by using tunable shells [23]. We then explored the possibility of electric field control of three-dimensional vortex states in core-shell ferroelectric nanoparticles [24]. The field-induced changes of the vortex structure are manifested in the appearance of an axial kernel in the form of a prolate nanodomain, the growth of the kernel, an increasing orientation of the polarization along the field, and the onset of a single-domain state. The in-field evolution of the polarization includes the formation of Bloch point structures (**BPS**) located at two diametrically opposite positions near the core surface. An interesting aspect is that the classical behavior of the vortex axis can simulate a "qubit" at room temperature, since some basic properties of qubits necessary for a quantum computation [25, 26] can be simulated by the vortex+kernel states "±1" revealed in Ref. [24]. However, one should bear in mind that the electrostatic interaction between the core-shell ferroelectric nanoparticles is different from the "true" entanglement of e.g. photons, because photons can be entangled at macroscopic distances, whereas coupling between nanoparticles decays at increasing distances due to the attenuation of the electrostatic field.

To the best of our knowledge, existing theoretical papers (cited above and many others) did not consider the **influence of elastic properties of a shell on** the core ferroelectric polarization and the morphology of its domain structure in the presence (or absence) of a flexoelectric coupling, which relates the electric polarization with an elastic strain gradient (or the strain with a polarization gradient). Motivated to fill this gap in knowledge, we simulate numerically the formation of complex three-dimensional domain structures in spherical nanoparticles consisting of a ferroelectric core covered with a paraelectric shell, and analyze how the domain structure and phase diagrams of the nanoparticles are influenced by the shell and core anisotropic elastic properties, electrostriction, flexoelectric coupling, and mismatch strains (arising from different lattice constants and thermal expansion coefficients of the core and shell materials).

The remainder of the paper has the following structure. The formulation of the problem is presented in **Section II**, which contains the method description and calculation details with emphasis on the mechanical state of the core covered by shells with different elastic properties. Results of numerical modeling and their analytical description are presented in **Sections III-IV**. The influence of electrostriction and flexoelectric coupling on the domain structure in a ferroelectric core covered with elastically soft shell is considered in **Section III**. The ferroelectric properties of a core covered with an elastically rigid shell is analyzed in **Section IV,** where special attention is paid to the influence of a mismatch strain between the core and the shell on the domain structure morphology in the core. Size-dependent phase diagrams of core-shell nanoparticles are discussed in **Section V**, and possible applications are discussed in **Section VI**. The obtained results are summarized in **Section VII.**

## II. PROBLEM STATEMENT, METHODS, AND PARAMETERS

**A. Methods, approximations, and limitations.** We use the LGD approach combined with electrostatic equations, because this method has proven to be successful in establishing the physical origin of anomalies in phase diagrams, determining polar and dielectric properties of ferroelectric nanoparticles [27, 28], and calculating the changes of their domain structure morphology with size reduction [29, 30]. The LGD approach allows for the consideration of various size and surface effects, such as correlation effects and depolarization fields arising in the case of incomplete polarization screening [31], surface bond contraction [32, 33], and intrinsic surface stresses and strains [34, 35, 36].

We perform finite element modeling (**FEM**) of the polarization, the internal electric field, and the elastic stress in a spherical BaTiO$_3$ core covered with a "tunable" paraelectric shell. The relative dielectric permittivity $\varepsilon_S$ of the shell is ultra-high and temperature-dependent. The core-shell nanoparticle is placed in a polymer or liquid medium with a relative dielectric permittivity $\varepsilon_e$. An external electric field is absent. The dielectric and elastic properties of the SrTiO$_3$ shell and the BaTiO$_3$ core are given in **Table AI** in **Appendix A.**

The main role of the shell is to provide an effective tunable screening of the core polarization [23]. Note that a 10-lattice constant thick ($\Delta R = 4$ nm) or thicker shell with $\varepsilon_S \geq 200$ can maintain a remanent polarization of a BaTiO$_3$ core with radius $R \geq 2$ nm, because of the effective dielectric screening in the shell. However, for shells significantly thinner than 10 lattice constants, different low-dimensional effects can change the dielectric and electronic properties. Two types of the shells are compared in this work: an elastically **"soft"** shell and a "**rigid**" shell, whose dielectric properties are the same as bulk SrTiO$_3$, but with very different elastic modulus values.

**The definition of a soft shell.** A tunable shell with high dielectric permittivity is considered to be elastically soft if its elastic stiffness is rather small. Soft matter, including liquid crystals, can play the role of a soft shell. Note that different concentrations of oxygen vacancies can be present in perovskites like SrTiO$_3$. These vacancies, being elastic dipoles [37], are effective sinks for elastic stresses [38]. In accordance with our estimates, the oxygen vacancies located in the shell can strongly reduce its effective elastic compliances, such that a SrTiO$_3$ shell with more than (1 – 3) vol. % of vacancies can be considered elastically soft (see **Fig. A1** in **Appendix A**).

**The definition of a rigid shell.** A "vacancy-free", i.e. stoichiometric, SrTiO$_3$ shell can be considered rigid, since the elastic stiffness and electrostriction tensor components of bulk crystalline SrTiO$_3$ are relatively high (see **Table AI** in **Appendix A**). Unlike the case of the soft shell, the rigid shell can include a mismatch strain, which originates from different lattice constants and thermal expansion coefficients between the core and shell materials. In this work, we vary the mismatch strain $\delta u$ between zero ("coherent" or "matched" case) and $u_m$ =2.2% (maximal tensile strain at room temperature for the BaTiO$_3$/SrTiO$_3$ interface).

**Elastic properties of the core.** Similar to the case of the SrTiO$_3$ shell, different synthesis paths can lead to a stoichiometric or oxygen-deficient BaTiO$_3$ core [39, 40]. It follows from **Table AI** in **Appendix A** that the elastic stiffness and electrostriction tensor components of stoichiometric bulk crystalline BaTiO$_3$ are of the same order as stoichiometric bulk crystalline SrTiO$_3$; therefore, the stoichiometric BaTiO$_3$ core can be regarded elastically rigid. We estimate that oxygen vacancies with a concentration of more than several volume percent can greatly reduce the effective elastic stresses in the oxygen-deficient BaTiO$_3$ core, making it elastically "soft" (see **Fig. A1** in **Appendix A**). Also, we note that vacancies located in the core can effectively screen elastic fields arising at domain walls and at the core-shell interface, and thus the oxygen-deficient core can be almost elastically isotropic. Below we consider an elastically isotropic core covered with a soft shell in comparison with an elastically rigid anisotropic core-shell pair.

**FEM simulations** are performed in COMSOL@MultiPhysics software, using electrostatics, solid mechanics, and general math (PDE toolbox) modules. The size of the computational region is not less than 40×40×40 nm$^3$, and is commensurate with the cubic unit cell constant (about 0.4 nm) of BaTiO$_3$ at room temperature. The minimal size of a tetrahedral element in a mesh with fine discretization is equal to the unit cell size, 0.4 nm, and the maximal size is (0.8 – 1.2) nm in the core, 1.6 nm in the shell, and 4 nm in the dielectric medium. The dependence on the mesh size is verified by increasing the minimal size to 0.8 nm. We verified that this only results in minor changes in the electric polarization, electric field, and elastic stress and strain, such that the spatial distribution of each of these quantities becomes less smooth (i.e. they contain numerical errors in the form of a small random noise). However, when using these larger cell sizes, all significant details remain visible, and more importantly, the system energy remains essentially the same with an accuracy of about 0.1%.

The mathematical formulation of the problem, comprising electrostatic equations and time-dependent Euler-Lagrange equations with boundary conditions, is given in detail in Ref. [23] and is repeated in **Appendix**

A with the addition of anisotropic elastic properties of the shell, electrostriction, flexoelectricity, and mismatch strains. The simulated system is shown in **Fig. 1a.** Examples of tetrahedral meshes of a core-shell nanoparticle are shown in **Fig. 1b-1c**.

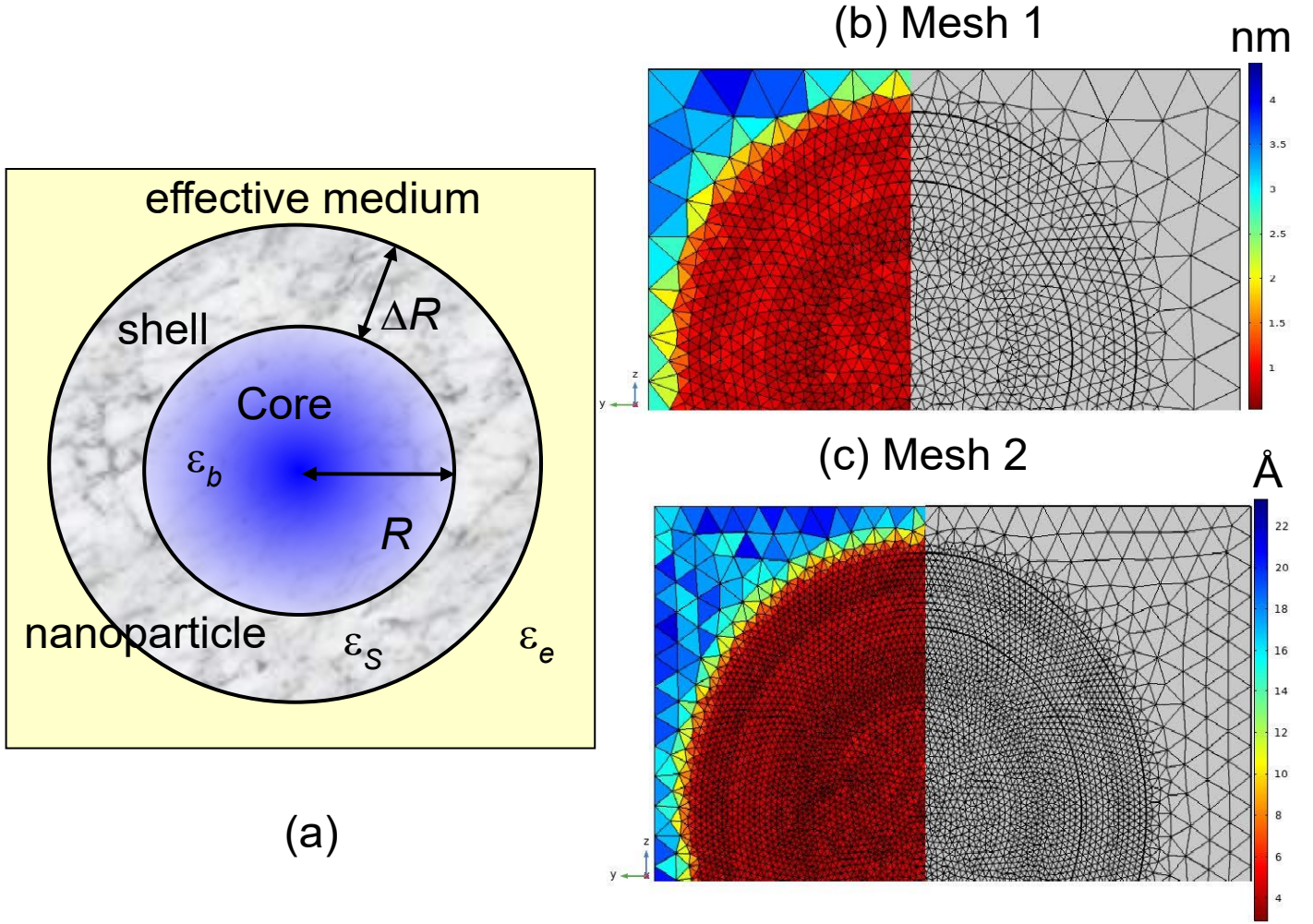

**Figure 1. (a)** A spherical ferroelectric nanoparticle (core) of radius $R$ with background relative dielectric permittivity $\varepsilon_b$, covered with a paraelectric shell of thickness $\Delta R$ with relative dielectric permittivity $\varepsilon_S$, placed in an isotropic dielectric effective medium with relative dielectric permittivity $\varepsilon_e$. Examples of self-adaptive fine **(b)** and hyper-fine **(c)** meshes with different element sizes, a color scale shows the element size in nanometers and angstroms, respectively.

To check the stability and convergence of the numerical algorithm, we use an entirely random distribution of polarization and strain as an initial configuration in the core. In order to obtain a rapid convergence, we use a 180º domain structure with straight domain walls oriented along different crystallographic directions (e.g. [100], [110], or [111]) as the initial distribution of the polarization, which corresponds to the equilibrium domain structure in a BaTiO$_3$ single crystal at room temperature. To facilitate the energy minimization starting from this artificial configuration, small-amplitude random fluctuations of polarization and strain are added to the 180º domain structure in the first time step. These fluctuations are very small in comparison with the values of the spontaneous polarization and strain for a bulk ferroelectric BaTiO$_3$ crystal. Initial values of polarization and strain in the paraelectric shell are zero values, but they are different from zero in the ferroelectric core. The calculation is stopped once the system relaxes to an equilibrium state in which the energy remains constant during subsequent iteration steps. In the vast majority of cases, the relaxation of an entirely random distribution of polarization and the relaxation of the 180º domain structure of a certain orientation leads to the same equilibrium domain structure in the core. When this is not the case, we chose the domain structure corresponding to the lowest energy for further analysis. It turns out that in contrast to the stoichiometric BaTiO$_3$ single crystal in the tetragonal phase, the axis of the vortex-like structure in the nanoparticles is different from the crystallographic axes [100], [010], and [001] (but often close to [110], [101], or [011]). This happens because the small-size (20 nm) core compressed by anisotropic elastic stresses has a significantly lower transition temperature between the tetragonal and orthorhombic phase in comparison with bulk BaTiO$_3$, such that the core becomes close to the orthorhombic (or even rhombohedral) phase at room temperature. An exception to this is the oxygen-deficient BaTiO$_3$ core, where the elastic anisotropy is almost absent.

### III. A FERROELECTRIC CORE COVERED WITH A SOFT TUNABLE SHELL

In this section, we analyze equilibrium distributions of the polarization and the electric and elastic fields in a ferroelectric core covered with a soft tunable shell. Typical equilibrium distributions of the polarization magnitude $P_r = \sqrt{P_1^2 + P_2^2 + P_3^2}$, its component $P_3$, the electric potential $\varphi$, and the radial stress $\sigma$ are shown in **Figs. 2-4**.

For **Fig. 2**, the flexoelectric coupling is zero in both the core and the shell, and the electrostriction anisotropy is small or absent in the core. One can see a thermodynamically stable polarization vortex with a kernel, and the axis of vortex rotation coincides with one of the core pseudo-cubic axis [001]. Similar structures can be stable in the nanoparticles when there is a large number of elastic defects (e.g. mobile oxygen vacancies), whose redistribution is accompanied by a significant decrease in local stresses due to the Vegard effect (see Refs. [37, 38] and **Appendix A** for details). This leads to the fact that a stable three-dimensional (3D) vortex has a polar anisotropy corresponding to the 4mm symmetry group with an equilibrium state in the form of a quasi-two-dimensional vortex with a prolate dipolar quasi-kernel, which is quasi-uniformly polarized. We studied the formation of a similar structure in our recent work [24]. **Figures 2a** and **2b** show that the polarization of the "vortex" part of the core is almost constant in amplitude and rotates in the same plane, while **Figs. 2c-2d** show a clear tetragonal anisotropy of the system: four symmetrical lobes of $P_r$ develop in the equatorial plane and a rounded square-shaped section of the polarized kernel can be seen in the $P_3$ component. In this case, the electrostatic potential reaches a significant value only near the "poles" of the kernel, defined as the points where the polarized kernel touches the surface (see **Fig. 2e**). The remainder of the core polarization is the vortex-like azimuthal distribution of the polarization. There are virtually no elastic stresses in the particle (see **Fig. 2f**), since the migration of elastic dipoles in the stress gradient compensates for the total stress and increases the "effective" elastic compliances of the material (see **Fig.A1** in **Appendix A**). Note, that BPS are absent at zero external field $\boldsymbol{E}_{ext} = 0$ whereas, at $\boldsymbol{E}_{ext} \neq 0$ two diametrically opposite Bloch points appear (i.e., point singularities of the polarization field at $P = 0$), which are located at the vicinity of the core surface where they intersect the kernel [24].

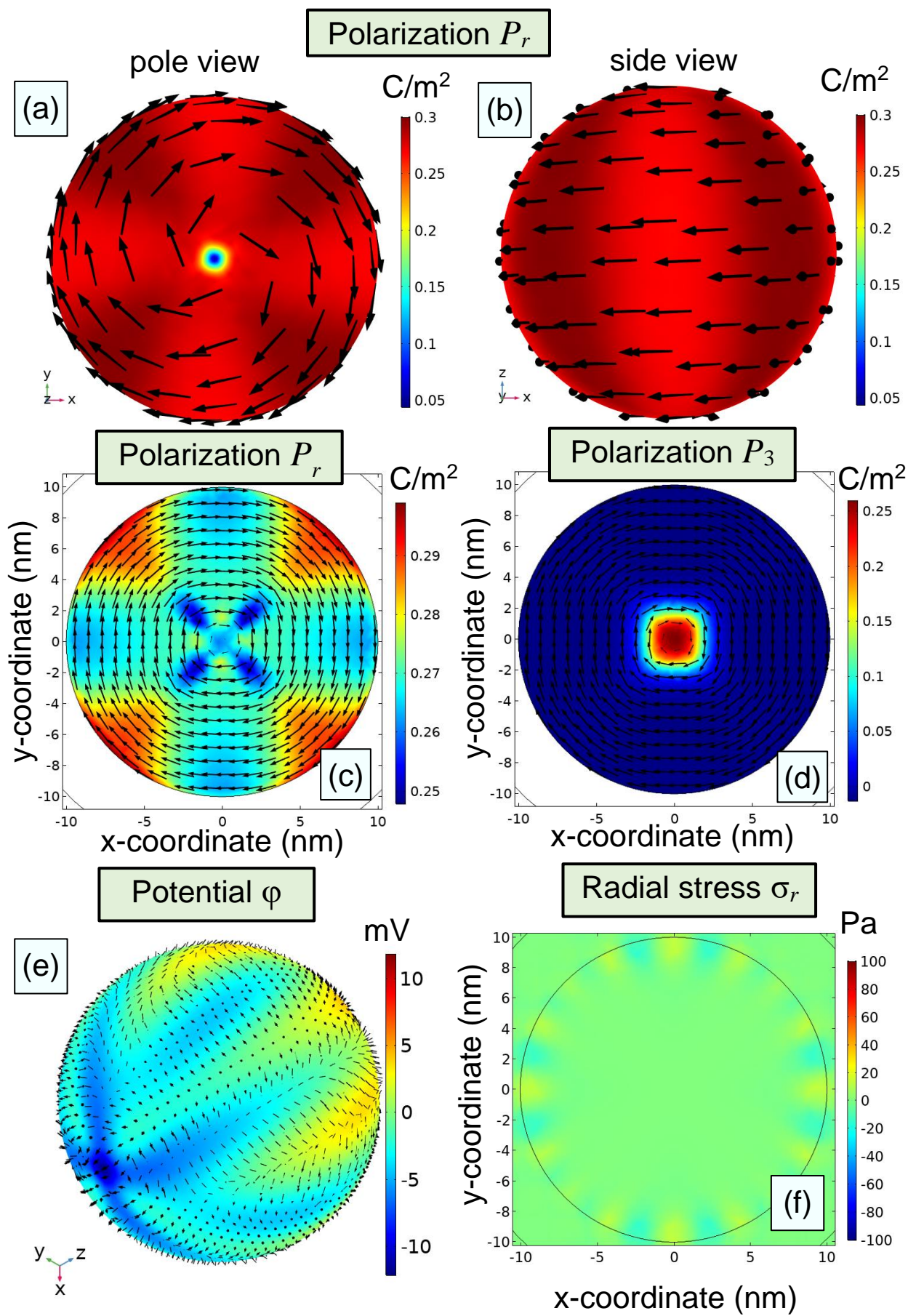

**Figure 2.** Ferroelectric BaTiO$_3$ core covered with a soft tunable shell. Flexoelectric coupling is absent in the core and the shell. Electrostriction anisotropy is small or absent in the core. **(a, b)** Distribution of the polarization magnitude $P_r$ at the core surface ($r = R$). **(c, d)** Distribution of the polarization magnitude $P_r$ and the $P_3$ component in the cross-section {001} perpendicular to the vortex axis pointed along [001]. Black arrows indicate the projection of the polarization vector onto the corresponding surface (a, b, c, and d). **(e)** Electrostatic potential $\varphi$ distribution at the core surface $r = R$. Black arrows indicate the electric field vector at the surface. **(f)** Radial stress in the cross-section {001} of the core. Core radius $R = 10$ nm, shell thickness $\Delta R = 4$ nm, and temperature $T = 298$ K. The tunable shell with high dielectric permittivity $\varepsilon_S = 300$ is regarded as being elastically soft, i.e. its elastic stiffness is negligibly small in comparison with the core values. For other parameters see **Table AI** in **Appendix A.**

The influence of anisotropic internal elastic stresses is demonstrated in **Fig. 3**. The figure uses a new coordinate system with coordinates $t = \frac{x+y}{\sqrt{2}}$, $s = \frac{y-x}{\sqrt{2}}$, and $z$. The coordinate "$t$" is chosen to be parallel to the central axis [110] of the vortex-like polarization structure. Here, the flexoelectric coupling is absent in the core and the shell, but there is a strong and highly anisotropic electrostriction in the core. One can see a regular thermodynamically stable vortex-like polarization structure, which, in contrast to the previous case, develops without a kernel. The actual structure of the polarization distribution is more complicated than a simple vortex or skyrmion (see **Figs. 3a-d**). There is a vortex-like structure near the center of the core, where the polarization rotates in one plane around a fixed axis (see **Figs. 3c-d**). Near the "poles", defined as the intersection points of the vortex axis with the surface, the polarization rotation degenerates into an elongated vortex connecting a pair of 180º domains of the tetragonal phase (see **Figs. 3a-b**). However, the orientation of the domains near the two poles is "crossed", namely the domains are rotated at 90º with respect to one another (see two blue segments in **Figs. 3a**). The surface-closures of these domains contain two diametrically opposite and perpendicular straight segments with a small polarization magnitude near the core surface (see **Figs. 3a**). The guide vectors of the segments are [001] and [-110], which are connected by a central line where $|P| \approx 0$ (see **Figs. 3b**). Stepkova et al. coined the term "Ising line" to describe line defects of this type [41]. Taking into account the importance of different BPS and line singularities for fundamental science and advanced memory applications, we will study the morphology of revealed BPS in more detail in **Section VI**.

"Pseudo-domains", in which several polarization components coexist, appear near the core equatorial plane; their fingerprints are visible on the surface map of the electric potential (see **Figs. 3e**), where the electric field, $\vec{E} = -\vec{\nabla}\varphi$, is shown by black arrows. These domains can be considered as phases with symmetries less than tetragonal. This conclusion is corroborated by the fact that strong elastic stresses arise in the considered vortex structure (see **Figs. 3f**). They are localized both in the near-surface core layer due to the influence of electrostriction anisotropy, and at the walls of pseudo-domains due to a sharp change in the polarization magnitude and direction. The sharp distribution of stresses in different layers of the nanoparticle core determines the different phase composition of the pseudo-domains, because the anisotropic compression or tension can induce the appearance of low-symmetric phases in ferroelectrics.

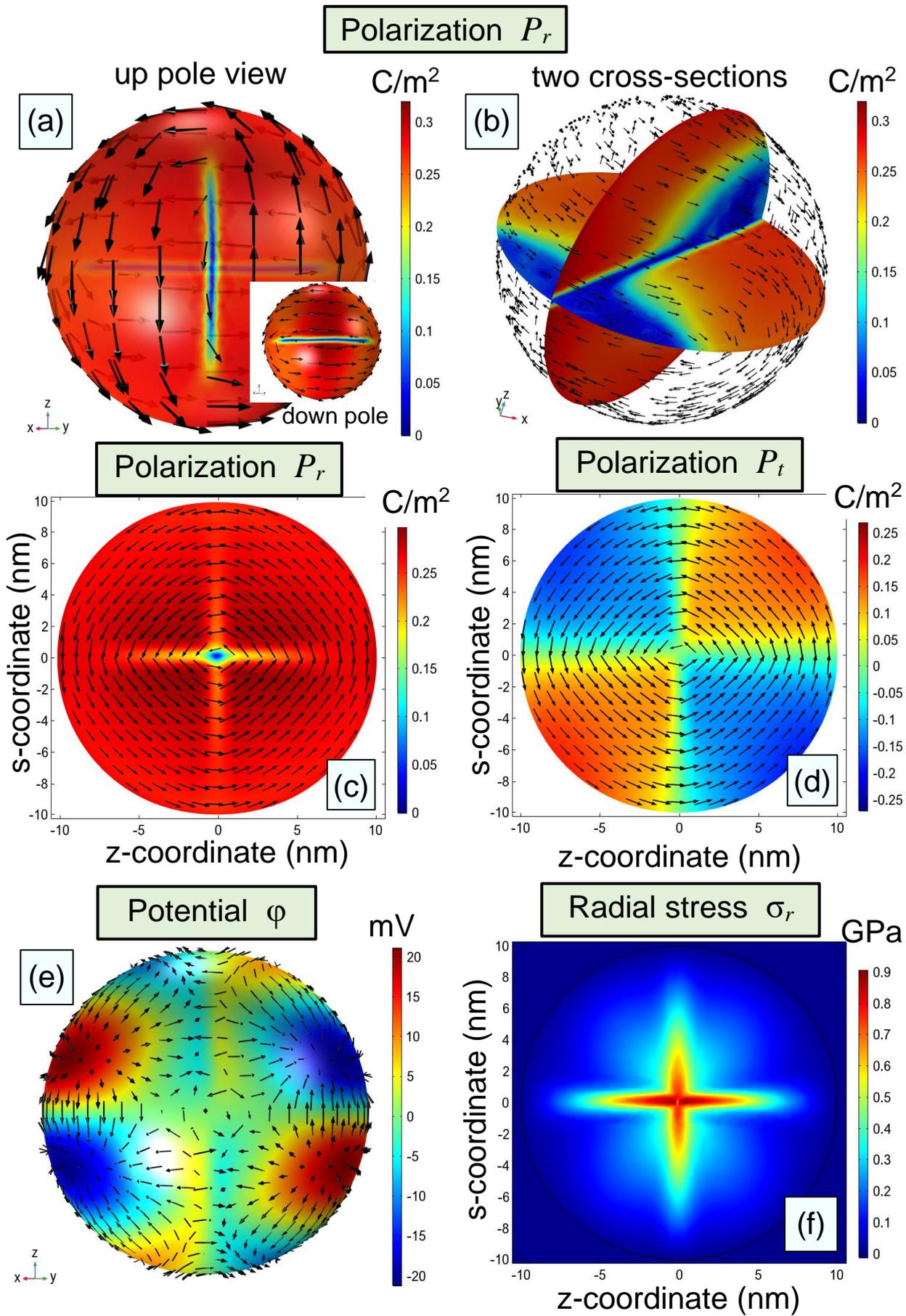

**Figure 3. Ferroelectric BaTiO$_3$ core covered with a soft tunable shell. Flexoelectric coupling is absent in the core and the shell. Electrostriction in the core is anisotropic and high.** Distribution of the polarization magnitude $P_r$ at **(a)** the core surface ($r = R$), and **(b)** on two perpendicular cross-sections. **(c, d)** Distribution of the polarization magnitude $P_r$ and the component $P_t$ on the cross-section

{110} perpendicular to the vortex axis pointed along [110]. Black arrows indicate the projection of the polarization vector onto the corresponding surface (a, b, c, and d). **(e)** Electrostatic potential $\varphi$ distribution at the core surface $r = R$. Black arrows indicate the electric field vector at the surface. **(f)** Radial stress in the cross-section {110} of the core. Core radius $R = 10$ nm, shell thickness $\Delta R = 4$ nm, and temperature $T = 298$ K. The tunable shell with high dielectric permittivity $\varepsilon_S = 300$ is regarded as being elastically soft, i.e. its elastic stiffness is negligibly small in comparison with the core values. For other parameters see **Table AI** in **Appendix A.**

In **Fig. 4**, flexoelectric coupling and electrostriction are anisotropic and high, in both the core and the shell. The figure uses the same coordinate system as in **Fig. 3,** where $t = \frac{x+y}{\sqrt{2}}$, $s = \frac{y-x}{\sqrt{2}}$, and $z$. The coordinate $t$ is chosen to be almost parallel to the axis of the vortex-like polarization structure, whose very small deviation from the [110] direction is caused by the flexoelectric coupling. One can see a thermodynamically stable vortex-like polarization structure, which resembles a double vortex structure in the equatorial plane and has a complex cross-type "curled" morphology, which is very different from the "straight" crossed domain walls shown in **Fig. 3**. Similar to the situation shown in **Fig. 3**, the surface-closures of the crossed domains form two diametrically opposite curved segments of different lengths with a small polarization magnitude near the core surface. In contrast to **Fig. 3,** the central line, where $|P|$ is small, is absent. These segments may contain Bloch points (see **Section VI** for more details).

For zero flexoelectric coupling, the internal electric field (the depolarization field), is very small due to the efficient minimization of charges by the polarization rotation inside the vortex (for details see **Fig. A4a-b** in **Appendix A**). This nearly solenoidal structure develops in the core covered by a soft shell, and the bound charges, $\rho_b = -\vec{\nabla} \cdot \vec{P}$, are virtually zero (see **Fig. A5a-b** in **Appendix A**). The condition div**E**≈0 follows from the zero divergence of the electric displacement **D** and polarization **P**. Small deviations from this condition are due to numerical errors.

To summarize, the room-temperature FEM of particles with the core covered by a **soft shell** shows that a single polarization vortex with a dipolar kernel can be stable in the core in the case of a relatively weak electrostriction coupling. The increase of anisotropic electrostriction coupling causes the disappearance of the vortex, and leads to the formation of 180º flux-closure domains, where the complex cross-type morphology is defined by the flexoelectric coupling in the core.

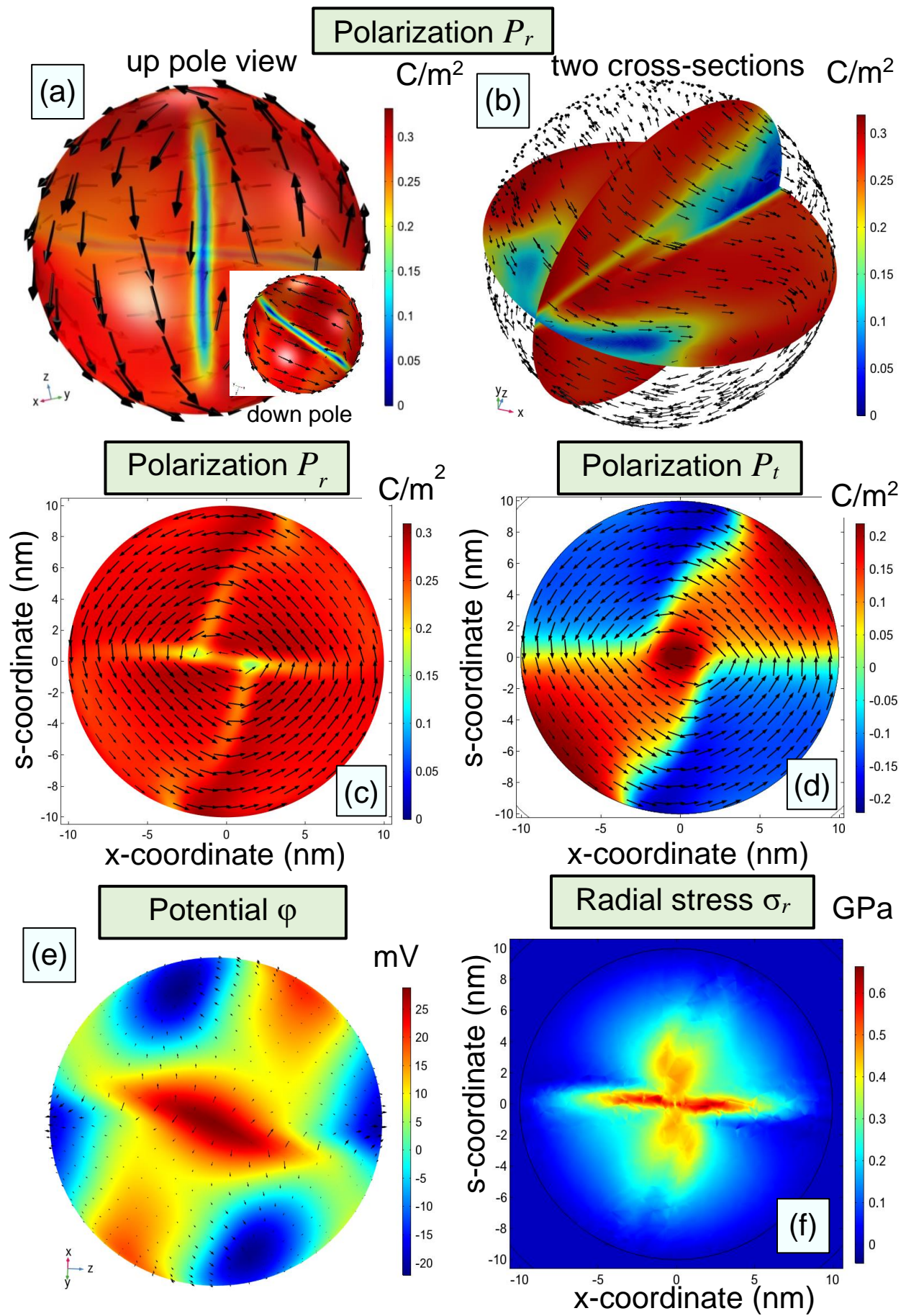

**Figure 4. Ferroelectric BaTiO₃ core covered with a soft tunable shell. Flexoelectric and electrostriction couplings are anisotropic and high in the core.** Distribution of the polarization magnitude $P_r$ at (**a**) the core surface ($r = R$), and (**b**) on two perpendicular cross-sections. (**c, d**) Distribution of the polarization magnitude $P_r$ and its components $P_t$ in the cross-section {110}, which is almost

perpendicular to the vortex axis. Black arrows indicate the projection of the polarization vector onto the corresponding surface (a, b, c, and d). **(e)** Electrostatic potential $\varphi$ distribution at the core surface $r = R$. Black arrows indicate the electric field vector at the surface. **(f)** Radial stress in the cross-section {110} of the core. Core radius $R = 10$ nm, shell thickness $\Delta R = 4$ nm, and $T = 298$ K. The tunable shell with high dielectric permittivity $\varepsilon_S = 300$ is regarded as being elastically soft, i.e., its elastic stiffness is negligibly small in comparison with the core values. For other parameters see **Table AI** in **Appendix A.**

## IV. A FERROELECTRIC CORE COVERED WITH A RIGID SHELL

Our results show that the thermodynamically stable vortex-like flux-closure polarization structure corresponding to a combination of several vortices can be stabilized in the BaTiO$_3$ core, when taking into account the realistic elastic and electrostriction properties of the SrTiO$_3$ shell (see **Table AI** in **Appendix A**). Typical distributions of the polarization magnitude $P_r = \sqrt{P_1^2 + P_2^2 + P_3^2}$, its rotated component $P_\omega$, the electric potential $\varphi$, and the radial stress $\sigma_r$ inside the BaTiO$_3$ core covered with a rigid SrTiO$_3$ shell are shown in **Figs. 5-7**.

Here we use the rotated coordinate frame with the following coordinates: $\xi = (x - y)/\sqrt{2}$, $\psi = (x + y - 2z)/\sqrt{6}$, and $\omega = (x + y + z)/\sqrt{3}$. As can be seen from **Figs. 5-7**, the polarization distribution has a vortex-like structure, where the vortex axis coincides with the $\omega$ axis; we refer to this as the polar axis of the vortex. The component $P_\omega$ is the polarization component along the domain structure axis, which should be equal to zero in a "pure" vortex state without a kernel. Two different projections are used to visualize the polarization structure, namely the top view along [111]-direction at the pole of the vortex-like structure (a) and the side view along the [001]-direction (b). These projections show the formation of three-pointed star-like vortex structures near the poles, where the superposition of these structures forms a six-pointed star. We note that the polarization distribution is not limited to a purely azimuthal rotation of the **P** vector in the plane perpendicular to the polar axis $\omega$. As follows from **Figs. 5-7**, the polar component $P_\omega$ changes its sign along the equator, resulting in a simple toroidal vortex splits into eight domains. The cross-shaped distribution of the polarization in the equatorial plane is due to the strong tetragonal anisotropy of the bulk ferroelectric.

The results in **Fig. 5** show the case where the flexoelectric coupling and misfit strain are zero in the core and shell, while the electrostriction anisotropy is high. Without taking into account the flexoelectric effect, the domain structure of the ferroelectric core covered by a rigid shell consists of six blurred domains. The boundaries between the domains only become relatively sharp in the region near the particle poles, defined as the points at the core surface where the polarization vector modulus drops to zero, i.e. the Bloch points (see **Fig. 5**). Three 120° domains separated by flat walls are observed near the poles. In spite of the similarity of these domains, their orientation and the domain walls are different at the poles; in fact, one group of domains is rotated by 60° with respect to the other. Moving away from each of the poles, the domain walls broaden and blur, resulting in regions that eventually transform into domains with a different orientation. Near the equatorial plane, all six domains are equivalent, such that the configuration of the polarization vector becomes vortex-like. In this case the symmetry of the walls is more complicated than that of 120° domains, because of the pronounced polarization component along the polar axis of the core. One can see a regular, thermodynamically stable vortex-like polarization structure with the shape of a six-ray star.

A case in which both, flexoelectric coupling and electrostriction anisotropy are high in the core and the shell, but without mismatch strain ($u_m = 0$) at the core-shell interface is shown in **Fig. 6**. It can be seen that the flexoelectric coupling increases the vorticity (compare **Fig. 6d** and **5d**). A contrasting example, where the flexoelectric coupling and electrostriction anisotropy are anisotropic and high in the core and the shell, and a tensile mismatch strain $u_m = 2.2\%$ exists at the core-shell interface, is shown in **Fig. 7**. In this latter case, it is seen that the mismatch strain compensates the curling of the flux-closure domains in the core confined by the elastically-anisotropic rigid shell. The flexoelectric coupling and mismatch strain result in a relatively strong electric field well-localized at the core surface, and consequently, the bound charges can be considered as surface charges (see **Fig. A4d** and **A5d** in **Appendix A**).

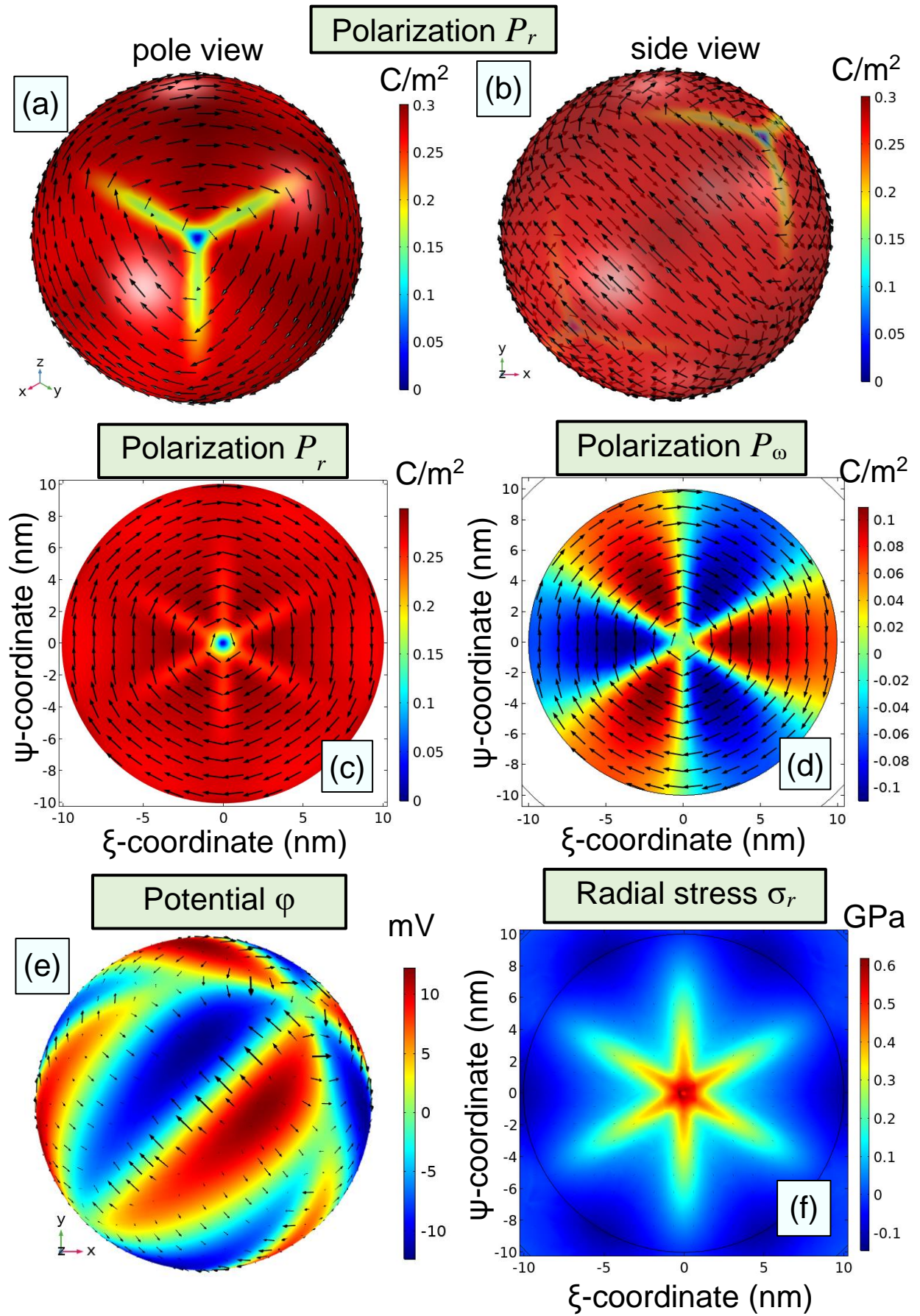

**Figure 5**. Ferroelectric BaTiO$_3$ core covered with a rigid SrTiO$_3$ shell. Flexoelectric coupling is absent. Electrostriction is anisotropic and high in the core and the shell. A misfit strain between the shell and core is absent. **(a, b)** Distribution of the polarization magnitude $P_r$ at the core surface ($r = R$). **(c, d)** Distribution of the polarization amplitude $P_r$ and the component $P_\omega$ on the cross-section {111} perpendicular to the vortex axis pointed along [111]. Black arrows indicate the projection of the polarization vector

onto the corresponding surface (a, b, c, and d). **(e)** Electrostatic potential $\varphi$ distribution at the core surface $r = R$. Black arrows indicate the electric field vector at the surface. **(f)** Radial stress in the cross-section {111} of the core. Core radius $R = 10$ nm, shell thickness $\Delta R = 4$ nm, $T = 298$ K. For other parameters see **Table AI** in **Appendix A.**

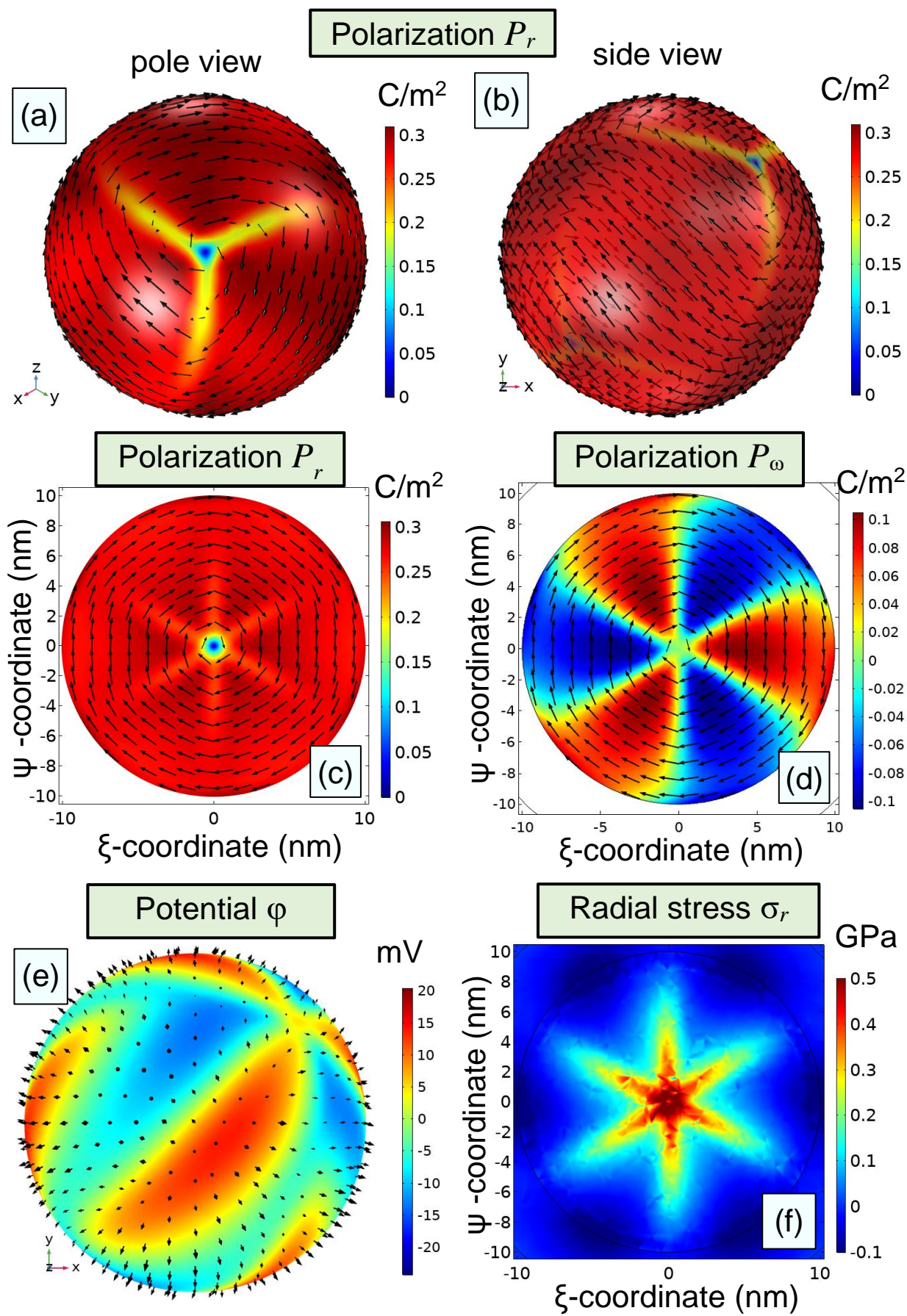

**Figure 6**. **Ferroelectric BaTiO₃ core covered with a rigid SrTiO₃ shell. Flexoelectric and electrostriction coupling are anisotropic and high in the core and the shell. A misfit strain between the shell and core is absent.** **(a, b)** Distribution of the polarization magnitude $P_r$ at the core surface ($r = R$). **(c, d)** Distribution of the polarization amplitude $P_r$ and of the component $P_\omega$ on the cross-section {111} perpendicular to the vortex axis pointed along [111]. Black arrows indicate the projection of the polarization vector onto the corresponding surface (a, b, c, and d). **(e)** Electrostatic potential $\varphi$ distribution at the core surface $r = R$. Black arrows indicate the electric field vector at the surface. **(f)** Radial stress in the cross-section {111} of the core. Core radius $R = 10$ nm, shell thickness $\Delta R = 4$ nm, $T = 298$ K. For other parameters see **Table AI** in **Appendix A.**

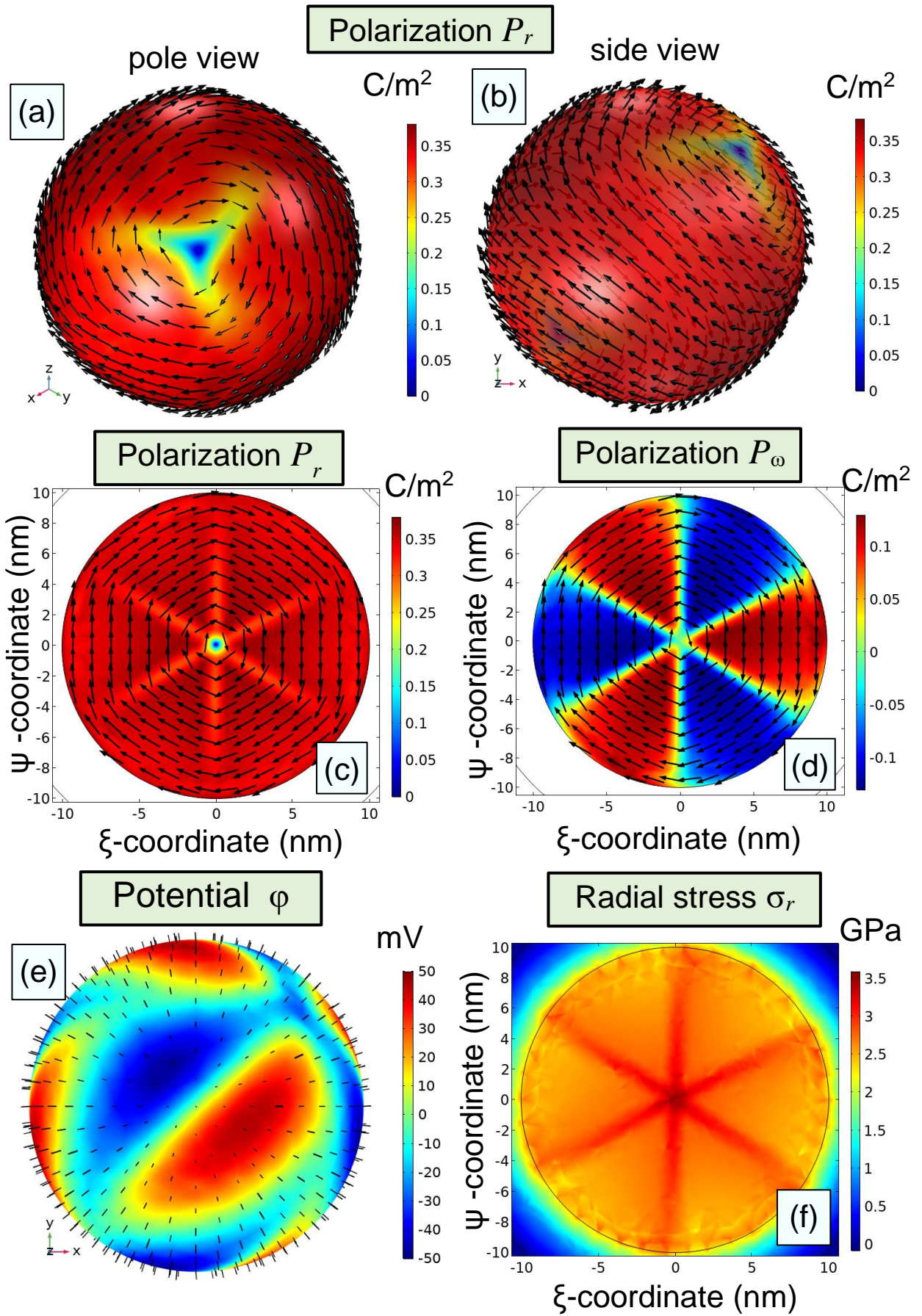

**Figure 7**. **Ferroelectric BaTiO$_3$ core covered with a rigid SrTiO$_3$ shell. Flexoelectric and electrostriction coupling are anisotropic and high in the core and the shell. A misfit strain between the shell and core (~2.2%) is taken into account.** **(a, b)** Distribution of the polarization magnitude $P_r$ at the core surface ($r = R$). **(c, d)** Distribution of the polarization amplitude $P_r$ and the component $P_\omega$ on the cross-section {111} perpendicular to the vortex axis pointed along [111]. Black arrows indicate the projection of the polarization

vector onto the corresponding surface (a, b, c, and d). **(e)** Electrostatic potential $\varphi$ distribution at the core surface $r = R$. Black arrows indicate the electric field vector at the surface. **(f)** Radial stress in the cross-section {111} of the core. Core radius $R = 10$ nm, shell thickness $\Delta R = 4$ nm, and $T = 298$ K. For other parameters see **Table AI** in **Appendix A.**

## V. PHASE DIAGRAMS AND THEIR DISCUSSION

For first-order ferroelectric phase transitions in BaTiO$_3$ crystals, one should distinguish the difference between the cubic paraelectric (**PE**), tetragonal (**FE$_T$**), orthorhombic (**FE$_O$**), and rhombohedral (**FE$_R$**) ferroelectric (**FE**) phases. The transition temperature between the PE and FE phases of the BaTiO$_3$ core is defined by the condition of free energies being equal in the phases, $G_{PE} = G_{FE}$. The boundaries between the PE and FE phases can depend on the core radius $R$, temperature $T$, flexoelectric tensor components $F_{ij}$, and mismatch strain $u_m$. Note that $G_{PE} = 0$ for the case $u_m = 0$, and $G_{PE} \sim u_m^2 > 0$ for $u_m \neq 0$. Below we discuss the results for particles with a fixed shell thickness, $\Delta R = 4$ nm, and a range of core sizes, $1$ nm $\leq R \leq 25$ nm, since the manifestation of size effects for bigger particles is rather weak as they tend to become single-domain in the central part of the core.

Our FEM calculations show that the BaTiO$_3$ core with $R > 1$ nm covered by a soft SrTiO$_3$ shell is mostly in the FE$_T$ phase in the vicinity of the PE phase. The FE$_T$, FE$_O$, and FE$_R$ phases coexist for $1.5$ nm $< R < 2$ nm, and the fraction of the core in the FE$_R$ phase increases as $R$ decreases. The core is almost completely in the FE$_R$ phase for $R < 1.5$ nm. These trends are in a good agreement with synchrotron XRD experiments reported by Zhu et al. [42], who observed the sequence of FE$_T$, FE$_O$, and FE$_R$ phases, as well as their coexistence and reappearance in BaTiO$_3$ nanospheres with sizes below 20 nm (see e.g. Table I in [42]). The appearance of the FE$_R$ phase with decreasing $R$ agrees with a previous study [43] in which a polarization gradient was not considered.

Counterintuitively, the core covered by a rigid shell is generally in the FE$_R$ phase in the immediate vicinity of the PE phase. Although the core domain structure can reveal features of the FE$_O$ phase in a minority of cases, it is never observed to be in the FE$_T$ phase. This result neither depends on the core radius, the flexoelectric coupling strength, nor, most surprisingly, on the type (compressive, zero, or tensile) of mismatch strain. A possible explanation of the effect could be related to the fact that the rigid SrTiO$_3$ shell is elastically anisotropic, and that anisotropy enforces the direction of the axis of vortex-type structures to be close to [111] or [110], but never along [001].

Typical phase diagrams of core-shell nanoparticles as a function of temperature $T$ and core radius $R$, calculated for the BaTiO$_3$ core covered with soft or rigid SrTiO$_3$ tunable shells, are presented in **Figs. 8-9**. The boundaries between the FE and PE phases are shown by fitting curves to the symbols presenting FEM data points. The $R$-dependence of the PE-FE transition temperatures $T_{pt}(R)$, calculated by FEM, can be fitted with the analytical expression [28-31, 44]:

$$T_{pt}(R) = T_b \left(1 - \frac{R_g^2}{R^2} - \frac{R_e R + R_m R_s}{R(R+R_s)}\right), \tag{1}$$

where $T_b = 384$ K is the bulk Curie temperature; and the fitting parameters $R_i$, with the subscript $i = g, e, m,$ and $s$, are given in **Table I.** The fit is accurate enough, and is in good agreement with FEM results. The fitting is particularly accurate for a rigid shell without misfit strain. The critical core radius, $R_{cr}$, determined from the condition $T_{pt}(R_{cr}) = 0$, is $R_{cr} = \frac{1}{2}\left(R_e + \sqrt{R_e^2 + 4R_g^2}\right)$ at $u_m = 0$, when $R_m = 0$. This expression, which is exact only at $u_m = 0$, is a good approximation for $|u_m| < 0.1\%$, since the last term in Eq.(1) can be neglected for $R < R_S$, where $R_S \gg R_{cr}$.

**Table I.** Fitting parameters for $T_{pt}(R)$ defined from Eq.(1)

| System description | Condition for $T_{pt}(R)$ determination | $R_e$ (nm) | $R_g$ (nm) | $R_m$ (nm) | $R_S$ (nm) |
|---|---|---|---|---|---|
| **Soft shell, Figure 8a** | | | | | |
| $F_{ij} = 0, u_m = 0$ * (black curve) | $G_{FE} = G_{PE} = 0$ ** | 0.109 | 0.66 | N/A | 0 |
| $F_{ij} \neq 0, u_m = 0$ (blue curve) | $G_{FE} = G_{PE} = 0$ | 0.084 (R)+ <br> 0.105 (T)+ | 0.66 (R) <br> 0.58 (T) | N/A | 0 |
| $F_{ij} = 0, u_m = 0$ (red curve) | $G_{FE} = G_{PE} + k_B T = k_B T$ | 0.067 | 0.65 | N/A | 0 |
| $F_{ij} \neq 0, u_m = 0$ (green curve) | $G_{FE} = G_{PE} + k_B T = k_B T$ | 0.044 (R) <br> 0.098 (T) | 0.64 (R) <br> 0.50 (T) | N/A | 0 |
| **Rigid shell, zero misfit, Figure 8b** | | | | | |
| $F_{ij} = 0, u_m = 0$ (black curve) | $G_{FE} = G_{PE} = 0$ | 0.126 | 0.66 | N/A | 0 |
| $F_{ij} \neq 0, u_m = 0$ (blue curve) | $G_{FE} = G_{PE} = 0$ | 0.127 | 0.67 | N/A | 0 |
| $F_{ij} = 0, u_m = 0$ (red curve) | $G_{FE} = G_{PE} + k_B T = k_B T$ | 0.077 | 0.65 | N/A | 0 |
| $F_{ij} \neq 0, u_m = 0$ (green curve) | $G_{FE} = G_{PE} + k_B T = k_B T$ | 0.069 | 0.66 | N/A | 0 |
| **Rigid shell, misfit strain, Figure 9** | | | | | |
| $|F_{ij}| \leq 6, u_m = -0.5\%$ (red curve) | $G_{FE} = G_{PE} \sim u_m^2$ | 4.40 | 0.711 | -0.0075 | 21.6 |
| $|F_{ij}| \leq 6, u_m = 0$ (black curve) | $G_{FE} = G_{PE} = 0$ | 0.127 | 0.67 | N/A | 0 |
| $|F_{ij}| \leq 6, u_m = 0.5\%$ (blue curve) | $G_{FE} = G_{PE} \sim u_m^2$ | -1.70 | 0.402 | 0.63 | 6.63 |

* $F_{ij}$ is a flexoelectric tensor in Voigt notations.

** Note that $G_{PE} = 0$ for the case $u_m = 0$, and $G_{PE} \sim u_m^2 > 0$ for $u_m \neq 0$.

+ "T" and "R" are the abbreviations for tetragonal (FE$_T$) and rhombohedral (FE$_R$) ferroelectric phases, respectively.

It is worth noting that the LGD approach in the form we used here, without inclusion of thermal fluctuations, is not applicable for sizes smaller than five lattice constants. The value of $R_{cr}$ becomes smaller than 1 nm (i.e. below the validity limit of the LGD approach) at $T < 200$ K (see insets to **Figs. 8-9**, from which one can determine $R_{cr}$ using the dependence of $T_{pt}(R)$ on $1/R$). Since Eq. (1) was derived for single-domain ferroelectric nanoparticles without any shell, its relevance for the core-shell nanoparticle with a vortex-type domain structure indicates that the domain formation only influences the values of $R_i$, but does not alter the "universal" functional form (1).

Unlike the case of a single-domain or homogeneously polarized core, mainly considered in Refs. [28-31], we were unable to derive approximate analytical expressions showing how the fitting parameters $R_i$ depend on the polarization gradient, electrostriction, flexoelectric tensor components, mismatch strain, and shell thickness.

However, we could establish the physical meaning of the different terms in Eq. (1). The first term, $\frac{R_g^2}{R^2}$, is related to the correlation size effect caused by the polarization gradient energy and a very small depolarization energy (negligible for the case of a pure vortex). The LGD approach gives an estimate for $R_g^2 \sim \frac{g_{44}}{\alpha_T T_b}$, where $g_{44}$ is the polarization gradient coefficient and $\alpha_T$ is the inverse Curie-Weiss constant (see **Table I**, **Appendix A**). The second $R$-dependent term, $\frac{R_e R + R_m R_s}{R(R+R_s)}$, arises from the joint action of the elastic self-clamping of the core, "external" core clamping (contraction or tension) by the shell via electrostriction, flexoelectric coupling, and mismatch strain at the core-shell surface. The value $R_e$ is proportional to $\frac{Q \sigma_{eff}}{\alpha_T T_b}$, where $Q$ is the combination of electrostriction coupling constants $(Q_{11} + 2Q_{12})$, and $\sigma_{eff}$ is the effective value of spontaneous stresses including those induced by the domain structure. The parameter $R_m \sim \frac{Q}{\alpha_T T_b} f(u_m)$ is proportional to the mismatch strain $u_m$, but not in a simple (e.g. linear) manner. The value $R_s$ is proportional, but not equal, to the shell thickness $\Delta R$. Note that the terms proportional to $\frac{1}{R}$ and $\frac{1}{R^2}$ can also originate from surface tension [28-31, 34-24] and surface bond contraction [32, 33], respectively.

A mismatch strain between the shell and core is absent ($u_m = 0$) for the diagrams in **Fig. 8.** Black and blue curves in **Fig. 8** are the FE-PE transition temperatures $T_{pt}(R)$ determined by using the condition of zero free energy, $G = 0$; red and green curves are the transition temperatures $T_{pt}(R)$ determined by using the condition $G = k_B T$, where $k_B T$ is the thermal energy (see right top legend to **Fig. 8** and **Fig. B1** in **Appendix B** for details). The temperature difference between these pairs of curves (about 10 K) illustrates the possibility to observe a thermal hysteresis in core-shell nanoparticles, as described in a scheme shown in the right bottom inset of **Fig. 8**. Every $T_{pt}(R)$ curve increases monotonically with increasing $R$ and then saturates to the bulk value.

Flexoelectric coupling is zero for black and red curves ($F_{ij} = 0$), and nonzero ($F_{ij} \neq 0$) for blue and green curves. There is a small splitting for the green and blue curves shown in **Fig. 8a**, which is due to the coexistence of the T, O, and R phases over the radius range $(1.5 - 2)$ nm. The splitting is the most pronounced for the green curves calculated for $G = k_B T$ and $F_{ij} \neq 0$. The influence of the flexoelectric coupling on the $T_{pt}(R)$ of the core covered by a rigid shell is very small: black and blue curves for $G = 0$, and red and green curves for $G = k_B T$, where $k_B T$ is the thermal energy, differing by less than 0.5% in **Fig. 8b** (see also **Fig. B1** in **Appendix B** for details). In contrast, the flexoelectric coupling increases the PE-FE transition temperature (up to 5-7 K) for the core covered by a soft shell. This is evident from a comparison of the black and blue curves for $G = 0$ with the red and green curves for $G = k_B T$ in **Fig. 8a**. A simple explanation of the temperature increase is the "flexibility" of the soft shell, where the flexoelectric coupling increases the de-localization of the stress gradients (compare e.g. **Fig. 3f** with **Fig. 4f**). Also, the flexoelectric coupling strongly decreases the value of $R_e$ for the case of the soft shell, and has a slight influence on $R_g$ values (see the values in **Table I**).

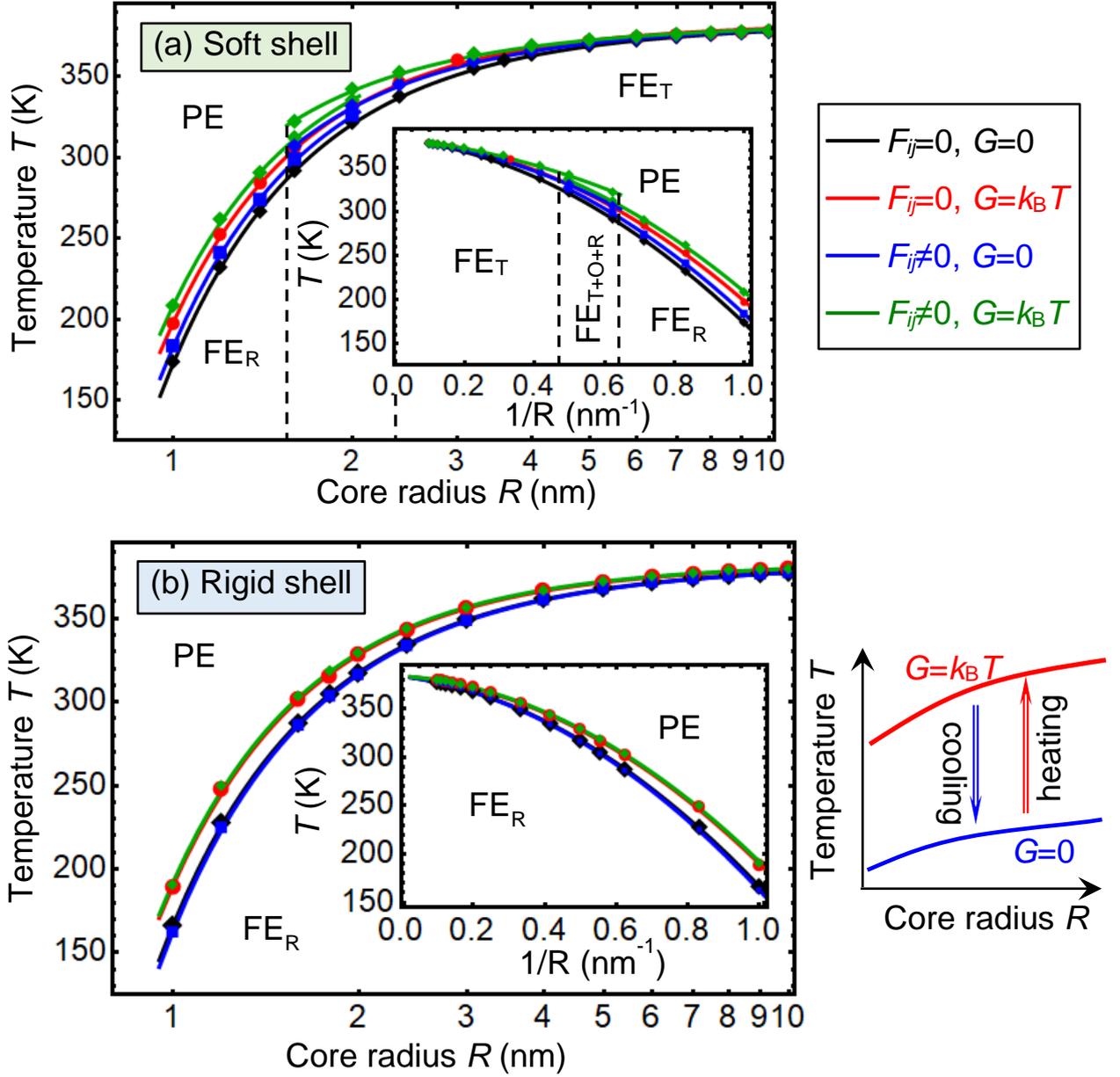

**Figure 8**. Phase diagrams of core-shell nanoparticles in coordinates of temperature $T$ and core radius $R$, calculated by FEM for a BaTiO$_3$ core covered with a soft (**a**) or rigid (**b**) SrTiO$_3$ tunable shell of thickness $\Delta R = 4$ nm. It is assumed that there is no mismatch strain between the shell and core, $u_m = 0$. The boundaries between the ferroelectric (FE) and paraelectric (PE) phases are shown by solid curves with symbols. The symbols are calculated by FEM and the curves are calculated using Eq.(1). Black and blue curves are the FE-PE transition temperatures $T_{pt}(R)$ defined from the condition of zero free energy $G = 0$; red and green curves are $T_{pt}(R)$ defined from the condition $G = k_B T$. Flexoelectric coupling is zero for black and red curves ($F_{ij} = 0$), and nonzero for blue and green curves ($F_{ij} \neq 0$). Insets show the dependence of $T_{pt}(R)$ on $1/R$. Material parameters of the BaTiO$_3$ core and the SrTiO$_3$ shell are listed in **Table AI** in **Appendix A.** Nonzero $F_{ij}$ values, listed in **Table AI,** are the following $F_{11} = 2 \ 10^{-11}$ m$^3$/C, $F_{12} = 1.8 \ 10^{-11}$ m$^3$/C, and $F_{44} = 6 \ 10^{-11}$ m$^3$/C.

Although the increase of $T_{pt}(R)$ caused by the flexoelectric effect is relatively small, it is important to understand its nature and compare the effect in core-shell nanoparticles with other geometries. Note that the flexoelectric coupling formally leads to the renormalization of the polarization gradient coefficient, $g'_{ijkl}$ in the gradient energy $g_{grad} = \frac{g_{ijkl}}{2} \frac{\partial P_i}{\partial x_j} \frac{\partial P_k}{\partial x_l}$ (see **Appendix A2** for details). The renormalization has different signs for the diagonal and non-diagonal components, but for the considered case of multiaxial ferroelectric perovskites the flexoelectric effect typically increases $g_{11}$ and decreases $g_{44}$. For a cubic symmetry of the paraelectric phase (as with BaTiO$_3$), the trend $g'_{11} > g_{11}$ and $g'_{44} < g_{44}$ is responsible for an increase of the charged domain walls' width, and a decrease of the uncharged domain structures' width, such as with vortex-like configurations. The formation of uncharged domain configurations, which are the most common and are significantly more preferable from an energetic viewpoint [45, 46], is affected by the flexoelectricity. In particular, the flexoelectricity induces the domain wall curvature and meandering in multiaxial ferroelectrics, and facilitates labyrinthine domain configurations in uniaxial ferroelectrics at $g'_{ijkl} < g^{cr}_{ijkl}$ (see e.g. Refs. [47, 48, 49]). In addition to influencing the wall shape, the flexoelectricity (due to the condition $g'_{44} < g_{44}$) increases (but not very strongly) the transition temperature from the ferroelectric to paraelectric phase (see e.g. [47, 50]). Another role of flexoelectricity comes from the inhomogeneous boundary conditions in strained nanoparticles [see e.g. Eq. (A.5) in **Appendix A1** for details]. The inhomogeneity, proportional to the flexoelectric coupling strength, can lead to the appearance of built-in inhomogeneous fields, so-called "flexo-electric" fields, which can blur out the FE-PE phase transition [47-50].

Phase diagrams of core-shell nanoparticles in coordinates of temperature $T$ and core radius $R$, calculated by FEM for compressive ($u_m = -0.5\%$, blue diamonds), zero ($u_m = 0$, black triangles), and tensile ($u_m = +0.5\%$, red circles, and $u_m = +2.2\%$, green dots) mismatch strains are shown in **Fig. 9.** Corresponding FE-PE transition temperatures $T_{pt}(R)$ are defined from the condition of the FE and PE free energies equality, $G_{FE} = G_{PE}$ (see the last two lines in **Table I**, and **Figs. B2-3** in **Appendix B** for details). Solid curves, which are interpolations using Eq. (1), correspond to first order FE-PE phase transition, except for the case of tensile mismatch strain, where the phase transition order changes in the tricritical point.

The effect of mismatch strain between the core and shell of the nanoparticle is similar to isotropic tension or compression. Furthermore, compressive strains ($u_m < 0$) significantly decrease $T_{pt}(R)$ for $R \leq 25$ nm (compare the black and blue curves in **Fig. 9**), while tensile strains ($u_m > 0$) significantly increase $T_{cr}(R)$ (compare the black curve with the red and green curves in **Fig. 9**). Note that this result principally differs from the situation in thin strained BaTiO$_3$ films, where $u_m < 0$ supports an out-of-plane polarization direction (corresponding to the so-called FEc phase) and increases the FEc-PE transition temperature, whereas $u_m > 0$ supports an in-plane polarization direction (corresponding to the so-called FEaa phase) and increases the FEaa-PE transition temperature [51, 52]. For a fixed shell thickness ($\Delta R = 4$ nm) and core radius $R > 50$ nm, the influence of the mismatch strain decreases gradually, and the curves calculated for different $u_m$ values converge with an increasing core radius (see inset to **Fig. 9**). Note that the quantitative difference between tensile and compressive strains of the same magnitude, $u_m = +0.5\%$ and $u_m = -0.5\%$, corresponds to an increase of the transition temperature by less than 70 K at $u_m = +0.5\%$, in comparison with a decrease by more than 100 K at $u_m = -0.5\%$. The reason for this asymmetry is the strong anisotropy of the elastic and electrostriction tensors, which is well-studied for thin ferroelectric films [51]. The influence of the flexoelectric coupling (at least for the flexoelectric coefficients $|F_{ij}| \leq 6$) is negligibly small.

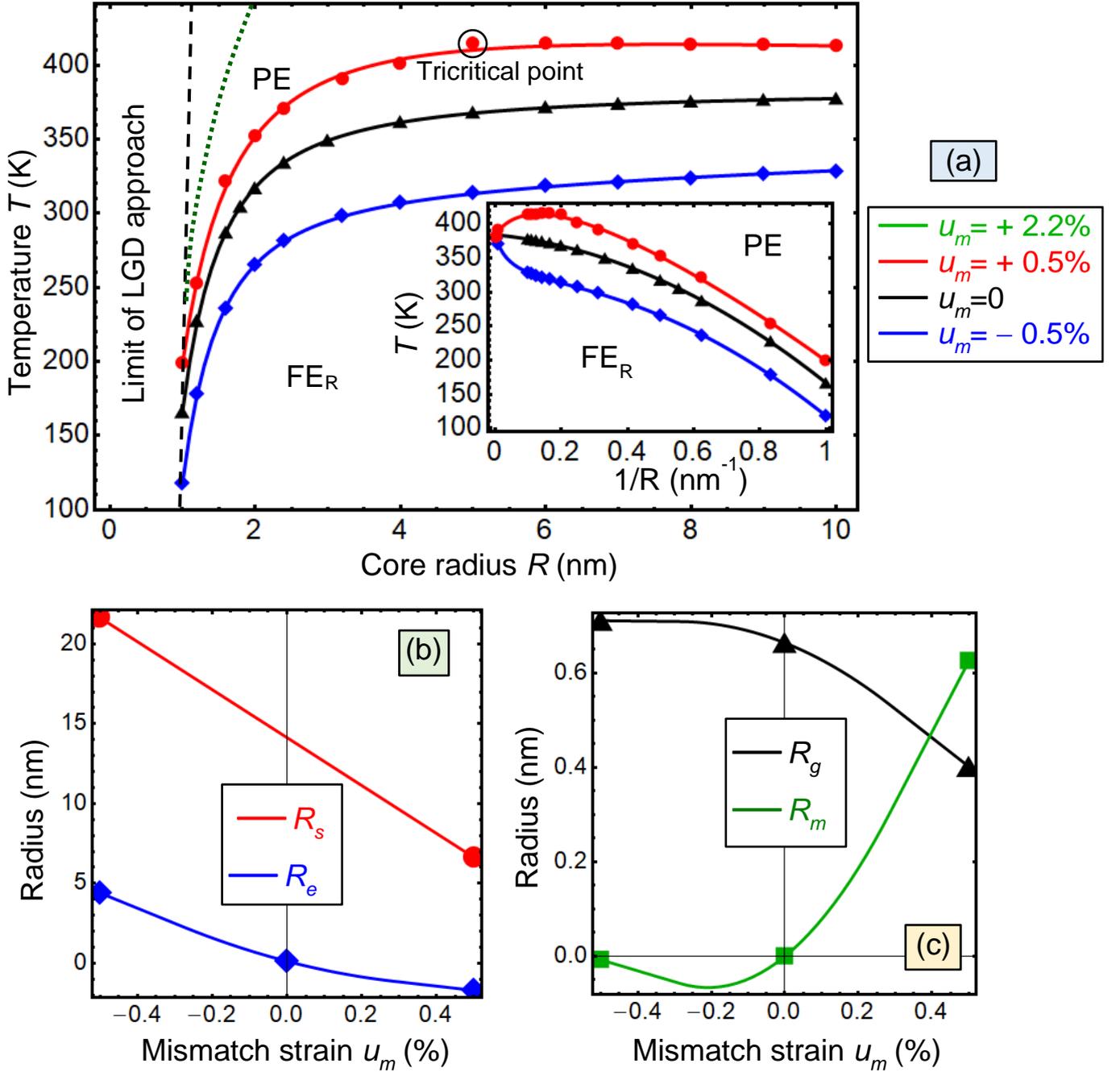

**Figure 9.** (a) Phase diagrams of core-shell nanoparticles in coordinates of temperature $T$ and core radius $R$, calculated by FEM for a BaTiO$_3$ core covered with a rigid SrTiO$_3$ shell with different values of mismatch strain ($u_m$) between the core and shell: $u_m = -0.5\%$ (blue diamonds), 0 (black triangles), +0.5% (red circles), and +2.2% (dots). Corresponding FE-PE transition temperatures $T_{pt}(R)$ are determined from the condition of the equality of FE and PE free energies. The inset shows the dependence of $T_{pt}(R)$ on $1/R$. Solid curves are interpolations given by Eq. (1). Shell thickness $\Delta R = 4$ nm, material parameters of BaTiO$_3$ core and SrTiO$_3$ shell are listed in **Table AI** in **Appendix A.** The dependence of the fitting parameter $R_e$ and $R_s$ (b), $R_g$ and $R_m$ (c) on the mismatch strain.

## VI. POSSIBLE APPLICATIONS OF THEORETICAL RESULTS

Ensembles of core-shell ferroelectric nanoparticles, whose polarization arranges in a vortex-like structure either with a conductive dipolar kernel [24] or with other types of conductive domain boundaries [53, 54], including those with BPS, can be considered as promising candidates for nanoelectronic devices, where the conductive parts can be nanochannels in versatile field effect transistors and logic units. If the nanoparticles are placed in a soft matter environment, the voltage applied between the gates can rotate or shift the particle (in order to rotate/move the conductive channel). The stability of the vortex-like structure and its ability to exhibit rotations are advantages for nano-device operation. The drawback of core-shell nanoparticles is a relatively low operation speed due to the sluggishness of spherical rotation and/or translational motion towards the gate(s).

Note that BPS play a crucial role in the switching of ferromagnetic vortex states, which are recognized candidates for advanced non-volatile RAM units with high storage density, low-power, and high operation speed. In such magnetic vortex states, BPS mediate an ultra-fast magnetic switching that has been calculated and realized in practice [55, 56, 57, 58]. The ferroelectric vortex states with BPS also may become attractive, because hypothetical possibilities of the ultra-fast ferroelectric switching have been predicted recently [59].

The domain morphology shown in **Figs. 3a-c**, **4a-c, 5a-c, 6a-c,** and **7a-c**, shows that the polarization magnitude $P_r$ is very small in two diametrically opposite points (or segments) located just under the core surface, and/or in the core center. This condition gives us a hope to confirm the existence of BPS in these regions by performing a more careful FEM analysis, and to better understand the influence of elastic anisotropy, electrostriction and flexoelectric coupling, and mismatch strain on the BPS morphology in core-shell ferroelectric nanoparticles. FEM analysis is applied to search the intersection regions of the polarization components' isosurfaces, $P_1 = 0$, $P_2 = 0$, and $P_3 = 0$ (see **Figure A6** in **Appendix A**). The triple intersection corresponds to the condition $|P| = P_r = 0$, and thus indicates the position of a Bloch point. Different BPS morphologies in a ferroelectric core covered with a soft or rigid shell in various elastic conditions are shown in **Figs. 10.** BPS are absent in the case of the stress-free core covered with an elastically isotropic soft shell with zero external electric field (compare **Fig. 10a** with Fig. 5 in Ref. [24], where two diametrically opposite Bloch points appeared at a small distance from the core surface at a nonzero external electric field).

Anisotropic electrostriction coupling strongly changes the morphology of the polarization isosurfaces in the core (see **Fig. A6b-f**), and flexoelectric coupling induces an additional curvature and twist of the isosurfaces (see **Fig. A6c** and **A6e-f**). The chains of aligned Bloch points in **Fig. 10b** (in the case of a soft shell) and **Fig. 10d** (in the case of a rigid shell) display one-dimensional topological line defects with $|P| = 0$. In a recent article, Stepkova et al. have coined the term "Ising line" to describe line defects of this type [41]. Both of these cases, **Fig. 10b** and **Fig. 10d**, are calculated without flexoelectric coupling ($F_{ij} = 0$) in the core and includes anisotropic electrostriction ($Q_{11} \neq Q_{12}$).

When a flexoelectric coupling is included ($F_{ij} \neq 0$), the Ising line (shown in **Fig. 10b**) disappears, and two Bloch points appear in its place (see **Fig. 10c**). They are located at the opposite sides of the domain wall very close to the surface; but not at diametrically opposite positions, which are sensitive to the sign and magnitude of the $F_{ij}$. We suspect that the asymmetric morphology of the two Bloch points is caused by the flexoelectric effect, this phenomenon will be the topic of our future study.

The shell rigidity very strongly flattens the twisted morphology of the polarization isosurfaces (compare **Fig. A6d** with **Fig. A6a-c**), while the inclusion of flexoelectric coupling leads to the reappearance of a slight twist (compare **Fig. A6e-f** with **Fig. A6c** in **Appendix A**). However, the twist and mutual shift of the isosurfaces induced by the flexoelectric coupling in a core covered with a rigid shell prevents the formation of an Ising line. The line defect transforms into a single Bloch point located in the core center (compare **Fig. 10e-f** with **Fig. 10d**). To the best of our knowledge, an analogue of Ising lines does not exist in ferromagnetism.

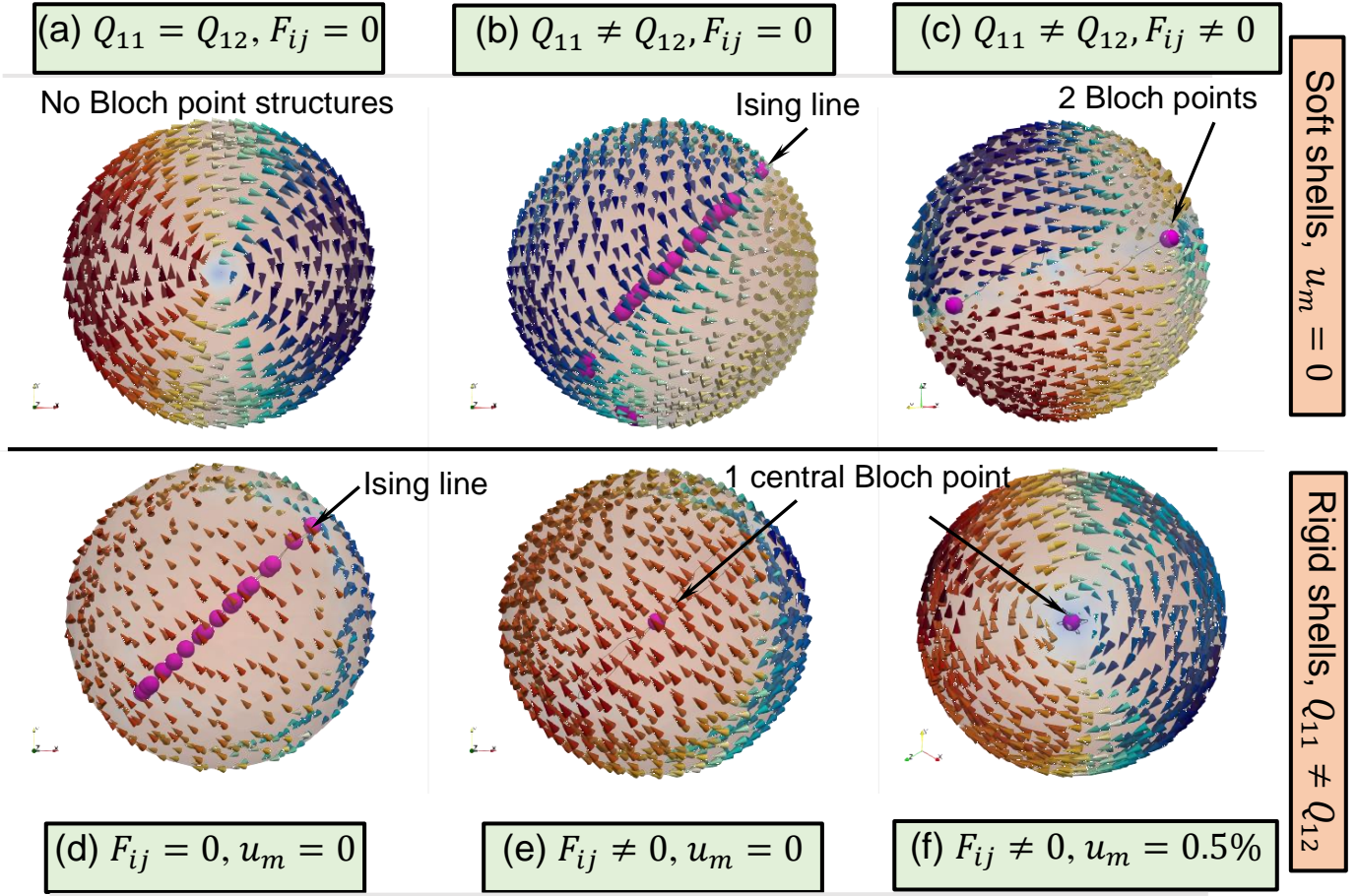

**Figure 10.** Bloch Point morphologies in a ferroelectric core covered with a soft (panels **(a-c)**) or a rigid (panels **(d-f)**) shell. The electrostriction anisotropy is small ($Q_{11}^{c,s} \approx Q_{12}^{c,s}$) for panel **(a)** and high ($Q_{11}^{c,s} \neq Q_{12}^{c,s}$) for panels **(b-f)**. The flexoelectric effect is absent ($F_{ij} = 0$) for panels **(a, b, d)** and present ($F_{ij} \neq 0$) for panels **(c, e, f)**. A mismatch strain is absent ($F_{ij} = 0$) for panels **(a, b, d)** and present ($F_{ij} \neq 0$) for panels **(c, e, f)**. The purple spheres show the position of Bloch points ($|P| = 0$), determined by the intersection points of the three isosurfaces $P_1 = 0$, $P_2 = 0$, and $P_3 = 0$. The structure **(a)** does not contain Bloch points. Bloch points in panels **(b)** and **(d)** display a one-dimensional topological line defect with $|P| = 0$ known as "Ising line". Two Bloch points in panels **(c)** are located at opposite sides of the domain wall. Although they are very close to the surface; these Bloch points are not located at diametrically opposite positions. The structures in panels **(e, f)** contain a single Bloch point located in the core center. Core radius $R = 10$ nm, shell thickness $\Delta R = 4$ nm, and $T = 298$ K. For other parameters see **Table AI** in **Appendix A.**

Several authors [60, 61, 62, 63, 64, 65] have studied numerically the electrocaloric effect (**ECE**) in ferroelectric nanoparticles using a phase field method combined with the LGD approach. In particular, using the "tetragonal core - cubic shell" model of the nanoparticle, Chen et al. [60] calculated the size dependence of the ECE in spherical single-domain BaTiO$_3$ nanoparticles and showed that a decrease of the nanoparticle size leads to a decrease of the adiabatic electrocaloric temperature $\Delta T_{EC}$. Chen et al. [60] also showed that a decrease of the nanoparticle size causes a blurring of $\Delta T_{EC}$ maxima and their mixing in the low-temperature region. These effects are associated with the increasing role of the nanoparticle shell with the particle size decrease. Li et al. [61] described the ECE in Bi$_4$Ti$_3$O$_{12}$ nanoparticles with a vortex-like domain structure and revealed a giant $\Delta T_{EC}$ (-16.6 K at 600°C) associated with a large change in the toroidal moment under the action of a curled electric field. Zeng et al. [62] studied the ECE in ferroelectric PbTiO$_3$ nanoparticles and related the giant positive or negative $\Delta T_{EC}$ to the change in the configuration of the vortex-like domain structure from clockwise to counter-clockwise, under the action of a curled electric field. Chen et al. [63] calculated the ECE during the transformation of the domain structure of PbTiO$_3$ nanoparticles from single-domain to vortex-like states, and backward, under the action of a curled electric field. Wang et al. [64] revealed the relationship between the changes in the vortex-like domain structure, and the negative or positive $\Delta T_{EC}$ under the action of an inhomogeneous electric field for ferroelectric PbTiO$_3$ nanoparticles. Ye et al. [65] showed the existence of a giant ECE in PbTiO$_3$ nanoparticles with a double vortex-like domain structure. They also studied the mismatch strain effect on the ECE, and demonstrated an increase of the $\Delta T_{EC}$ under compression and a decrease under tension of the nanoparticle.

This brief overview demonstrates the possibility to reveal a giant ECE in various ferroelectric nanoparticles, where the conditions for observing the effect were almost always determined in an empirical way, except for the case of single-domain nanoparticles [66]. Since results obtained in this work for core-shell nanoparticles with a complex domain structure can be well fitted by an analytical expression (1), we can make analytical estimates for $\Delta T_{EC}$ and the EC coefficient $\Sigma$, and establish the role of the size effect. Following Ref. [66], the values $\Delta T_{EC}$ and $\Sigma$ can be calculated as:

$$\Delta T_{EC}(\vec{r}) = -\int_0^{E_{ext}} \frac{T}{\rho(\vec{r})C_P(\vec{r})}\left(\frac{\partial P(\vec{r})}{\partial T}\right)_E dE, \qquad \Sigma(\vec{r}) = \frac{d\Delta T_{EC}(\vec{r})}{dE_{ext}}, \qquad (2)$$

where $E_{ext}$ is an external field applied to the core-shell nanoparticle via an effective media, $\rho$ is the mass density, and $C_P$ is the heat capacity of the nanoparticle core or shell, depending on the point $\vec{r}$. Following Ref. [66], the spatially averaged values $\langle \overline{\Delta T_{EC}} \rangle$ and $\langle \overline{\Sigma} \rangle$ can be estimated as

$$\langle \overline{\Delta T_{EC}(E_{ext})} \rangle \approx \frac{T}{\eta \rho C_P}\left(\frac{\alpha_T}{2}[P_r^2(E_{ext}) - P^2(0)] + \frac{\beta_T}{4}[P_r^4(E_{ext}) - P_r^4(0)] + \frac{\gamma_T}{6}[P_r^6(E_{ext}) - P_r^6(0)]\right), \qquad (3a)$$

$$\langle \overline{\Sigma(E_{ext}, T)} \rangle \approx \frac{d\langle \overline{\Delta T_{EC}(E_{ext})} \rangle}{dE_{ext}}. \qquad (3b)$$

When deriving expressions (3), we used the fact that the average core polarization is almost zero at $E_{ext} \to 0$, and so $\bar{P} = \overline{P^3} = \overline{P^5} \to 0$ in this case. Expressions (3) are valid for a quasi-static electric field. They contain the even powers of the polarization magnitude, $P_r^2$, $P_r^4$, and $P_r^6$, and a size-dependent dielectric factor $\eta$. The expression for the factor $\eta$ is conditioned by the dielectric weakening or enhancement of the external electric field inside the core [66]. The minimization of the "effective" Landau-type energy leads to an algebraic equation for $P_r$ [66]:

$$\alpha_T(T - T_{pt})P_r + \beta P_r^3 + \gamma P_r^5 = \eta E_{ext}. \quad (4a)$$

Equation (4a) allows one to calculate the dependence of $P_r$ on $E_{ext}$ if the coefficients $\alpha_T$, $\beta(T)$, $\gamma(T)$, and the dielectric factor $\eta$ are known. The coefficients $\alpha_T$, $\beta(T)$, and $\gamma(T)$ are listed for BaTiO$_3$ in **Table CI** in **Appendix C.** In accordance with our estimates, listed in **Appendix C,** the factor $\eta$ is given by the expression:

$$\eta = \frac{9(R+\Delta R)^3 \varepsilon_e \varepsilon_s}{2R^3(\varepsilon_e - \varepsilon_s)(\varepsilon_s - \varepsilon_b) + (R+\Delta R)^3(2\varepsilon_e + \varepsilon_s)(\varepsilon_b + 2\varepsilon_s)}. \quad (4b)$$

By using expression (1) for $T_{pt}(R)$ in combination with Eqs.(2)-(4), we can make a prediction about the polarization magnitude $P_r(E_{ext})$ and on how $\langle\overline{\Delta T_{EC}}\rangle$ changes depending on the external electric field, the shell, thickness, and the core radius. Results for the BaTiO$_3$ core covered with a rigid SrTiO$_3$ shell are shown in **Fig. 11**. For the demonstration of ECE we choose a rigid shell with and without mismatch strain, because the influence of a mismatch effect appears to be the strongest among all the effects considered in this work.

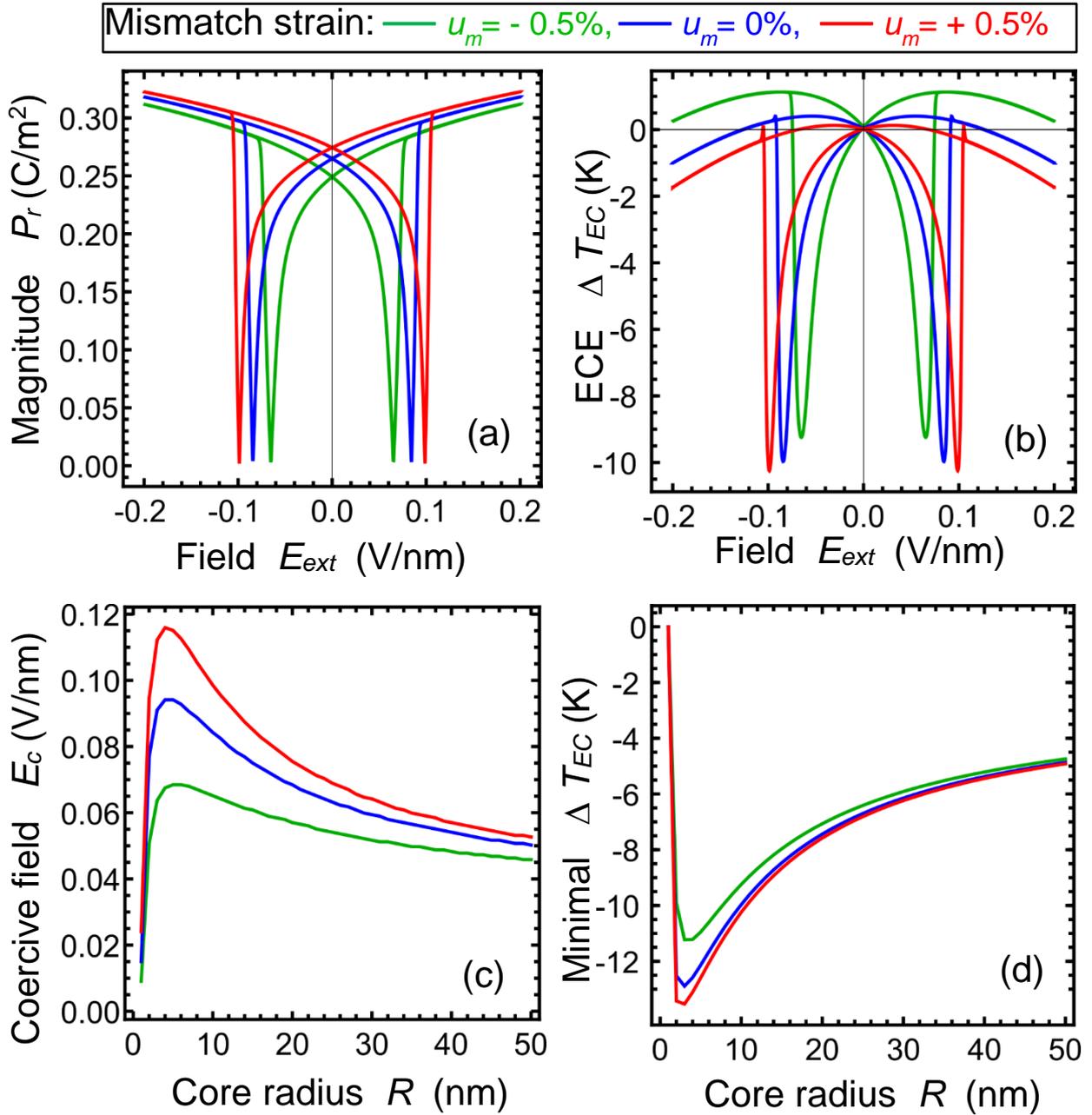

**Figure 11.** Dependence of the polarization magnitude $P_r$ (**a**) and the EC temperature change $\Delta T_{EC}$ (**b**) on a quasi-static external electric field. The curves are calculated from Eqs.(1)-(2) for a BaTiO$_3$ core with radius $R$ =10 nm covered with a rigid SrTiO$_3$ shell for different values of mismatch strain between the core and shell: $u_m = -0.5\%$ (green curves), 0 (blue curves), and +0.5% (red curves). The coercive field $E_c$ (**c**) and minimal negative values of $\Delta T_{EC}$ (**d**) as a function of the core radius $R$. Shell thickness $\Delta R = 4$ nm, and $T = $ 293 K, BaTiO$_3$ density $\rho = 6.02 \times 10^3$ kg/m$^3$, and specific heat $C_p = 4.6 \times 10^2$ J/(kg·K) at room temperature. For other parameters see **Table CI** in **Appendix C**.

The field dependence of $P_r$ (shown in **Fig. 11a** for a 10 nm core radius) has the form of a butterfly-type hysteresis loop and drops to zero at the coercive field $E_c$, whose value is about 0.08 V/nm for zero mismatch strain. $E_c$ decreases to 0.06 V/nm under compressive strain $u_m = -0.5\%$, and increases to 0.1 V/nm in the case of tensile strain $u_m = +0.5\%$. The field dependence of $\Delta T_{EC}$ (shown in **Fig. 11b**) also has the form of a butterfly-type hysteresis loop and reaches maximal negative values at $E_c$. These values, ranging from −8 K for $u_m = -0.5\%$ to $-12$ K for $u_m = +0.5\%$, are relatively high in comparison with $\Delta T_{EC} = -4$ K for a stress-free bulk BaTiO$_3$. **Figures 11c** and **11d** illustrate the size effect of $E_c$ and $\Delta T_{EC}$ with a pronounced maximum of $E_c$ (about 0.12 V/nm) and a minimum $\Delta T_{EC}$ (about −18 K), which is reached for a tensile misfit strain $u_m = +0.5\%$ and the core radius (2 – 4) nm. Corresponding values calculated for a compressive strain $u_m = +0.5\%$ are significantly smaller (compare red and green curves in **Fig. 11c** and **11d**). Note that a negative ECE providing effective cooling (~ −20 K) could be very promising for advanced applications of ferroelectric nanocomposites in energy convertors and cooling systems.

Let us underline the significant asymmetry of the domain morphology and ferroelectric properties (transition temperature, polarization magnitude, coercive field) and ECE with respect to the sign of the mismatch strain. This result is in a qualitative agreement with experimental results of Barnakov et al. [67], who studied the ferroelectric properties of BaTiO$_3$ nanocubes coated with metal carboxylates in two forms – one which was crystalline and provided a lattice mismatch, and the other that was non-crystalline without mismatch conditions. The observed polar effects differed by many orders of magnitude for these two coatings.

## VII. CONCLUSION

Within the framework of the LGD approach we have explored the impact of the elastic anisotropy, electrostriction and flexoelectric couplings, and mismatch strain on the domain structure morphology in spherical core-shell ferroelectric nanoparticles. We have performed FEM for a multiaxial ferroelectric core covered with an elastically-isotropic soft or elastically-anisotropic rigid paraelectric shell, with or without mismatch strains induced by the difference of the core and shell lattice constants.

Our FEM performed for the core covered by a **soft shell** shows that, at room temperature, a single polarization vortex with a dipolar kernel can be stable in the core with a relatively weak electrostriction coupling. The vortex disappears as the anisotropic electrostriction coupling increases, and evolves into 180º flux-closure domains, where complex cross-type morphology is controlled by the flexoelectric coupling in the core.

In the case of the core covered by a **rigid shell,** FEM shows that at room temperature the anisotropic elastic properties of the shell can stabilize vortex-like 120º flux-closure domains, which gradually "cross" in the equatorial plane of the core. The flexoelectric coupling leads to a noticeable curling of the flux-closure domain walls. The mismatch strain compensates the curling of the flux-closure domains in the core confined by the elastically-anisotropic rigid shell.

Using FEM and derived analytical expressions, we calculated the phase diagrams of core-shell ferroelectric nanoparticles as a function of the core radius and temperature. Phase diagrams for a core covered with an elastically-isotropic **soft shell** show a relatively small but noticeable increase of the PE-FE transition temperature induced by the flexoelectric coupling. Phase diagrams for a core covered with an elastically-anisotropic **rigid shell** demonstrate a relatively strong influence of mismatch strain and a negligible effect of flexoelectric coupling.

We identified a significant asymmetry of the domain morphology and ferroelectric properties with respect to the sign of the mismatch strain that originates at the core-shell interface (compare with experiment [67]). Specifically, tensile strains enhance the properties, and compressive strains deteriorate them. Using size and mismatch effects, we can select optimal parameters to reach high negative values of an electrocaloric response from an ensemble of noninteracting core-shell nanoparticles, which is important for energy convertors and cooling systems. This leads to the conclusion that the obtained analytical results can be used for size-optimization of core-shell nanoparticles for advanced applications in nanoelectronics and nano-coolers.

**APPENDIX A** contains a mathematical formulation of the problem in the framework of Landau-Ginzburg-Devonshire theory, and parameters for both the $BaTiO_3$ (core) and the soft and rigid shell materials used in the FEM. It also shows the potential impact of the flexoelectric effect and Vegard strains created by the oxygen vacancies on the effective elastic compliances of the core.

**APPENDIX B** contains the details of the phase diagrams calculations.

**APPENDIX C** contains the details of the electrocaloric effect calculations.

**Acknowledgements.** A.N.M. acknowledges EOARD project 9IOE063 and related STCU partner project P751. R.H. acknowledges funding from the French National Research Agency through contract ANR-18-CE92-0052 "TOPELEC". V.Y.R. acknowledges the support of COST Action CA17139.

**Authors' contribution.** A.N.M., V.Yu.R., and D.R.E. generated the research idea. A.N.M. and V.Yu.R. stated the problem. A.N.M. performed analytical estimates and wrote the manuscript draft. E.A.E. wrote the codes and performed FEM calculations. R.H. analyzed the morphology of domain structures. Y.M.F. tested the codes. H.V.S. performed numerical calculations of electrocaloric properties. R.H., V.Yu.R., and D.R.E. worked intensively on the results interpretation and manuscript improvement.

# REFERENCES

1   Y. L. Tang, Y. L. Zhu, X. L. Ma, A. Y. Borisevich, A. N. Morozovska, E. A. Eliseev, W. Y. Wang, Y. J. Wang, Y.B. Xu, Z.D. Zhang, S.J. Pennycook, Observation of a periodic array of flux-closure quadrants in strained ferroelectric $PbTiO_3$ films, Science **348**, 547 (2015).

2   S. V. Kalinin, Y. Kim, D. Fong, and A. N. Morozovska, Surface-screening mechanisms in ferroelectric thin films and their effect on polarization dynamics and domain structures, Rep. Prog. Phys. **81**, 036502 (2018).

3   S. Zhang, X. Guo, Y. Tang, D. Ma, Y. Zhu, Y. Wang, S. Li, M. Han, D. Chen, J. Ma, B. Wu, and X. Ma. Polarization rotation in ultrathin ferroelectrics tailored by interfacial oxygen octahedral coupling, ACS Nano, **12**, 36813 (2018).

4   Y. Nahas, S. Prokhorenko, J. Fischer, B. Xu, C. Carrétéro, S. Prosandeev, M. Bibes, S. Fusil, B. Dkhil, V. Garcia, and L. Bellaiche, Inverse transition of labyrinthine domain patterns in ferroelectric thin films, Nature **577**, 47 (2020).

5   S. Cherifi-Hertel, H. Bulou, R. Hertel, G. Taupier, K. D. H. Dorkenoo, C. Andreas, J. Guyonnet, I. Gaponenko, K. Gallo, and P. Paruch, Non-ising and chiral ferroelectric domain walls revealed by nonlinear optical microscopy, Nature Communications **8**, 15768 (2017).

6   J. Hlinka, and P. Ondrejkovic. Skyrmions in ferroelectric materials. Solid State Physics, **70**, 143-165, Chapter 4 in "Recent Advances in Topological Ferroics and Their Dynamics" Edited by Robert L. Stamps and Helmut Schulthei. Academic Press (2019).

7   P. Chen, X. Zhong, J.A. Zorn, M. Li, Y. Sun, A.Y. Abid, C. Ren, Yuehui Li, X. Li, X. Ma, J. Wang, K. Liu, Z. Xu, C.Tan, L. Chen, P. Gao, X. Bai, Atomic Imaging of Mechanically Induced Topological Transition of Ferroelectric Vortices, Nat. Comm. 11, article number: **1840** (2020).

8   Z. Li, Y. Wang, G. Tian, P. Li, L. Zhao, F. Zhang, J. Yao, H. Fan, X. Song, D. Chen, and Z. Fan. High-density array of ferroelectric nanodots with robust and reversibly switchable topological domain states. Sci. Adv. **3**, e1700919 (2017).

9   D. Karpov, Z. Liu, T. dos Santos Rolo, R. Harder, P.V. Balachandran, D. Xue, T. Lookman, and E. Fohtung. Three-dimensional imaging of vortex structure in a ferroelectric nanoparticle driven by an electric field. Nat. Commun. **8**, 280 (2017).

10  J. M. Gregg, Exotic Domain States in Ferroelectrics: Searching for Vortices and Skyrmions, Ferroelectrics, **433**, 74 (2012).

11  F. Xue, L. Li, J. Britson, Z. Hong, C. A. Heikes, C. Adamo, D.G. Schlom, X. Pan, and L.-Q. Chen, Switching the curl of polarization vectors by an irrotational electric field, Phys. Rev. B **94**, 100103 (2016).

12      L. Baudry, A. Sené, I. A. Luk'Yanchuk, L. Lahoche, and J. F. Scott, Polarization vortex domains induced by switching electric field in ferroelectric films with circular electrodes, Phys. Rev. B **90**, 024102 (2014).

13      J. Liu, W. Chen, and Y. Zheng, Shape-induced phase transition of vortex domain structures in ferroelectric nanodots and their controllability by electrical and mechanical loads, Theor. Appl. Mech. Lett. **7**, 81 (2017).

14      W. J. Chen, and Y. Zheng, Vortex switching in ferroelectric nanodots and its feasibility by a homogeneous electric field: effects of substrate, dislocations and local clamping force, Acta Mater. **88**, 41 (2015).

15      C. M. Wu, W. J. Chen, Y. Zheng, D. C. Ma, B. Wang, J. Y. Liu, and C. H. Woo. Controllability of vortex domain structure in ferroelectric nanodot: Fruitful domain patterns and transformation paths. Sci. Rep. **4**, 3946 (2014).

16      W. M. Xiong, Q. Sheng, W. J. Chen, C. M. Wu, B. Wang, and Y. Zheng. Large controllability of domain evolution in ferroelectric nanodot via isotropic surface charge screening. Appl. Phys. A **122**, 783 (2016).

17      D. Zhu, J. Mangeri, R. Wang, and S. Nakhmanson. Size, shape, and orientation dependence of the field-induced behavior in ferroelectric nanoparticles. J. Appl. Phys. **125**, 134102 (2019).

18      J.-J. Wang, B. Wang, and L.-Q. Chen. Understanding, Predicting, and Designing Ferroelectric Domain Structures and Switching Guided by the Phase-Field Method. Ann. Rev. Mater. Res. **49**, 127 (2019).

19      J. Mangeri, Y. Espinal, A. Jokisaari, S. P. Alpay, S. Nakhmanson, and O. Heinonen. Topological phase transformations and intrinsic size effects in ferroelectric nanoparticles. Nanoscale **9**, 1616 (2017).

20      J. Mangeri, S. P. Alpay, S. Nakhmanson, and O. G. Heinonen. Electromechanical control of polarization vortex ordering in an interacting ferroelectric-dielectric composite dimer. Appl. Phys. Lett. **113**, 092901 (2018).

21      K. C. Pitike, J. Mangeri, H. Whitelock, T. Patel, P. Dyer, S. P. Alpay, and S. Nakhmanson. Metastable vortex-like polarization textures in ferroelectric nanoparticles of different shapes and sizes. J. Appl. Phys. **124**, 064104 (2018).

22      X. Chen, and C. Fang. Study of electrocaloric effect in barium titanate nanoparticle with core–shell model. Physica B **415**, 14 (2013).

23      A. N. Morozovska, E. A. Eliseev, Y. M. Fomichov, Y. M. Vysochanskii, V. Yu. Reshetnyak, and D. R. Evans. Controlling the domain structure of ferroelectric nanoparticles using tunable shells. Acta Mater., **183**, 36 (2020).

24      A. N. Morozovska, E. A. Eliseev, R. Hertel, Y. M. Fomichov, V. Tulaidan, V. Yu. Reshetnyak, and D. R. Evans. Electric Field Control of Three-Dimensional Vortex States in Core-Shell Ferroelectric Nanoparticles. Acta Materialia, **200**, 256 (2020).

25      D. P. DiVincenzo, Quantum computation. Science **270**, 255 (1995).

26      M. A. Nielsen, I.L. Chuang, Quantum Computation and Quantum Information, Cambridge University Press. ISBN 978-1-107-00217-3 (2010).

27      S. Lin, T. Lü, C. Jin, and X. Wang. Size effect on the dielectric properties of $BaTiO_3$ nanoceramics in a modified Ginsburg-Landau-Devonshire thermodynamic theory. Phys. Rev. B **74**, 134115 (2006).

28   E. A. Eliseev, A. V. Semchenko, Y. M. Fomichov, M. D. Glinchuk, V. V. Sidsky, V. V. Kolos, Yu M. Pleskachevsky, M. V. Silibin, N. V. Morozovsky, and A. N. Morozovska, Surface and finite size effects impact on the phase diagrams, polar, and dielectric properties of (Sr, Bi)Ta$_2$O$_9$ ferroelectric nanoparticles, J. Appl. Phys. **119**, 204104 (2016).

29   E. A. Eliseev, Y. M. Fomichov, S. V. Kalinin, Y. M. Vysochanskii, P. Maksymovich and A. N. Morozovska. Labyrinthine domains in ferroelectric nanoparticles: Manifestation of a gradient-induced morphological phase transition. Phys. Rev. B **98**, 054101 (2018).

30   A. N. Morozovska, Y. M. Fomichov, P. Maksymovych, Y. M. Vysochanskii, and E. A. Eliseev. Analytical description of domain morphology and phase diagrams of ferroelectric nanoparticles. Acta Mater. **160**, 109 (2018).

31   A. N. Morozovska, M. D. Glinchuk, E. A. Eliseev. Phase transitions induced by confinement of ferroic nanoparticles. Phys. Rev. B **76**, 014102 (2007).

32   H. Huang, C. Q. Sun, P. Hing. Surface bond contraction and its effect on the nanometric sized lead zirconate titanate. J. Phys.: Condens. Matter **12,** L127 (2000).

33   H. Huang, C. Q. Sun, Z. Tianshu, P. Hing. Grain-size effect on ferroelectric Pb(Zr$_{1-x}$Ti$_x$)O$_3$ solid solutions induced by surface bond contraction. Phys. Rev. B **63**, 184112 (2001).

34   W. Ma. Surface tension and Curie temperature in ferroelectric nanowires and nanodots. Appl. Phys. A **96**, 915 (2009).

35   J. J. Wang, X. Q. Ma, Q. Li, J. Britson, Long-Qing Chen, Phase transitions and domain structures of ferroelectric nanoparticles: Phase field model incorporating strong elastic and dielectric inhomogeneity, Acta Mater. **61**, 7591 (2013).

36   J. J. Wang, E. A. Eliseev, X. Q. Ma, P. P. Wu, A. N. Morozovska, and L.-Q. Chen. Strain effect on phase transitions of BaTiO$_3$ nanowires. Acta Mater. **59**, 7189 (2011).

37   D. A. Freedman, D. Roundy, and T. A. Arias. Elastic effects of vacancies in strontium titanate: Short- and long-range strain fields, elastic dipole tensors, and chemical strain. Phys. Rev. B **80**, 064108 (2009).

38   Y. Kim, A. S. Disa, T. E. Babakol, and J. D. Brock. Strain screening by mobile oxygen vacancies in SrTiO$_3$. Appl. Phys. Lett. **96**, 251901 (2010).

39   A. Sundaresan, C.N.R. Rao. Ferromagnetism as a universal feature of inorganic nanoparticles. Nano Today, **4**, 96–106. (2009). https://doi:10.1016/j.nantod.2008.10.002

40   S. C. Johnson. Oxygen vacancies in BaTiO$_3$ have a surprising impact on dielectric and ferroelectric properties, AIPScilight (2018). https://doi.org/10.1063/1.5033331

41   V. Stepkova, P. Marton, and J. Hlinka. Ising lines: Natural topological defects within ferroelectric Bloch walls. Phys. Rev. B **92,** 094106 (2015).

42   J. Zhu, W. Han, H. Zhang, Z. Yuan, X. Wang, L. Li, and C. Jin. Phase coexistence evolution of nano BaTiO$_3$ as function of particle sizes and temperatures. J. Appl. Phys. **112**, 064110 (2012).

43    S. Lin, T. Lü, C. Jin, and X. Wang. Size effect on the dielectric properties of $BaTiO_3$ nanoceramics in a modified Ginsburg-Landau-Devonshire thermodynamic theory. Phys. Rev. B **74,** 134115 (2006).

44    A. N. Morozovska and M. D. Glinchuk. Flexo-chemo effect in nanoferroics as a source of critical size disappearence at size-induced phase transitions. J. Appl. Phys. **119**, 094109 (2016).

45    M. Y. Gureev, A. K. Tagantsev, and N. Setter Phys. Rev. B **83,** 184104 (2011).

46    E. A. Eliseev, A. N. Morozovska, G. S. Svechnikov, P. Maksymovych, S. V. Kalinin. Domain wall conduction in multiaxial ferroelectrics: impact of the wall tilt, curvature, flexoelectric coupling, electrostriction, proximity and finite size effects. Phys. Rev.B. **85**, 045312 (2012).

47    I. S. Vorotiahin, E. A. Eliseev, Q. Li, S. V. Kalinin, Y. A. Genenko and A. N. Morozovska. Tuning the Polar States of Ferroelectric Films via Surface Charges and Flexoelectricity. Acta Materialia **137** (15), 85 (2017).

48    E. A. Eliseev, A. N. Morozovska, C. T. Nelson, and S. V. Kalinin. Intrinsic structural instabilities of domain walls driven by gradient couplings: meandering anferrodistortive-ferroelectric domain walls in $BiFeO_3$. Phys. Rev. B **99**, 014112 (2019).

49    M. J. Han, E. A. Eliseev, A. N. Morozovska, Y. L. Zhu, Y. L. Tang, Y. J. Wang, X. W. Guo, X. L. Ma. Mapping gradient-driven morphological phase transition at the conductive domain walls of strained multiferroic films. Phys. Rev. B **100**, 104109 (2019).

50    E. A. Eliseev, I. S. Vorotiahin, Y. M. Fomichov, M. D. Glinchuk, S. V. Kalinin, Y. A. Genenko, and A. N. Morozovska. Defect driven flexo-chemical coupling in thin ferroelectric films. Physical Review B **97,** 024102 (2018).

51    N. A. Pertsev, A. G. Zembilgotov, and A. K. Tagantsev. Effect of mechanical boundary conditions on phase diagrams of epitaxial ferroelectric thin films. Phys. Rev. Lett. **80,** 1988 (1998).

52    C. Ederer, and N. A. Spaldin. Effect of epitaxial strain on the spontaneous polarization of thin film ferroelectrics. Phys. Rev. Lett. **95**, 257601 (2005).

53    N. Balke, B. Winchester, W. Ren, Y.H. Chu, A.N. Morozovska, E.A. Eliseev, M. Huijben, R.K. Vasudevan, P. Maksymovych, J. Britson, S. Jesse, I. Kornev, R. Ramesh, L. Bellaiche, L.Q. Chen, and S.V. Kalinin, Enhanced electric conductivity at ferroelectric vortex cores in $BiFeO_3$, Nat. Phys. **8**, 81 (2012).

54    B. Winchester, N. Balke, X.X. Cheng, A.N. Morozovska, S. Kalinin, and L.Q. Chen, Electroelastic fields in artificially created vortex cores in epitaxial $BiFeO_3$ thin films, Appl. Phys. Lett. **107**, 052903 (2015).

55    B. Van Waeyenberge, A. Puzic, H. Stoll, K. W. Chou, T. Tyliszczak, R. Hertel, M. Fähnle, H. Brückl, K. Rott, G. Reiss, and I. Neudecker, Magnetic vortex core reversal by excitation with short bursts of an alternating field, Nature 444, 461 (2006).

56    R. Hertel, S. Gliga, M. Fähnle, and C. M. Schneider, Ultrafast nanomagnetic toggle switching of vortex cores, Phys. Rev. Lett. **98,** 117201 (2007).

57      S.-K. Kim, Y.-S. Choi, K.-S. Lee, K.Y. Guslienko, and D.-E. Jeong, Electric-current-driven vortex-core reversal in soft magnetic nanodots, Appl. Phys. Lett. **91,** 082506 (2007).

58      R. Hertel, Ultrafast domain wall dynamics in magnetic nanotubes and nanowires, J. Phys.: Condens. Matter. **28**, 483002 (2016).

59      V.I. Yukalov and E.P. Yukalova. Ultrafast polarization switching in ferroelectrics, Phys. Rev. Research **1,** 033136 (2019).

60      X. Chen, and C. Fang. Study of electrocaloric effect in barium titanate nanoparticle with core–shell model. Physica B: Condensed Matter **415**, 14 (2013).

61      B. Li, J. B. Wang, X. L. Zhong, F. Wang, Y. K. Zeng, and Y. C. Zhou. Giant electrocaloric effects in ferroelectric nanostructures with vortex domain structures. RSC Advances **3**, 7928 (2013).

62      Y. K. Zeng, B. Li, J. B. Wang, X. L. Zhong, W. Wang, F. Wang, and Y. C. Zhou. Influence of vortex domain switching on the electrocaloric property of a ferroelectric nanoparticle. RSC Advances **4**, 30211 (2014).

63      Z. Y. Chen, Y. X. Su, Z. D. Zhou, L. S. Lei, and C. P. Yang. The influence of the electrical boundary condition on domain structures and electrocaloric effect of $PbTiO_3$ nanostructures. AIP Advances **6,** 055207 (2016).

64      F. Wang, L. F. Liu, B. Li, Y. Ou, L. Tian, and W. Wang. Inhomogeneous electric-field–induced negative/positive electrocaloric effects in ferroelectric nanoparticles. EPL (Europhysics Letters) **117**, 57002 (2017).

65      C. Ye, J. B. Wang, B. Li, X. L. Zhong, Giant electrocaloric effect in a wide temperature range in $PbTiO_3$ nanoparticle with double-vortex domain structure. Sci. Rep. **8**, 293 (2018).

66      A. N. Morozovska, E. A. Eliseev, M. D. Glinchuk, H. V. Shevliakova, G. S. Svechnikov, M. V. Silibin, A. V. Sysa, A. D. Yaremkevich, N. V. Morozovsky, and V. V. Shvartsman. Analytical description of the size effect on pyroelectric and electrocaloric properties of ferroelectric nanoparticles. Phys. Rev. Materials **3**, 104414 (2019).

67      Yu. A. Barnakov, I. U. Idehenre, S. A. Basun, T. A. Tyson, and D. R. Evans. Uncovering the Mystery of Ferroelectricity in Zero Dimensional Nanoparticles. Royal Society of Chemistry, Nanoscale Adv. **1,** 664 (2019).

# SUPPLEMENTARY MATERIALS

**APPENDIX A. Mathematical formulation of the problem and computation details**

**A1. Mathematical formulation of the problem**

We consider a ferroelectric nanoparticle core of radius $R$ with a three-component ferroelectric polarization vector $\mathbf{P}$. The core is regarded as insulating, without any free charges. It is covered with a semiconducting tunable shell of thickness $\Delta R$ that is characterized by a strongly temperature-dependent "tunable" relative dielectric permittivity tensor $\varepsilon_{ij}^S$. The core-shell nanoparticle is placed in a dielectric medium (polymer, gas, liquid, air, or vacuum) with an effective dielectric permittivity, $\varepsilon_e$. The word "effective" implies the presence of other particles in the medium, which can be described in an effective medium approach. For the sake of clarity, we consider the medium as being isotropic and temperature-independent, i.e. $\varepsilon_{ij}^e = \delta_{ij}\varepsilon_e$, in contrast to anisotropic and/or tunable shells. The considered physical model corresponds to a nanocomposite consisting of core-shell nanoparticles in a dielectric medium, with a small volume fraction of ferroelectric nanoparticles (less than 10%) in the composite. The core-shell geometry is shown in **Fig. 1** of the main text.

Since the ferroelectric polarization contains background and soft mode contributions, the electric displacement vector has the form $\mathbf{D} = \varepsilon_0 \varepsilon_b \mathbf{E} + \mathbf{P}$ inside the core. In this expression $\varepsilon_b$ is a relative background permittivity of the core [1], $\varepsilon_0$ is the universal dielectric constant, and $\mathbf{P}$ is a ferroelectric polarization containing the spontaneous and field-induced contributions, $\mathbf{P} = \mathbf{P}_S + \varepsilon_0 \hat{\chi}_f \mathbf{E} + \varepsilon_0 \hat{\chi}_{ff} \mathbf{E}^3 + \varepsilon_0 \hat{\chi}_{fff} \mathbf{E}^5 + \ldots$, where $\mathbf{P}_S$ is the spontaneous polarization at $\mathbf{E} = 0$. Note that the expression $\mathbf{D} = \varepsilon_0 \varepsilon_b \mathbf{E} + \mathbf{P}$ is different from the usual textbook definition, $\mathbf{D} = \varepsilon_0 \mathbf{E} + \mathbf{P}$, where $\mathbf{P}$ is the total polarization. Usually $4 < \varepsilon_b < 10$, and so $\varepsilon_b$ can be significantly smaller than the linear susceptibility $\chi_f$, whose temperature-dependent values strongly increase in the vicinity of ferroelectric-paraelectric phase transition. As a rule, $\chi_f > (30 - 100)$ even when far from the phase transition; this is due to the dominant contribution from soft mode-related optic phonons. In the case of a linear response to a small external electric field the electric displacement in the core is $\mathbf{D} \approx \varepsilon_0 \hat{\varepsilon}_f \mathbf{E} + \mathbf{P}_S$, where $\hat{\varepsilon}_f = \hat{\chi}_f + \hat{\varepsilon}_b$. The expression $D_i = \varepsilon_0 \varepsilon_{ij}^S E_j$ is valid in the shell and $D_i = \varepsilon_0 \varepsilon_e E_i$ in the isotropic effective medium.

The electric field components $E_i$ are derived from the electric potential $\varphi$ in a conventional way, $E_i = -\partial \varphi / \partial x_i$. The potential $\varphi$ satisfies the Poisson equation in the ferroelectric core (subscript "$f$"):

$$\varepsilon_0 \varepsilon_b \left( \frac{\partial^2}{\partial x_1^2} + \frac{\partial^2}{\partial x_2^2} + \frac{\partial^2}{\partial x_3^2} \right) \varphi_f = \frac{\partial P_i}{\partial x_i}, \qquad 0 \leq r \leq R, \tag{A.1a}$$

and a Debye-type equation in the shell (subscript "$s$"):

$$\frac{\partial}{\partial x_i}\left( \varepsilon_{ij}^s \frac{\partial \varphi_s}{\partial x_j} \right) = -\frac{\varphi_s}{R_d^2}, \qquad R < r < R + \Delta R, \tag{A.1b}$$

where $R_d = \sqrt{\frac{\varepsilon_0 k_B T}{e^2 n}}$ is the "net" screening length defined by the concentration of free carriers $n$ in the shell. The "dressed" screening length $R_d^* = \sqrt{\frac{\varepsilon_0 \varepsilon_s k_B T}{e^2 n}}$ can be introduced for the shell with an isotropic relative permittivity, $\varepsilon_{ij}^S = \delta_{ij}\varepsilon_s$. A small internal conductivity of the shell is required for the most effective penetration of external electric field to the core. Using a proper wide-gap semiconductor or a paraelectric as the shell, we can estimate that $R_d^* \gg 10$ nm for room temperature with a typical concentration of intrinsic carriers in the shell.

The electric potential φ in the external region outside the shell satisfies the Laplace equation (subscript "$e$"):

$$\varepsilon_0 \varepsilon_e \left(\frac{\partial^2}{\partial x_1^2} + \frac{\partial^2}{\partial x_2^2} + \frac{\partial^2}{\partial x_3^2}\right)\varphi_e = 0, \qquad r > R + \Delta R, \tag{A.1c}$$

Equations (A.1) are supplemented with the continuity conditions for electric potential and normal components of the electric displacements at the particle surface and core-shell interface:

$$(\varphi_e - \varphi_s)|_{r=R+\Delta R} = 0, \quad \boldsymbol{n}(\boldsymbol{D}_e - \boldsymbol{D}_s)|_{r=R+\Delta R} = 0, \tag{A.1d}$$

$$(\varphi_s - \varphi_f)|_{r=R} = 0, \quad \boldsymbol{n}(\boldsymbol{D}_s - \boldsymbol{D}_f)|_{r=R} = 0. \tag{A.1e}$$

Either charges are absent or the applied voltage is fixed at the boundaries of the computation region:

$$\left.\frac{\partial \varphi_e}{\partial x_l} n_l\right|_{x=\pm\frac{L}{2}} = 0, \quad \left.\frac{\partial \varphi_e}{\partial x_l} n_l\right|_{y=\pm\frac{L}{2}} = 0, \quad \varphi_e|_{z=+\frac{L}{2}} = 0, \quad \varphi_e|_{z=-\frac{L}{2}} = V_e \tag{A.1f}$$

Here $V_e$ is the applied voltage difference and $L$ is the size of computation region. For FEM we use a cube with an edge size $L \gg 2(\Delta R + R)$, and set $V_e = 0$ for the purposes of this work.

The LGD free energy functional $G$ additively includes a Landau expansion on powers of 2-4-6 of the polarization, $G_{Landau}$; a polarization gradient energy contribution, $G_{grad}$; an electrostatic contribution, $G_{el}$; the elastic, electrostriction, flexoelectric contributions, $G_{es+flexo}$; an electrochemical (Vegard strain) energy, $G_{VS}$; and a surface energy, $G_S$. It has the form [2, 3, 4]:

$$G = G_{Landau} + G_{grad} + G_{el} + G_{es+flexo} + G_{VS} + G_S, \tag{A.2a}$$

$$G_{Landau} = \int_{0<r<R} d^3r \left[a_i P_i^2 + a_{ij} P_i^2 P_j^2 + a_{ijk} P_i^2 P_j^2 P_k^2\right], \tag{A.2b}$$

$$G_{grad} = \int_{0<r<R} d^3r \frac{g_{ijkl}}{2} \frac{\partial P_i}{\partial x_j} \frac{\partial P_k}{\partial x_l}, \tag{A.2c}$$

$$G_{el} = -\int_{0<r<R} d^3r \left(P_i E_i + \frac{\varepsilon_0 \varepsilon_b}{2} E_i E_i\right) - \frac{\varepsilon_0}{2} \int_{R<r<R+\Delta R} \varepsilon_{ij}^S E_i E_j d^3r - \frac{\varepsilon_0}{2} \int_{r>R+\Delta R} \varepsilon_{ij}^e E_i E_j d^3r, \tag{A.2d}$$

$$G_{es+flexo} = -\int_{0<r<R} d^3r \left(\frac{s_{ijkl}}{2} \sigma_{ij}\sigma_{kl} + Q_{ijkl}\sigma_{ij} P_k P_l + F_{ijkl}\sigma_{ij} \frac{\partial P_l}{\partial x_k}\right)$$

$$- \int_{R<r<R+\Delta R} d^3r \left(\frac{s_{ijkl}^S}{2} \sigma_{ij}\sigma_{kl} + \varepsilon_0^2(\varepsilon_s - 1)^2 Q_{ijkl}^S \sigma_{ij} E_k E_l\right), \tag{A.2e}$$

$$G_{VS} = \int_{R<r<R+\Delta R} d^3r \left( k_B T \left[ N_V^+ \ln\left(\frac{N_V^+}{N_V}\right) + (N_V - N_V^+) \ln\left(\frac{N_V - N_V^+}{N_V}\right) \right] - N_V^+ W_{ij}^V \sigma_{ij} - \left(Z_V^{eff} N_V^+ - n\right)\varphi \right)$$
(A.2f)

$$G_S = \frac{1}{2} \int_{r=R} d^2r \, a_{ij}^{(S)} P_i P_j.$$
(A.2g)

The coefficient $a_i$ linearly depends on temperature $T$:

$$a_i(T) = \alpha_T [T - T_C(R)],$$
(A.3a)

where $\alpha_T$ is the inverse Curie-Weiss constant and $T_C(R)$ is the ferroelectric Curie temperature renormalized by electrostriction and surface tension. Actually, the surface tension induces additional surface stresses $\sigma_{ij}$ proportional to the surface tension coefficient μ and equal to $\sigma_{11} = \sigma_{22} = \sigma_{33}|_{r=R} = \frac{-2\mu}{R}$ for a spherical nanoparticle of radius $R$. The stresses affect the Curie temperature and ferroelectric polarization behavior due to the electrostriction coupling. Thus, the renormalized Curie temperature, $T_C(R)$, acquires the following form [2, 3]:

$$T_C(R) = T_C \left( 1 - \frac{Q}{\alpha_T T_C} \frac{2\mu}{R} \right)$$
(A.3b)

where $T_C$ is a Curie temperature of a bulk ferroelectric. $Q$ is the sum of the electrostriction tensor diagonal components, which is positive for most ferroelectric perovskites with cubic m3m symmetry in the paraelectric phase, namely $0.004 < Q < 0.04$m$^4$/C$^2$ [2-4]. Recent experiments tell us that μ is relatively small, not more than (2 – 4) N/m for most perovskites.

Tensor components $a_{ij}$ are regarded as temperature-independent. The tensor $a_{ij}$ is positively defined if the ferroelectric material undergoes a second order transition to the paraelectric phase and negative otherwise. The higher nonlinear tensor $a_{ijk}$ and the gradient coefficients tensor $g_{ijkl}$ are positively defined and regarded as temperature-independent. The following designations are used in Eq.(A.2e): $\sigma_{ij}$ is the stress tensor, $s_{ijkl}$ is the elastic compliances tensor, $Q_{ijkl}$ is the electrostriction tensor, and $F_{ijkl}$ is the flexoelectric tensor.

The Vegard strain energy is given by Eq.(A.2f) [4], where $N_V$ is the concentration of oxygen vacancies, which can be charged or neutral. The charged vacancies with concentration $N_V^+$ are mobile, and $N_V^+ < N_V$. The concentration $N_V$ is significantly smaller than the maximal possible concentration $N_0$. The introduction of a maximal possible concentration $N_0$ takes into account steric effects [5] and limits the vacancy accumulation in the vicinity of domain walls, surfaces, and interfaces. The entropy of charged vacancies (the first term of Eq.(A.2f)) corresponds to the approximation of an infinitely thin quasi-level, where $T$ is the absolute temperature, $k_B$ is the Boltzmann constant, $W_{ij}^V$ is the Vegard strain tensor (also known as the elastic dipole) [6, 7, 8]. In Eq.(A.2f) we neglect the difference between the Vegard tensor in the core and shell since both are cubic perovskites. The electrostatic energy, $\left(Z_V^{eff} N_V^+ - en\right)\varphi$, exists for charged vacancies with a concentration $N_V^+$ and free electrons with a concentration $n$. $Z_V^{eff}$ is an effective vacancy charge. For oxygen vacancies $0 \leq Z_V^{eff} \leq +2e$, where $e$ in the absolute value of electron charge.

Allowing for the Khalatnikov mechanism of polarization relaxation [9], minimization of the free energy (A.2) with respect to polarization leads to three coupled time-dependent Euler-Lagrange equations for polarization components inside the core, $\frac{\delta G}{\delta P_i} = -\Gamma \frac{\partial P_i}{\partial t}$, where $i = 1, 2, 3$. The explicit form of the equations for a ferroelectric crystal with m3m parent symmetry is:

$$\Gamma \frac{\partial P_1}{\partial t} + 2P_1(a_1 - Q_{12}(\sigma_{22} + \sigma_{33}) - Q_{11}\sigma_{11}) - Q_{44}(\sigma_{12}P_2 + \sigma_{13}P_3) + 4a_{11}P_1^3 + 2a_{12}P_1(P_2^2 + P_3^2) +$$

$$6a_{111}P_1^5 + 2a_{112}P_1(P_2^4 + 2P_1^2P_2^2 + P_3^4 + 2P_1^2P_3^2) + 2a_{123}P_1P_2^2P_3^2 - g_{11}\frac{\partial^2 P_1}{\partial x_1^2} - g_{44}\left(\frac{\partial^2 P_1}{\partial x_2^2} + \frac{\partial^2 P_1}{\partial x_3^2}\right) =$$

$$-F_{11}\frac{\partial \sigma_{11}}{\partial x_1} - F_{12}\left(\frac{\partial \sigma_{22}}{\partial x_1} + \frac{\partial \sigma_{33}}{\partial x_1}\right) - F_{44}\left(\frac{\partial \sigma_{12}}{\partial x_2} + \frac{\partial \sigma_{13}}{\partial x_3}\right) + E_1 \quad \text{(A.4a)}$$

$$\Gamma \frac{\partial P_2}{\partial t} + 2P_2(a_1 - Q_{12}(\sigma_{11} + \sigma_{33}) - Q_{11}\sigma_{22}) - Q_{44}(\sigma_{12}P_1 + \sigma_{23}P_3) + 4a_{11}P_2^3 + 2a_{12}P_2(P_1^2 + P_3^2) +$$

$$6a_{111}P_2^5 + 2a_{112}P_2(P_1^4 + 2P_2^2P_1^2 + P_3^4 + 2P_2^2P_3^2) + 2a_{123}P_2P_1^2P_3^2 - g_{11}\frac{\partial^2 P_2}{\partial x_2^2} - g_{44}\left(\frac{\partial^2 P_2}{\partial x_1^2} + \frac{\partial^2 P_2}{\partial x_3^2}\right) =$$

$$-F_{11}\frac{\partial \sigma_{22}}{\partial x_2} - F_{12}\left(\frac{\partial \sigma_{11}}{\partial x_2} + \frac{\partial \sigma_{33}}{\partial x_2}\right) - F_{44}\left(\frac{\partial \sigma_{12}}{\partial x_1} + \frac{\partial \sigma_{23}}{\partial x_3}\right) + E_2 \quad \text{(A.4b)}$$

$$\Gamma \frac{\partial P_3}{\partial t} + 2P_3(a_1 - Q_{12}(\sigma_{11} + \sigma_{22}) - Q_{11}\sigma_{33}) - Q_{44}(\sigma_{13}P_1 + \sigma_{23}P_2) + 4a_{11}P_3^3 + 2a_{12}P_3(P_1^2 + P_2^2) +$$

$$6a_{111}P_3^5 + 2a_{112}P_3(P_1^4 + 2P_3^2P_1^2 + P_2^4 + 2P_2^2P_3^2) + 2a_{123}P_3P_1^2P_2^2 - g_{11}\frac{\partial^2 P_3}{\partial x_3^2} - g_{44}\left(\frac{\partial^2 P_3}{\partial x_1^2} + \frac{\partial^2 P_3}{\partial x_2^2}\right) =$$

$$-F_{11}\frac{\partial \sigma_{33}}{\partial x_3} - F_{12}\left(\frac{\partial \sigma_{11}}{\partial x_3} + \frac{\partial \sigma_{33}}{\partial x_3}\right) - F_{44}\left(\frac{\partial \sigma_{13}}{\partial x_1} + \frac{\partial \sigma_{23}}{\partial x_2}\right) + E_3 \quad \text{(A.4c)}$$

The temperature-dependent Khalatnikov coefficient $\Gamma$ [10] determines the relaxation time of the polarization $\tau_K = \Gamma/|\alpha|$, where $\alpha(T) = \alpha_T[T - T_C]$. Consequently, $\tau_K$ typically varies in the range ($10^{-9} - 10^{-6}$) seconds for temperatures far from $T_C$. As argued by Hlinka et al. [11], we assumed that $g'_{44} = -g_{12}$ in Eqs.(A.4).

The boundary condition for polarization at the core-shell interface $r = R$ accounts for the flexoelectric effect:

$$a_{ij}^{(S)}P_j + \left(g_{ijkl}\frac{\partial P_k}{\partial x_l} - F_{klij}\sigma_{kl}\right)n_j\bigg|_{r=R} = 0 \quad \text{(A.5)}$$

where **n** is the outer normal to the surface, $i=1, 2, 3$. In our FEM studies, we use the so-called "natural" boundary conditions corresponding to $a_{ij}^{(S)} = 0$. Under the condition of the negligibly small term $F_{klij}\sigma_{kl}n_j\big|_{r=R} \approx 0$, which corresponds to the absence of either normal stress and/or zero flexoelectric coupling and specific properties of $g_{ijkl}$, the condition $\frac{\partial P_n}{\partial \mathbf{r}}\big|_{r=R} = 0$ becomes incompatible with the condition $P_n|_{r=R} = 0$. This means that the regions with **P**=0 (if any exist for the condition) can be located near the surface, but not directly at the surface.

Elastic stresses satisfy the equation of mechanical equilibrium in the computation region,

$$\frac{\partial \sigma_{ij}}{\partial x_j} = 0, \qquad -L/2 < \{x, y, z\} < L/2. \tag{A.6a}$$

Elastic equations of state follow from the variation of the energy (A.2e) with respect to elastic stress, $\frac{\delta G}{\delta \sigma_{ij}} = -u_{ij}$, namely:

$$s_{ijkl}\sigma_{ij} + Q_{ijkl}P_k P_l + F_{ijkl}\frac{\partial P_l}{\partial x_k} = u_{ij} - \beta_T^c \Delta T \delta_{ij}, \qquad 0 < r \leq R, \tag{A.6b}$$

$$s_{ijkl}^S \sigma_{ij} + \varepsilon_0^2(\varepsilon_s - 1)^2 Q_{ijkl}^S E_k E_l + W_{ij}^V N_V^+ = u_{ij} - u_{ij}^m - \beta_T^s \Delta T \delta_{ij}, \qquad R < r \leq R + \Delta R, \tag{A.6c}$$

$$s_{ijkl}^e \sigma_{ij} = u_{ij}, \qquad r > R + \Delta R, \tag{A.6d}$$

where $u_{ij}$ is the strain tensor components related to displacement components $U_i$ in the following way: $u_{ij} = (\partial U_i/\partial x_j + \partial U_j/\partial x_i)/2$. The terms $\beta_T^{c,s} \Delta T \delta_{ij}$ originate from the linear thermal strains in the core (superscript C) and shell (superscript S), where $\beta_T^{c,s}$ are the coefficients of linear thermal expansion and $\Delta T = T - T_g$ is the difference between the surrounding temperature and the system growth/deposition temperature. The strain $u_{ij}^m$ is proportional the core and shell lattice constants mismatch taken at the shell deposition temperature, i.e. $u_{ij}^m = \frac{a_c - a_s}{a_c}\delta_{ij}$.

The mechanical boundary conditions for the elastic sub-problem are listed below. The elastic displacement components $U_i$ and normal stresses $\sigma_{ij}$ are continuous functions at the core-shell interface ($r = R$):

$$U_i|_{r=R-0} = U_i|_{r=R+0}, \qquad \sigma_{ij}n_j\big|_{r=R-0} = \sigma_{ik}n_k|_{r=R+0}, \tag{A.6e}$$

as well as at the interface between the shell and the external media ($r = R + \Delta R$):

$$U_i|_{r=R+\Delta R-0} = U_i|_{r=R+\Delta R+0}, \qquad \sigma_{ij}n_j\big|_{r=R+\Delta R-0} = \sigma_{ik}n_k|_{r=R+\Delta R+0}. \tag{A.6f}$$

All forces are absent at the surface of the computational region:

$$\sigma_{kl}n_l\big|_{x=\pm\frac{L}{2}} = 0, \quad \sigma_{kl}n_l\big|_{y=\pm\frac{L}{2}} = 0, \quad \sigma_{kl}n_l\big|_{z=\pm\frac{L}{2}} = 0 \tag{A.6g}$$

Here we consider a tunable shell of paraelectric strontium titanate (SrTiO$_3$), which has an isotropic and strongly temperature-dependent dielectric permittivity, $\varepsilon_{ij}^S = \delta_{ij}\varepsilon_s$, with the following expression

$$\varepsilon_s(T) = \frac{1}{\varepsilon_0 \alpha_T T_q^{(E)}}\left(\coth\left(\frac{T_q^{(E)}}{T}\right) - \coth\left(\frac{T_q^{(E)}}{T_0^{(E)}}\right)\right)^{-1}, \tag{A.7}$$

where the Curie-Weiss parameter $\alpha_T = 0.75\times10^6$ m/(F K) and characteristic temperatures $T_0^{(E)} = 30$ K and $T_q^{(E)} = 54$ K [12]. It should be noted that $\varepsilon_s(T) \approx 3000$ at $T = 50$ K and $\varepsilon_s(T) \approx 300$ at T = 298 K allow the spontaneous polarization of the ferroelectric core to be effectively screened by the tunable shell at room and lower temperatures. Other parameters are listed in **Table AI.**

**Table AI.** LGD coefficients and other material parameters of a BaTiO$_3$ core covered with a SrTiO$_3$ shell

| Coefficient | Numerical value |
|---|---|
| $\varepsilon_{b, e}$ | $\varepsilon_b$ = 7 (core background)    $\varepsilon_e$ = 10 (surrounding) |
| $a_i$  (C$^{-2}$·mJ) | $a_1$ = 3.34(T–381)×10$^5$,   $\alpha_T$ = 3.34×10$^5$        ($a_1$ = –2.94×10$^7$ at 298°K) |
| $a_{ij}$  (C$^{-4}$·m$^5$J) | $a_{11}$ = 4.69(T–393)×10$^6$–2.02×10$^8$, $a_{12}$ = 3.230×10$^8$, <br> (at 298°K $a_{11}$ = –6.71×10$^8$, $a_{12}$ = 3.23×10$^8$) |
| $a_{ijk}$ (C$^{-6}$·m$^9$J) | $a_{111}$ = –5.52(T–393)×10$^7$+2.76×10$^9$, $a_{112}$ = 4.47×10$^9$, $a_{123}$ = 4.91×10$^9$ <br> (at 298°K $a_{111}$ = 82.8×10$^8$, $a_{112}$ = 44.7×10$^8$, $a_{123}$ = 49.1×10$^8$) |
| $Q_{ij}$ (C$^{-2}$·m$^4$) | $Q_{11}$=0.11, $Q_{12}$= –0.043, $Q_{44}$=0.059 |
| $s_{ij}$  (×10$^{-12}$ Pa$^{-1}$) | $s_{11}$=8.3, $s_{12}$= –2.7, $s_{44}$=9.24 |
| $g_{ij}$  (×10$^{-10}$C$^{-2}$m$^3$J) | $g_{11}$=1.0, $g_{12}$= 0.3, $g_{44}$= 0.2 |
| $F_{ij}$ (×10$^{-11}$C$^{-1}$m$^3$) <br> $f_{ij}$ (V) | $F_{11}$ = 2, $F_{12}$ = 1.8, $F_{44}$ = 6 (these values are used as estimates, exact values are unknown)  $f_{11}$ = 6.6, $f_{12}$ = 6.4, $f_{44}$ = 6.5 |
| $v_{ijklm}$ | 0 (since its characteristic values are unknown for BaTiO$_3$ and other perovskites) |
| $a_i^{(s)}$ | 0 (that corresponds to the so-called natural boundary conditions) |
| $\beta_T^{(c)}$(10$^{-6}$K$^{-1}$) | 9.8 (thermal expansion coefficient) |
| $a_{cubic}^{(c)}$ | 4.035 Å lattice constant at 1000 °C |
| $R$ (nm) | 10 (vary from 2 to 20 nm) |
| **Electric parameters of the SrTiO$_3$ tunable shell** | |
| $\varepsilon_s(T)$ | $\left(\varepsilon_0 \alpha_T T_q^{(E)}\right)^{-1} \left(\coth\left(T_q^{(E)}/T\right) - \coth\left(T_q^{(E)}/T_0^{(E)}\right)\right)^{-1}$ |
| $\alpha_T$ (10$^6$ m/(F K)) | 0.75 |
| $T_{0,q}^{(E)}$ (K) | $T_0^{(E)}$ = 30 K,  $T_q^{(E)}$ = 54 K |
| **Elastic parameters of the "soft" shell (bulk compound)** | |
| $Q_{ij}^{(s)}$ (C$^{-2}$·m$^4$) | $Q_{11}^{(s)}$ = 0.051, $Q_{12}^{(s)}$ = –0.016, $Q_{44}^{(s)}$ = 0.020 |
| $s_{ij}^{(s)}$ (10$^{-12}$ Pa$^{-1}$) | $\left\|s_{ij}^{(s)}\right\|$ > 10$^{-8}$ Pa$^{-1}$ |
| $\beta_T^{(s)}$(10$^{-6}$K$^{-1}$) | 10.8 (thermal expansion coefficient) |
| $a_{cubic}^{(s)}$ | 3.946 Å lattice constant at 1000 °C |
| **Elastic parameters of the "rigid" SrTiO$_3$shell (bulk crystalline)** | |
| $Q_{ij}^{(s)}$ (C$^{-2}$·m$^4$) | $Q_{11}^{(s)}$ = 0.051, $Q_{12}^{(s)}$ = –0.016, $Q_{44}^{(s)}$ = 0.020 |
| $s_{ij}^{(s)}$ (10$^{-12}$ Pa$^{-1}$) | $s_{11}^{(s)}$ = 3.52, $s_{12}^{(s)}$ = –0.85, $s_{44}^{(s)}$ = 7.87 |
| $W_{ij}^V$ (Å$^3$) | $W_{11}^V = 16.33, W_{22}^V = -8.05, W_{33}^V = -8.05$ |
| $\Delta R$ (nm) | 4 (vary from 4 to 10 nm) |
| $R_d$ (nm) | >100 nm (shell is a paraelectric material) |

### A2. The impact of oxygen vacancies

The gain $\delta G$ of the electrochemical part of the free energy has the form:

$$\delta G = -Q_{ijkl}^{(m)} P_k P_l \sigma_{ij} - \frac{s_{ijkl}^{(m)}}{2} \sigma_{ij} \sigma_{kl} - N_V W_{ij}^V \sigma_{ij} + k_B T \left[ N_V^+ \ln\left(\frac{N_V^+}{N_V}\right) + (N_V - N_V^+) \ln\left(\frac{N_V - N_V^+}{N_V}\right) \right] - $$
$$\left(Z_V^{eff} N_V^+ - en\right)\varphi, \tag{A.8}$$

where the superscript $m = s$ represents the shell and $m = c$ represents the core.

The continuity equation for the vacancy concentration $N_V^+$ is:

$$\frac{d}{dt}N_V^+ + div\boldsymbol{J} = 0, \qquad 0 < r \leq R + \Delta R. \tag{A.9a}$$

The current $\boldsymbol{J}$ is proportional to the gradients of the electrochemical potential levels $\xi$ according to

$$\boldsymbol{J} = -\eta N_V^+ grad(\xi), \tag{A.9b}$$

where $\eta$ is the mobility coefficient which is a constant. The electrochemical potential level $\xi$ is defined as

$$\xi = \frac{\partial G}{\partial N_V^+} = \xi_0 + k_B T \ln\left(\frac{N_V^+}{N_V - N_V^+}\right) - W_{ij}^V \sigma_{ij} - Z_V^{eff} \varphi, \tag{A.9c}$$

where $\xi_0$ is the equilibrium value (e.g. the Fermi level defined at $\varphi = 0$).

In the static case $\boldsymbol{J} = 0$ and $\xi = \xi_0$, and the substitution of the latter condition in Eq.(A.9c) yields:

$$k_B T \ln\left(\frac{N_V^+}{N_V - N_V^+}\right) = W_{ij}^V \sigma_{ij} + Z_V^{eff} \varphi, \quad N_V^+ = N_V \left(1 + exp\left[-\frac{W_{ij}^V \sigma_{ij} + Z_V^{eff} \varphi}{k_B T}\right]\right)^{-1}. \tag{A.10a}$$

From Eqs.(A.6b-c) and Eq.(A.10a) we obtain:

$$s_{ijkl}^{(m)} \sigma_{kl} + W_{ij}^V N_V \left(1 + exp\left[-\frac{W_{kl}^V \sigma_{kl} + Z_V^{eff} \varphi}{k_B T}\right]\right)^{-1} \approx \delta u_{ij}^{(m)}, \tag{A.10b}$$

where $\delta u_{ij}^s = u_{ij} - u_{ij}^m - \beta_T^s \Delta T \delta_{ij} - \varepsilon_0^2 (\varepsilon_s - 1)^2 Q_{ijkl}^S E_k E_l$ and $\delta u_{ij}^c = u_{ij} - \beta_T^c \Delta T \delta_{ij} - Q_{ijkl}^c P_k P_l$.

Assuming the local electroneutrality condition we can regard $\varphi \approx 0$, and also assuming $|W_{ij}^V \sigma_{ij}| \ll k_B T$, we can expand the expression (A.10a) for $N_V^+$ as $N_V^+ \approx N_V \left(\frac{1}{2} + \frac{W_{kl}^V \sigma_{kl}}{4 k_B T}\right)$. Being interested only in the renormalization of elastic compliances caused by mobile charged vacancies, we can rewrite the left-hand side of Eq.(A.10b) as

$$s_{ijkl}^{(m)} \sigma_{kl} + W_{ij}^V N_V^+ \cong \left(s_{ijkl}^{(m)} + W_{ij}^V W_{kl}^V \frac{N_V}{4 k_B T}\right) \sigma_{kl} + stress - independent\ terms. \tag{A.10c}$$

It is seen from Eq.(A.10c) that the effect of vacancy migration is mainly the renormalization of elastic compliances:

$$s_{ijkl}^{(R)} = s_{ijkl}^{(m)} + W_{ij}^V W_{kl}^V \frac{N_V}{4 k_B T}. \tag{A.11a}$$

Note that Eq.(A.11a) can substantially overestimate or underestimate the vacancies' role, because the condition $|W_{ij}^V \sigma_{ij}| \ll k_B T$ is frequently violated, and the inequality $|W_{ij}^V \sigma_{ij}| \geq k_B T$ becomes possible e.g. with decreasing temperature. In the case where $W_{ij}^V \sigma_{ij} \ll -k_B T$, we obtain $s_{ijkl}^{(m)} \sigma_{kl} \approx \delta u_{ij}^{(m)}$ from Eq.(A.10b). This means that the vacancies do not affect the stress field, which is possible for small concentrations $N_V$ and/or small Vegard tensor components. In the opposite case, $W_{ij}^V \sigma_{ij} \gg k_B T$, we obtain $s_{ijkl}^{(m)} \sigma_{kl} \approx \delta u_{ij}^{(m)} - W_{ij}^V N_V$ from Eq.(A.10b). This means that the mobile oxygen vacancies can completely shield the stress field [8], if $\left|\delta u_{ij}^{(m)}\right| \leq \left|W_{ij}^V N_V\right|$.

Let us return to the case when the estimate (A.11a) is valid and correctly reflects the system tendency to reach elastic equilibrium. For the case of an m3m symmetry cubic material and diagonal Vegard strain, the nontrivial renormalization components are:

$$s_{11}^{(R)} = s_{11}^{(m)} + (W_{11}^V)^2 \frac{N_V}{4k_BT}, \quad s_{22}^{(R)} = s_{22}^{(m)} + (W_{22}^V)^2 \frac{N_V}{4k_BT}, \quad s_{33}^{(R)} = s_{33}^{(m)} + (W_{33}^V)^2 \frac{N_V}{4k_BT} \quad \text{(A.11b)}$$

$$s_{12}^{(R)} = s_{12}^{(m)} + W_{11}^V W_{22}^V \frac{N_V}{4k_BT}, \quad s_{13}^{(R)} = s_{12}^{(m)} + W_{11}^V W_{33}^V \frac{N_V}{4k_BT}, \quad s_{23}^{(R)} = s_{23}^{(m)} + W_{22}^V W_{33}^V \frac{N_V}{4k_BT} \quad \text{(A.11c)}$$

$$s_{44}^{(R)} = s_{44}^{(m)}, \quad s_{55}^{(R)} = s_{55}^{(m)}, \quad s_{66}^{(R)} = s_{66}^{(m)} \quad \text{(A.11d)}$$

Expressions (A.11) must be averaged over all possible orientations of the elastic dipole in a cubic perovskite ABO$_3$ lattice, where an oxygen vacancy can occupy several equivalent sites corresponding to the oxygen octahedron vertices. The averaging is, in fact, an averaging over six possible "orientations" of the anisotropic Vegard tensor. The result for the diagonal components is

$$\langle s_{ii}^{(R)} \rangle = s_{ii}^{(m)} + [(W_{11}^V)^2 + (W_{22}^V)^2 + (W_{33}^V)^2] \frac{N_V}{12k_BT}, \quad \text{(A.12a)}$$

where $ii = 11, 22, 33$. The result for non-diagonal components is

$$\langle s_{ij}^{(R)} \rangle = s_{ij}^{(m)} + (W_{11}^V W_{22}^V + W_{33}^V W_{22}^V + W_{11}^V W_{33}^V) \frac{N_V}{12k_BT}, \quad \text{(A.12b)}$$

where $ij = 12, 23, 13$. The shear components, Eqs.(A.11d), remains unchanged. Note that the combination of elastic compliances, $\langle s_{eff}^{(R)} \rangle = \langle s_{11}^{(R)} \rangle + 2\langle s_{12}^{(R)} \rangle$, coupled with a radial stress, $\langle \sigma_{rr}^{(R)} \rangle \cong \langle \sigma_{11}^{(R)} \rangle + \langle \sigma_{22}^{(R)} \rangle + \langle \sigma_{33}^{(R)} \rangle$, is:

$$\langle s_{eff}^{(R)} \rangle = \langle s_{11}^{(R)} \rangle + 2\langle s_{12}^{(R)} \rangle = s_{11}^{(m)} + 2s_{12}^{(m)} + (W_{11}^V + W_{22}^V + W_{33}^V)^2 \frac{N_V}{12k_BT}, \quad \text{(A.12c)}$$

For the case of oxygen vacancies in SrTiO$_3$, the Vegard strain tensor is diagonal with components taken from [8] and listed in **Table AI**. It is instructive to measure the vacancy concentration $N_V$ in percent of molar concentration $N_m$=1.56·10$^{28}$ m$^{-3}$ (the inverse volume of the unit cell). The change of "effective" elastic compliances $\langle s_{ii}^{(R)} \rangle \approx s_{ii}^{(m)} + (W_{11}^V)^2 \frac{N_V}{8k_BT}$ caused by the Vegard strains created by oxygen vacancies with concentration (2 – 5)% can reach one order of magnitude at room temperature**.** The change of effective elastic compliances $\langle s_{ij}^{(R)} \rangle \approx s_{ij}^{(m)} - (W_{11}^V)^2 \frac{N_V}{16k_BT}$ are smaller in SrTiO$_3$, since $W_{11}^V \approx -2W_{22}^V \approx -2W_{33}^V$ in accordance with **Table AI**.

The numerical solution of the nonlinear Eq.(A.10b) with respect to the unknown stress $\sigma_{kl}$ in response to thermal strains $\delta u_{ij}^{(m)}$ demonstrates a significant decrease of the radial stress $\langle \sigma_{rr}^{(R)} \rangle$ and a simultaneous increase of the effective elastic compliances $\langle s_{eff}^{(R)} \rangle$, with an increase of vacancy concentration (see **Fig. A1**). The increase is much stronger than predicted by the approximate expression (A.12a), which yields $|W_{11}^V + W_{22}^V + W_{33}^V| \cong 0.23$ Å$^3$ for SrTiO$_3$ [8]. Note that the increase exists for a positive sign of the Vegard tensor anisotropy factor $W = \pm \frac{1}{3}\sqrt{(W_{11}^V - W_{22}^V)^2 + (W_{22}^V - W_{33}^V)^2 + (W_{33}^V - W_{11}^V)^2}$ (see red curves in **Fig. A1**), and is absent for a negative $W$ (see blue curves in **Fig. A1**). The asymmetry with respect to the sign of $W$ is explained by a positive sign of thermal strains, $\beta_T^{(m)} \Delta T \delta_{ij}$, in both core and shell materials. For negative values of $\beta_T^{(m)} \Delta T \delta_{ij}$, which is a very rare case, an increase in both the elastic stress compensation and effective compliances is possible for negative $W$ only.

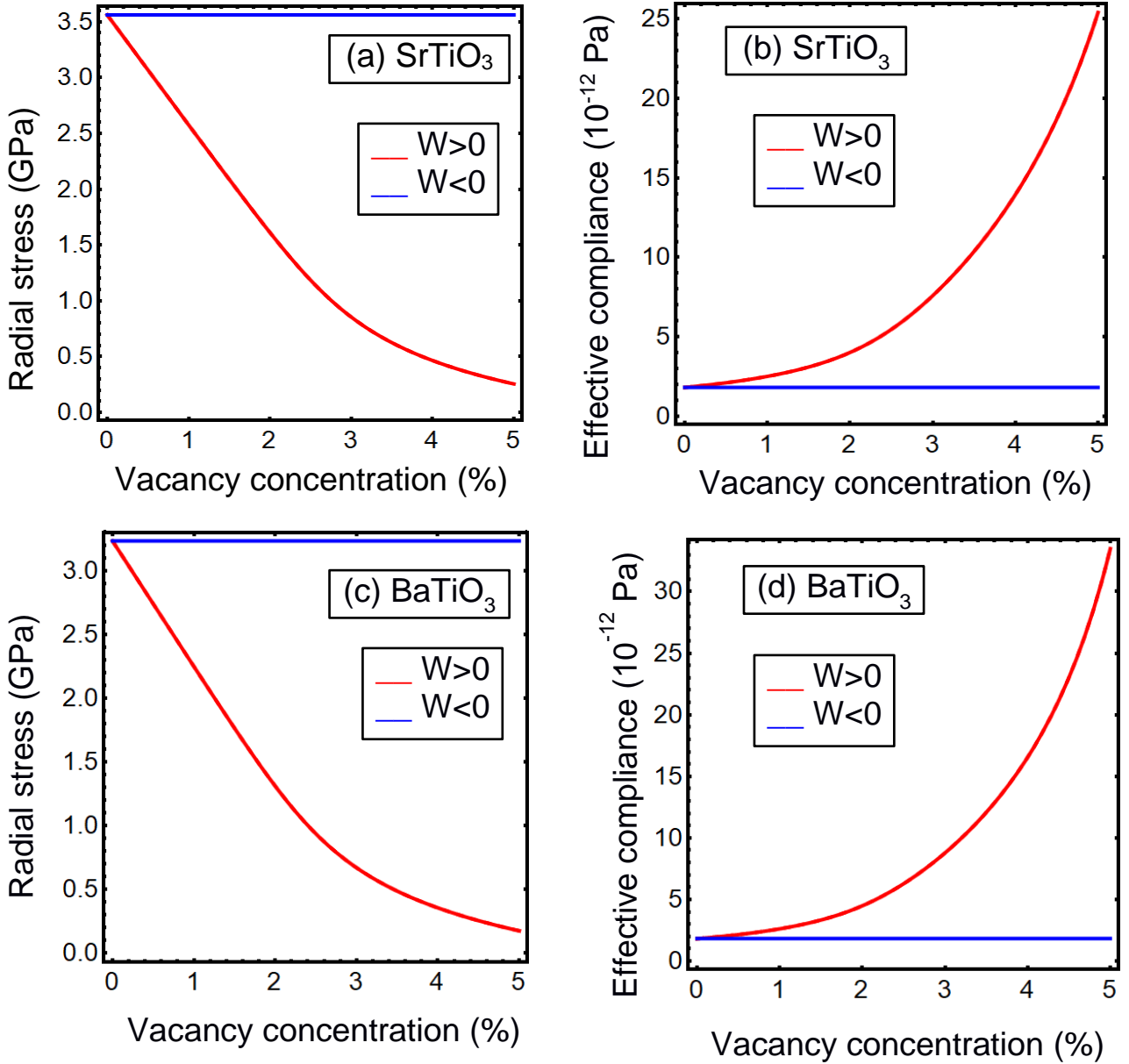

**Figure A1.** The radial stress $\langle \sigma_{rr}^{(R)} \rangle$ **(a, c)** and effective elastic compliances $\langle s_{eff}^{(R)} \rangle$ **(b, d)** calculated numerically as a function of the vacancy concentration (in %) from Eqs.(A.10b) for SrTiO$_3$ **(a, b)** and BaTiO$_3$ **(c, d)**. A mismatch strain is absent. Other parameters are listed in **Table A1**.

### A3. The impact of flexoelectricity and gradient effects

The polarization gradient energy and flexoelectric coupling energy are

$$g_{grad} = \frac{g_{ijkl}}{2} \frac{\partial P_i}{\partial x_j} \frac{\partial P_k}{\partial x_l}, \qquad g_{flexo} = \frac{f_{ijkl}}{2} u_{ij} \frac{\partial P_k}{\partial x_l}, \qquad (A.13)$$

where the elastic strains $u_{ij} = s_{ijkl}\sigma_{kl} + Q_{ijkl}P_k P_l + F_{ijkl}\frac{\partial P_k}{\partial x_l}$. The substitution of the electrostrictive part of the strains in the case when the stress-related term $s_{ijkl}\sigma_{kl}$ can be regarded as small (it is the simplest model) leads to:

$$\delta g_{grad+flexo} = \frac{g_{ijkl}}{2}\frac{\partial P_i}{\partial x_j}\frac{\partial P_k}{\partial x_l} + \frac{f_{ijkl}}{2}\left(Q_{ijmn}P_m P_n + F_{ijmn}\frac{\partial P_m}{\partial x_n}\right)\frac{\partial P_k}{\partial x_l}$$

$$\cong \left(\frac{g_{ijkl}}{2} + \frac{f_{qskl}}{2}F_{qsij}\right)\frac{\partial P_i}{\partial x_j}\frac{\partial P_k}{\partial x_l} \equiv \frac{1}{2}\left(g_{ijkl} + s_{qsmn}f_{qskl}f_{mnij}\right)\frac{\partial P_i}{\partial x_j}\frac{\partial P_k}{\partial x_l} \equiv \frac{1}{2}g'_{ijkl}\frac{\partial P_i}{\partial x_j}\frac{\partial P_k}{\partial x_l}. \quad (A.14)$$

Since the term $\frac{f_{ijkl}}{2}Q_{ijmn}P_m P_n \frac{\partial P_k}{\partial x_l}$ has (almost) a zero average it can be omitted, then a renormalized gradient coefficient $g'_{ijkl} = g_{ijkl} + s_{qsmn}f_{qskl}f_{mnij}$ can be introduced. The renormalization has different signs for the diagonal and non-diagonal components, but for the cases of interest $g'_{11} = g_{11} + s_{qqmm}f_{qq11}f_{mm11}$ and $g'_{44} = g_{44} + s_{qqmm}f_{qq44}f_{mm44}$, it (typically) increases $g_{11}$ and decreases $g_{44}$. For a cubic symmetry of the parent phase, the trend $g'_{11} > g_{11}$ and $g'_{44} < g_{44}$ is responsible for an increase of intrinsic width of the charged domain walls/structures/configurations, and a decrease of the width of uncharged domain walls. The formation of uncharged domain configurations, which are the most common and are significantly more preferable from an energetic viewpoint [13, 14], is affected by the flexoelectricity. In particular, the flexoelectricity induces the domain wall curvature and meandering in multiaxial ferroelectrics, and facilitates labyrinthine domain configurations in uniaxial ferroelectrics at $g'_{ijkl} < g^{cr}_{ijkl}$ (see e.g. Refs. [15, 16, 17]). In addition to influencing the wall shape, the flexoelectricity (due to the condition $g'_{44} < g_{44}$) increases (but not very strongly) the transition temperature from the ferroelectric to paraelectric phase (see e.g. [15, 18]). Another role of the flexoelectricity comes from the inhomogeneous boundary conditions in strained nanoparticles [see Eq.(A.5) and e.g. Ref. [18] for details]. The inhomogeneity, which is proportional to the flexoelectric coupling strength, can lead to the appearance of built-in inhomogeneous flexoelectric fields with specific geometries.

### A.3. FEM results: energy contributions and supplementary figures

**Table AII.** Energy contributions (in $10^{-18}$ J) of different polarization states in a BaTiO$_3$ ferroelectric core ($R$=10 nm) covered by different shells ($\Delta R$=4 nm), $T$ = 293 K

| Description of the elastic sub-problem in a core-shell nanoparticle | Total energy, $G$ | Landau energy, $G_{Landau}$ | Polarization gradient energy, $G_{grad}$ | Depolarization field energy*, $G_{dep}$ | Flexo-Elastic energy, $G_{es+flexo}$ | Figure Number |
|---|---|---|---|---|---|---|
| Electrostriction and flexoeffect are absent in the core covered by a soft shell | -8.754 | -9.174 | 0.415 | 0.005 | <0.001 | Fig.2 |

| Description | | | | | | |
|---|---|---|---|---|---|---|
| Flexoeffect is absent but anisotropic electrostriction is present in the core covered by a soft shell | -6.658 | -8.483 | 0.655 | 0.080 | 1.091 | Fig.3 |
| Flexoeffect and anisotropic electrostriction are present in the core covered by a soft shell [§] | -6.912 [(r)] | -8.443 | 0.722 | 0.075 | 0.734 | Fig.4 |
| Mismatch strain and flexoeffect are absent in the core covered by a rigid shell | -5.990 | -7.601 | 0.462 | 0.047 | 1.102 | Fig.5 |
| Mismatch strain is absent, but flexoeffect is present in the core covered by a rigid shell | -6.035 | -7.618 | 0.471 | 0.070 | 1.043 | Fig.6 |
| Mismatch strain and flexoeffect are present in the core covered by a rigid shell | -186.6 | -5.444 | 1.291 | 0.718 | -183.1, | Fig.7 |

[§] with axis close to [011]

[*] $G_{dep} = -\int_{0<r<R} d^3r \frac{P_i E_i}{2}$, other energies are introduced by Eqs.(A.2)

**Soft shell case.** Without taking into account the influence of the flexoelectric effect, the domain structure of the ferroelectric core in a **soft shell** can be imagined as follows. Near each of the core poles, two 180º domains separated by a flat wall are observed (see **Fig. A2**). Inside the core the domain polarization is $P_2$; and the analogs of the flux-closure domains with polarization $P_1$ appear when the bound polarization charges approach the core surface near the poles. In addition, the width of the closure domain increases with distance from the pole such that the distributions of $P_1$ and $P_2$ turn out to be almost equivalent near the equatorial plane, where the structure resembles a vortex without clear boundaries between the regions with different polarization directions. At the same time, the polarization in the surface layers unfolds parallel to the surface (in fact, the component $P_3$ is along the z axis, on which both poles of the domain structure lie). In other words, near the center of the core there is a vortex-like structure, where the polarization rotates in one plane about a fixed axis. Near the poles, which are defined as the points of exit of the vortex axis to the surface, the rotation of the polarization degenerates into a pair of 180º domains of the tetragonal phase. In addition, the domains near the two poles are completely different. Several polarization components coexist near the equator, and these "pseudo-domains" correspond to a certain pseudo-phase with symmetry below tetragonal.

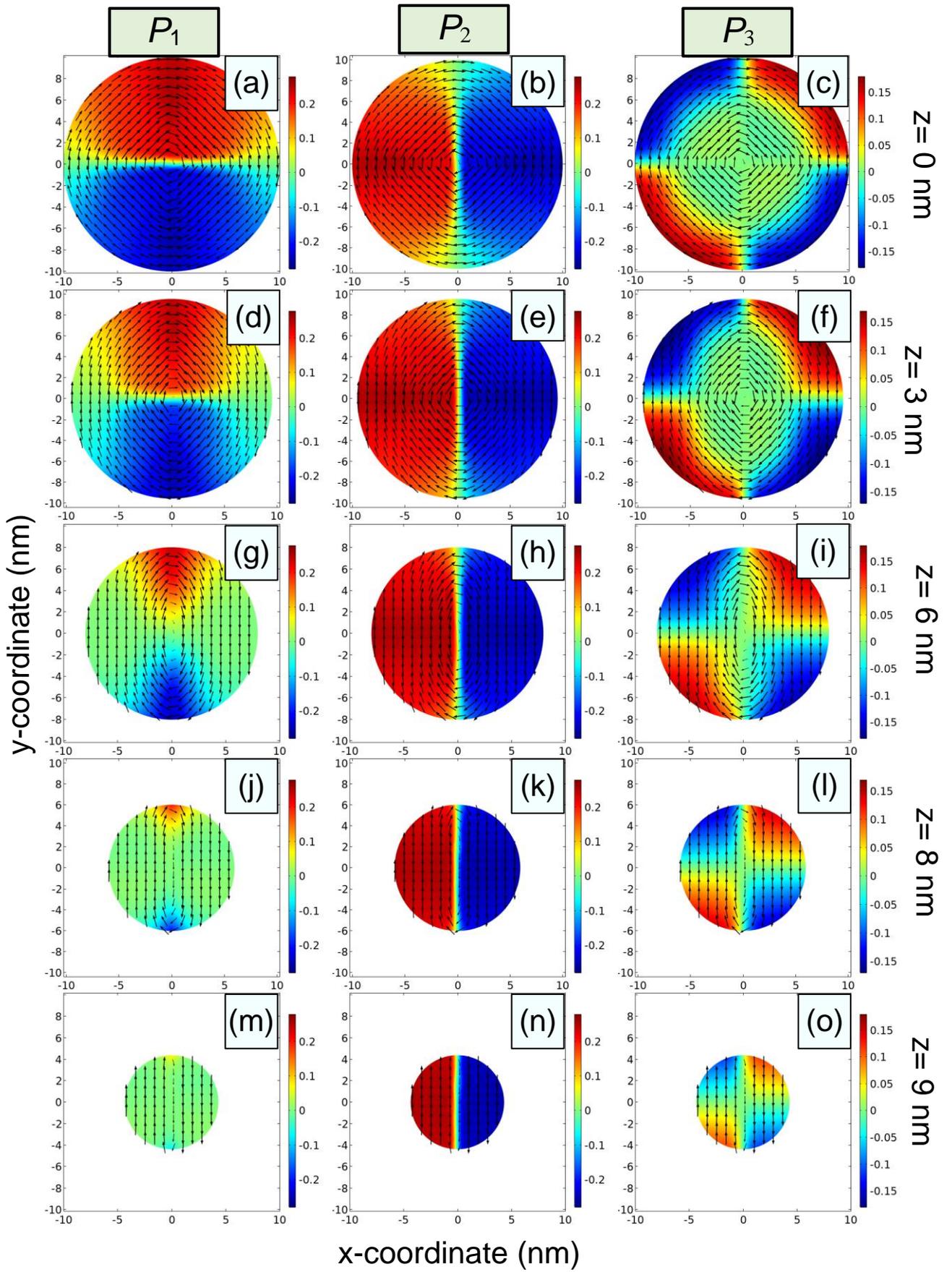

**Figure A2**. **The case of a soft shell is considered. Flexoelectric coefficients are zero.** Distribution of polarization components $P_1$ (a, d, g, j, m), $P_2$ (b, e, h, k, n), and $P_3$ (c, f, i, l, o) in the cross-sections XY for different values of z coordinate: z = 0 nm (a,b,c), 3 nm (d, e, f), 6 nm (g, h, i), 8 nm (j, k, l), and 9 nm (m, n, o), which is the distance

from the equatorial plane. Core radius $R = 10$ nm, shell thickness $\Delta R = 4$ nm, and $T = 298$ K. The biggest cross-section corresponds to the equatorial plane, and the smallest cross-section is near the pole.

**Rigid shell case.** Without taking into account the flexoelectric effect, the domain structure of the ferroelectric core covered by a rigid shell consists of six blurred domains. The boundaries between the domains become relatively sharp only near the particle "poles", which are defined as the points at the core surface where the polarization vector modulus drops to zero. The picture is shown in **Fig. A3**, where we use the rotated coordinate frame with the following coordinates, $\xi = (x - y)/\sqrt{2}$, $\psi = (x + y - 2z)/\sqrt{6}$, and $\omega = (x + y + z)/\sqrt{3}$. The axis $\omega$ is pointed along the six-feature vortex-like configuration. Three 120º domains separated by flat walls are observed near the poles. However, the domains and their walls are different for different poles; in fact, one group of domains is rotated by 60º relative to the other. Moving away from each of the poles, the domain walls broaden and blur, such that these regions evolve into domains with a different orientation. Near the equatorial plane, all six domains are equivalent, so that the configuration of the polarization vector becomes vortex-like; but the symmetry of the walls is more complicated than that of 120º domains, since the polarization component along the polar axis of the core is quite large.

For zero flexoelectric coupling the internal electric field, which is in fact a depolarization field, is very small. This field is small due the polarization rotation inside the vortex (see **Fig. A4a-c**). Actually, **P** rotates in such way that $div\mathbf{P} \approx 0$ in the core covered by a soft shell. The bound charges, whose density $\rho_b$ is equal to $-div\mathbf{P}$, are almost absent (see **Fig. A5a-c**). The condition $div\mathbf{E} \approx 0$ follows from the zero divergence of electric displacement and polarization, and tiny deviations from $div\mathbf{D} = 0$ are caused by computational error.

The joint action of flexoelectric coupling and mismatch strain cause the appearance of a relatively strong electric field, which is well-localized at the core surface (see **Fig. A4d**). Consequently, the bound charges with density $\rho_b = -div\mathbf{P}$ can be considered as surface charges (see **Fig. A5d**).

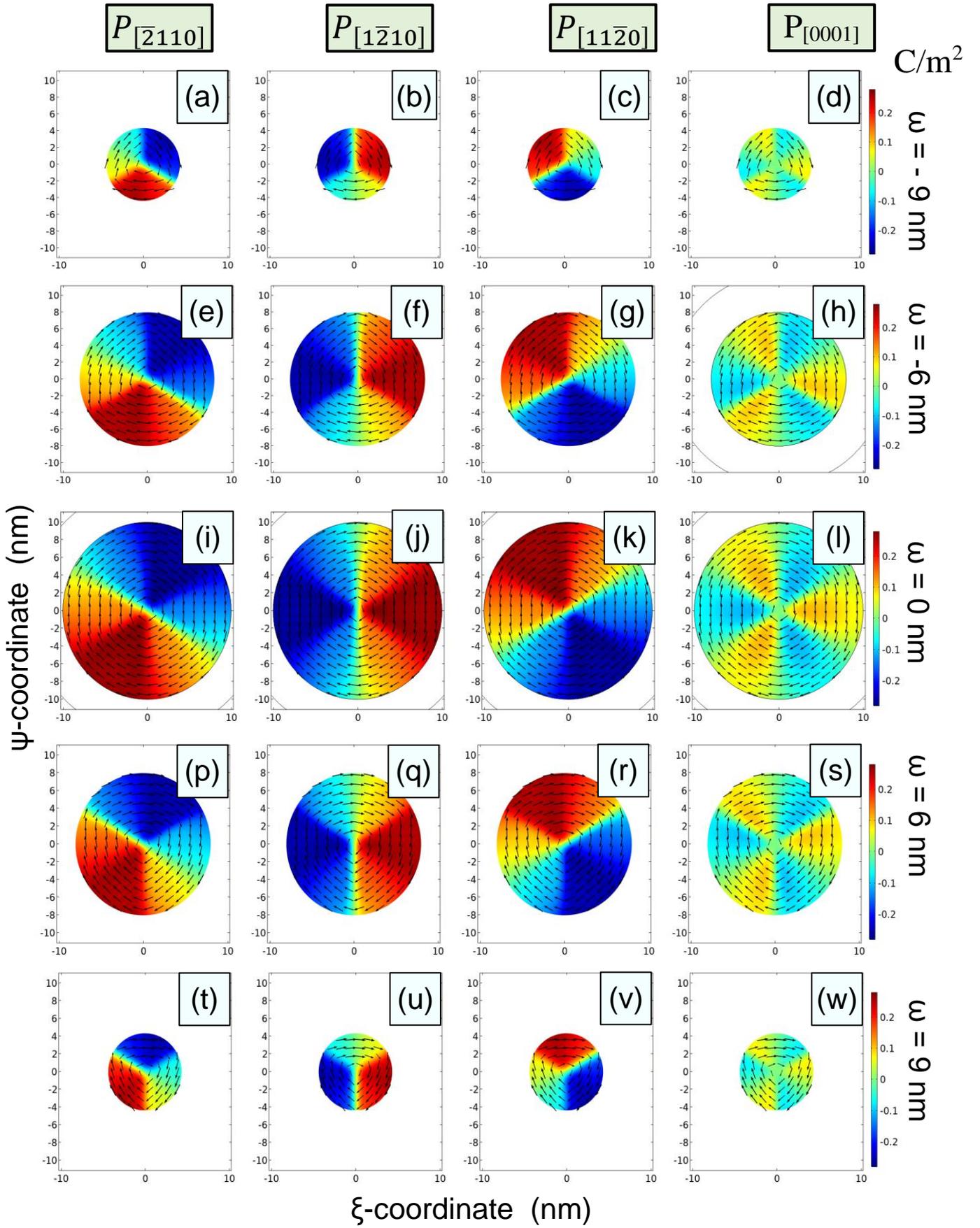

**Figure A3. The case of a rigid shell is considered. Flexoelectric coefficients are zero.** The distribution of polarization projection on the directions [$\bar{2}110$] (a, e, i, p, t), [$1\bar{2}10$] (b, f, j, q, u), [$11\bar{2}0$] (c, g, k, r, v), and [0001] (d, h, l, s, w) in the cross-section planes {111} for different values of ω coordinate: $\omega = -9$ nm (a, b, c, d), -6 nm

(e, f, g, h,), 0 nm (i, j, k, l), 6 nm (p, q, r, s), and 9 nm (t, u, v, w), which is the distance from equatorial plane. Core radius $R = 10$ nm, shell thickness $\Delta R = 4$ nm, and $T = 298$ K. The biggest cross-section corresponds to the equatorial plane, and the smallest cross-section is near the pole.

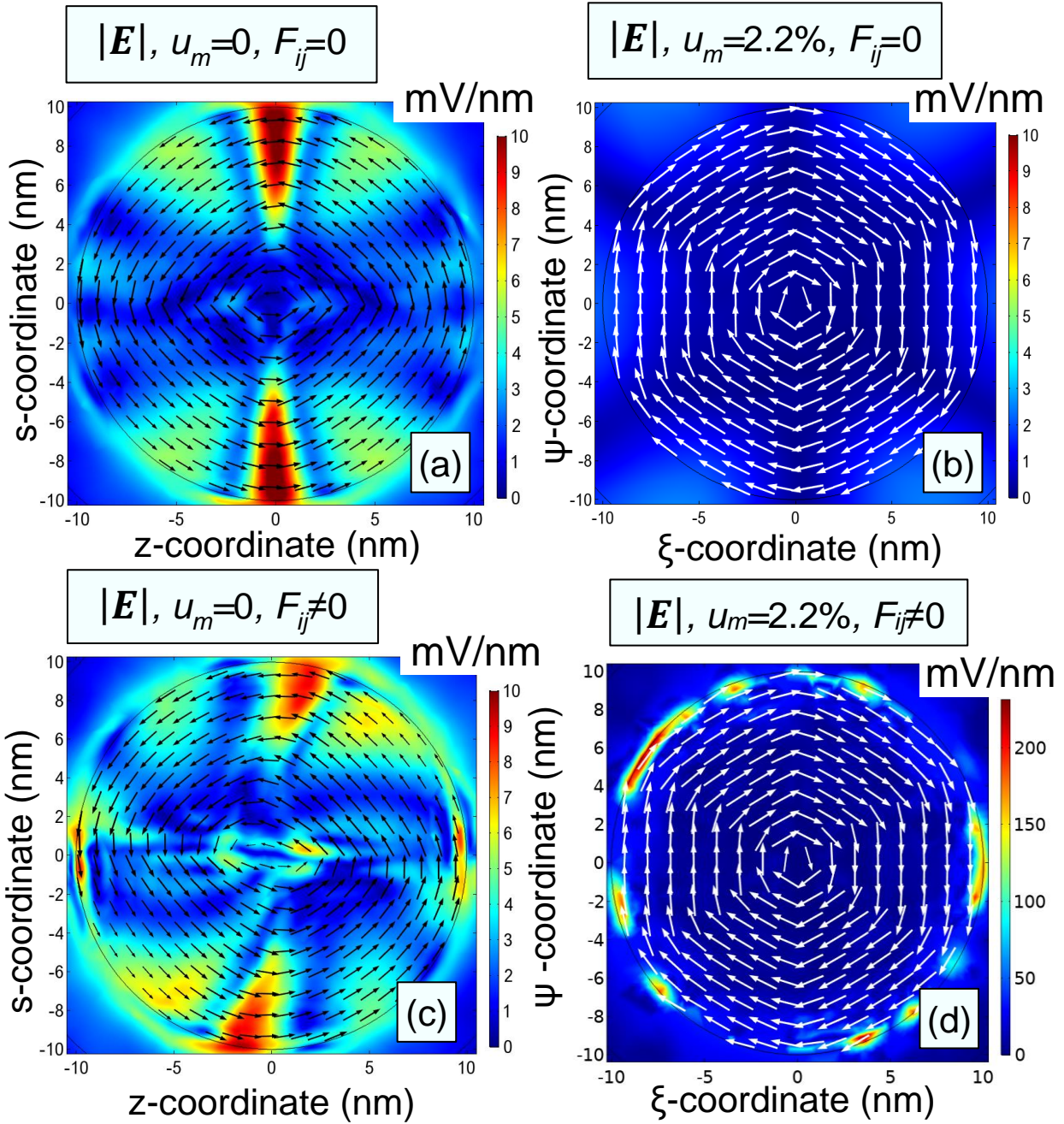

**Figure A5**. **Ferroelectric core covered with a soft tunable shell without any mismatch strain (a, b) and rigid SrTiO$_3$ shell with mismatch strain 2.2% (c,d).** The distribution of the absolute value of the internal electric field in the cross-sections {110} **(a,b)** and {111} **(c,d)** perpendicular to the vortex axis pointed along [110] **(a,b)** and [111] **(a,b)**, respectively. Note the difference in scale in panel **(d)**, showing that the field arising at the surfaces is significantly larger than in the other cases. Black/white arrows indicate the projection of the polarization vector onto the corresponding surface (a, b, c, d). Flexoelectric coefficients are either set to zero (a, b) or taken from Table

I. Core radius $R = 10$ nm, shell thickness $\Delta R = 4$ nm, $T$=298 K. The dielectric and elastic properties of the SrTiO$_3$ shell and all other parameters are listed in **Table I.**

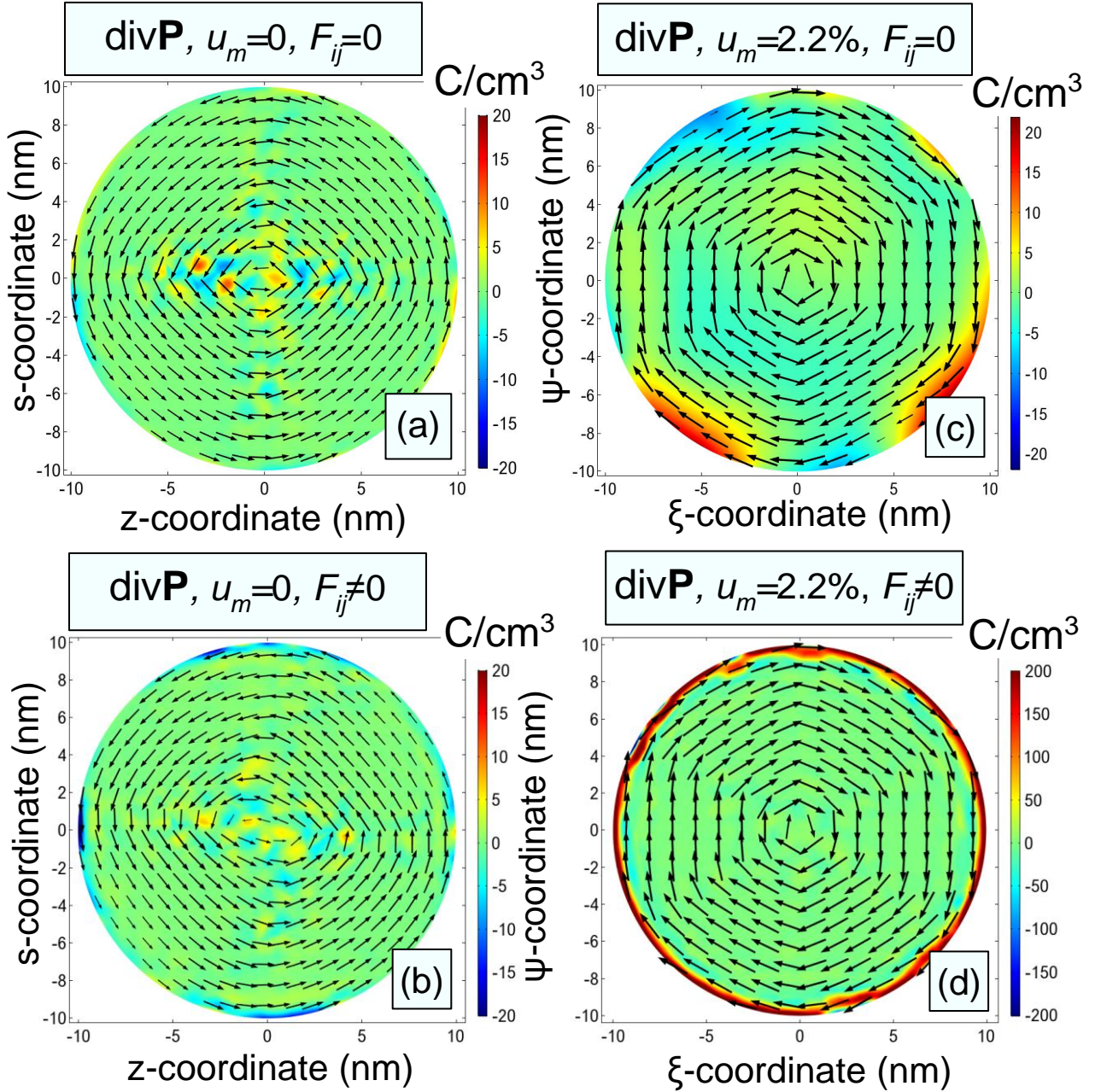

**Figure A5. Ferroelectric core covered with a soft tunable shell without any mismatch strain (a, b) and a rigid SrTiO$_3$ shell with mismatch strain 2.2%.** The distribution of the divergence of the electric polarization in the cross-sections {110} **(a,b)** and {111} **(c,d)**, which are perpendicular to the vortex axis pointed along [110] **(a,b)** and [111] **(a,b)**, respectively. Note the difference in scale in panel **(d)**, showing that the density of bound charges at the surface is significantly larger than in the other cases. Black arrows indicate the projection of the polarization vector onto the corresponding surface (a, b, c, d). Flexoelectric coefficients are either set to zero (a, c) or taken from **Table AI** (c, d). Core radius $R = 10$ nm, shell thickness $\Delta R = 4$ nm, $T = 298$ K. The dielectric and elastic properties of the SrTiO$_3$ shell and all other parameters are listed in **Table AI**.

## A.4. Domain structure morphology

The analysis of the simulated polarization structures involves the search for topological defects, i.e., Bloch points and Ising lines. These defects are characterized by regions in which the magnitude of ***P*** is zero. To identify them in the polarization configurations calculated with FEM, we use three isosurfaces, $P_1 = 0, P_2 = 0,$ and $P_3 = 0,$ and search for their intersection points [19]. These crossings of the isosurfaces indicate the position of Bloch Points, displayed as purple spheres in **Fig. 10**. Ising lines are formed when the isosurfaces intersect on a line or a line segment, rather than at a single point. In numerical simulations, Ising lines are more difficult to identify than Bloch points. Because of the discretized representation of the polarization vector field, defects of this type do not appear as one-dimensional continuous objects, but as a set of aligned Bloch Points whose density depends on the numerical accuracy and on the size of the discretization cells.

In spite of the method's inability to identify their one-dimensional nature, Ising lines are usually easy to recognize, as can be seen, e.g., in Fig. **10b, 10d**. More complicated situations may occur in cases when the isosurfaces meet at very small angles, such that two (or even three) of them are almost parallel to each other (see, e.g., **Fig. A6b**). This results in regions in which each of the components $P_1$, $P_2$, and $P_3$ is nearly zero, and it may be practically impossible to discern such regions from "real" Bloch points or Ising lines, where all of the components are "exactly" zero. The quotation marks are used here to emphasize the underlying problem, which consists in the physically ill-defined task of establishing a clear-cut distinction between these two cases.

Different domain wall and BPS morphologies in a ferroelectric core covered with a soft or rigid shell in various elastic conditions are shown in **Figs. A6.** A twisted morphology of the polarization isosurfaces without BPS (at zero external electric field) is found in the case of the stress-free core covered with an elastically isotropic soft shell (see **Fig. A6a** and compare it with Fig. 5 in Ref. [20], where two diametrically opposite Bloch points appear at a small distance from the core surface at a nonzero external electric field).

Anisotropic electrostriction coupling strongly changes the morphology of polarization isosurfaces in the core (see **Fig. A6b**), and flexoelectric coupling induces an additional curvature and twist of the isosurfaces (see **Fig. A6c**). The shell rigidity very strongly flattens the twisted morphology of polarization isosurfaces (compare **Fig. A6d** with **Fig. A6a-c**), while the inclusion of flexoelectric coupling leads to a slight reappearance of the twist (compare **Fig. A6e-f** with **Fig. A6c**). However, the twist and the mutual shift of the isosurfaces induced by the flexoelectric coupling in a core covered with a rigid shell prevents the Ising line formation. The line defect transforms into a single Bloch point located in the core center (see **Fig. A6e-f**).

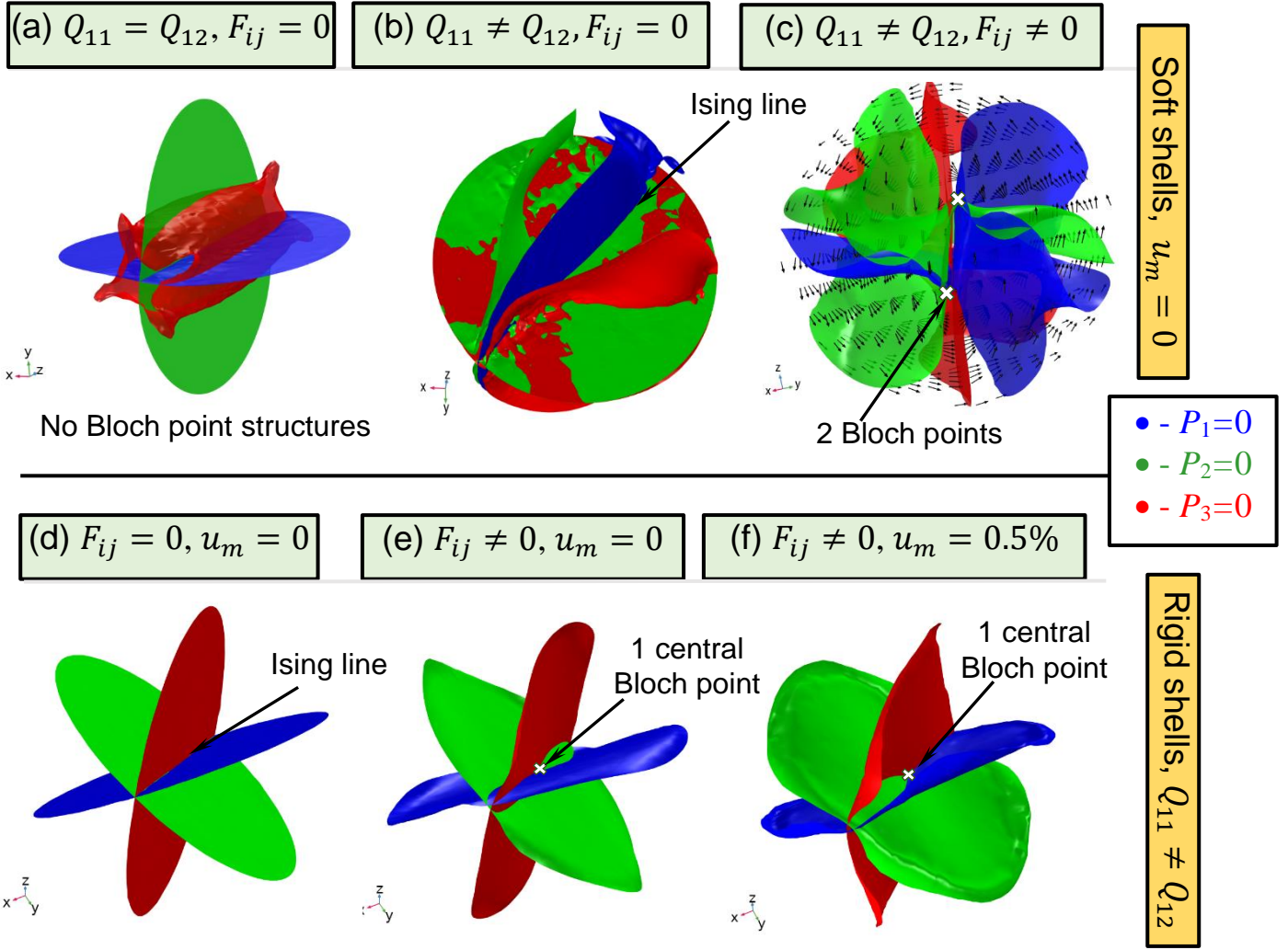

**Figure A6.** Domain walls and BPs morphologies in a ferroelectric core covered with a soft ($|s_{ij}^s| > 10^{-8}$ Pa$^{-1}$ in panels **(a-c)**) or rigid ($|s_{ij}^s|$ $10^{-11}$ Pa$^{-1}$ in panels **(d-f)**) shell. Electrostriction anisotropy is small ($Q_{11}^{c,s} \approx Q_{12}^{c,s}$) for panel **(a)** and high ($Q_{11}^{c,s} \neq Q_{12}^{c,s}$) for panels **(b-f)**. Flexoelectric effect is absent ($F_{ij} = 0$) for panels **(a, b, d)** and present ($F_{ij} \neq 0$) for panels **(c, e, f)**. Mismatch strain is absent ($F_{ij} = 0$) for panels **(a, b, d)** and present ($F_{ij} \neq 0$) for panels **(c, e, f)**. The blue, green, and red isosurfaces denote the regions where the $P_1$, $P_2$, and $P_3$ are zero, respectively. The intersection points of these isosurfaces (denoted with a white cross) show the position of Bloch points ($|\mathbf{P}| = 0$). Core radius $R = 10$ nm, shell thickness $\Delta R = 4$ nm, and $T = 298$ K. For other parameters see **Table AI**.

## APPENDIX B. Supplementary figures for analysis of the phase diagrams

The minimal value of the total free energy as the function of temperature for the nanoparticles with different core radii are shown in **Figs. B1-B3. Figure B1** displays a comparison of the energies for nanoparticles covered by soft and rigid shells with and without a flexoelectric effect, and without a mismatch strain between the core and shell. **Figure B2** shows a comparison of the energies for nanoparticles covered by a rigid shell with compressive mismatch strain and without a flexoelectric effect.

**Figure B3** compares the energies for nanoparticles covered by a rigid shell with tensile mismatch strain and without a flexoelectric effect.

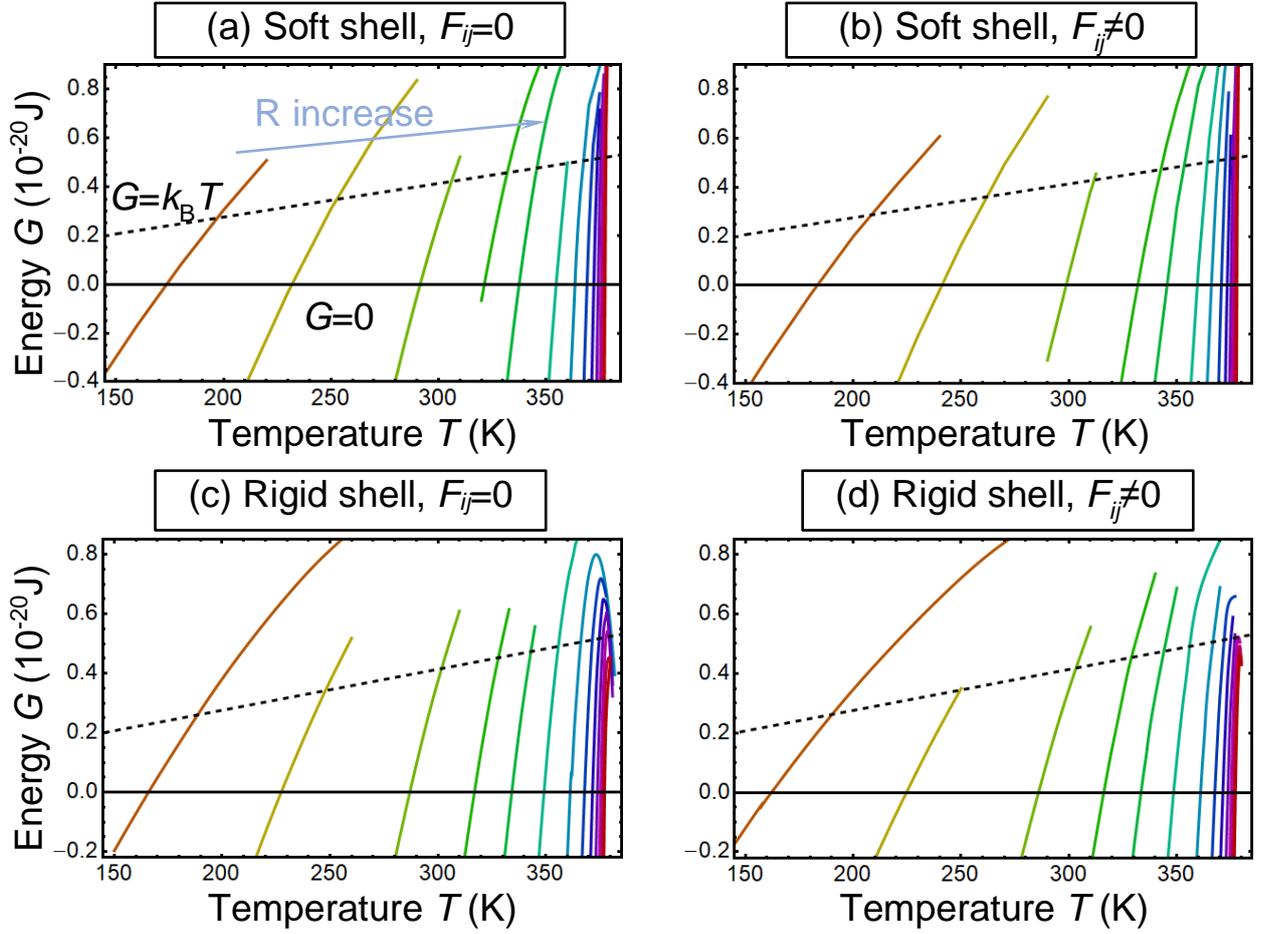

**Figure B1**. Minimal value of the total free energy as the function of temperature for the nanoparticles with different core radii $R = 1, 1.2, 1.6, 2, 2.4, 3.2, 4, 5, 6, 7, 8, 9$, and 10 nm (solid curves from the left to right, colored from brown to red), black dashed line shows the thermal energy level $G = k_B T$. Flexoelectric coupling is zero for panels (a) and (c). The mismatch strain between the shell and core is not taken into account, $u_m = 0$. The dielectric and elastic properties of the SrTiO$_3$ shell and all other parameters are listed in **Table AI.**

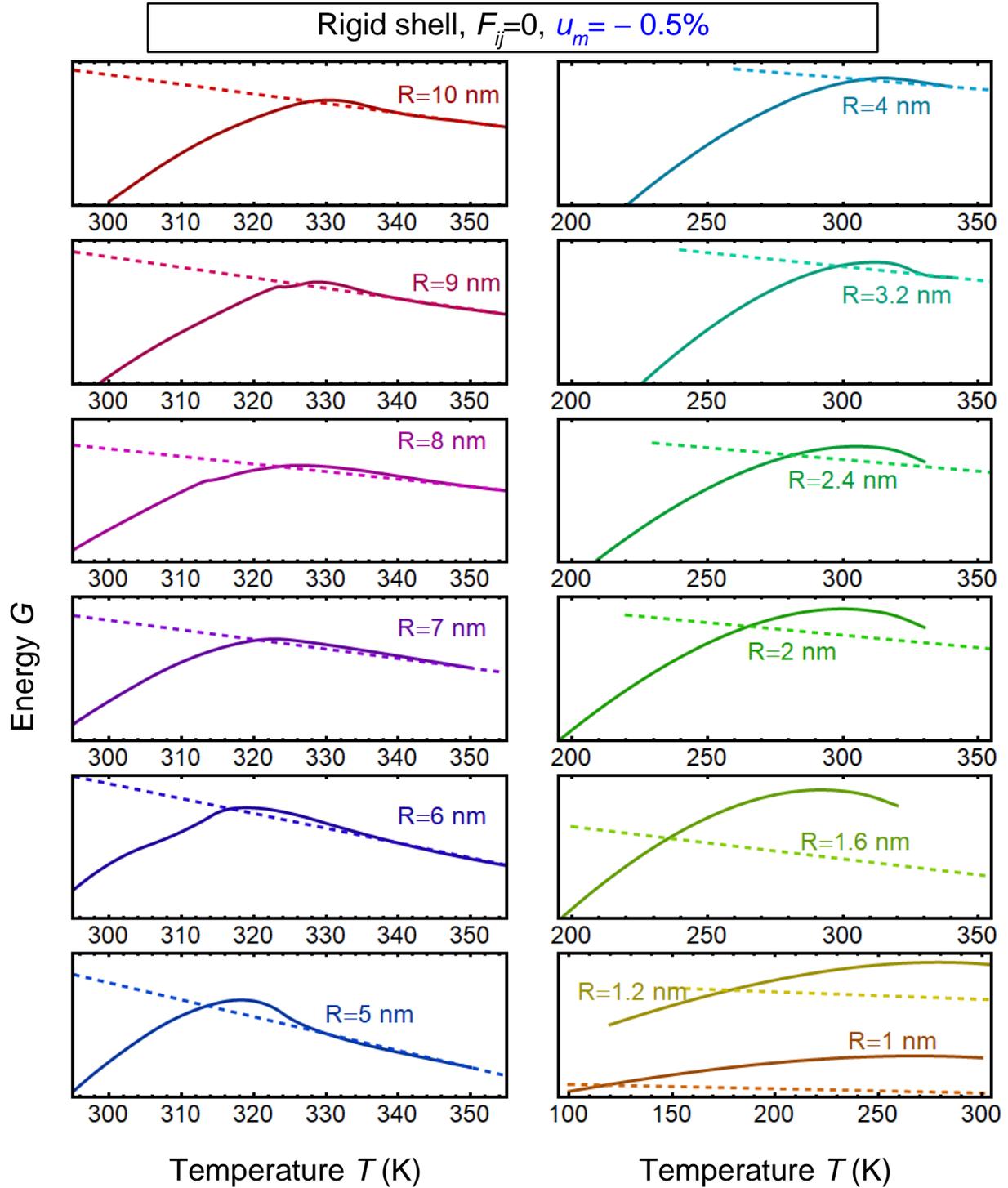

**Figure B2.** Minimal value of the total free energy as a function of temperature for nanoparticles with different core radii $R$ = 1, 1.2, 1.6, 2, 2.4, 3.2, 4, 5, 6, 7, 8, 9, and 10 nm (specified near the curves). Solid and dashed curves correspond to the energy of ferroelectric and paraelectric phases, respectively. The case of a rigid shell imposing **compression strain** on the core is considered here: the mismatch strain is $u_m$ = -0.5%; flexoelectric coefficients are set to zero. The dielectric and elastic properties of the SrTiO$_3$ shell and all other parameters are listed in **Table AI.**

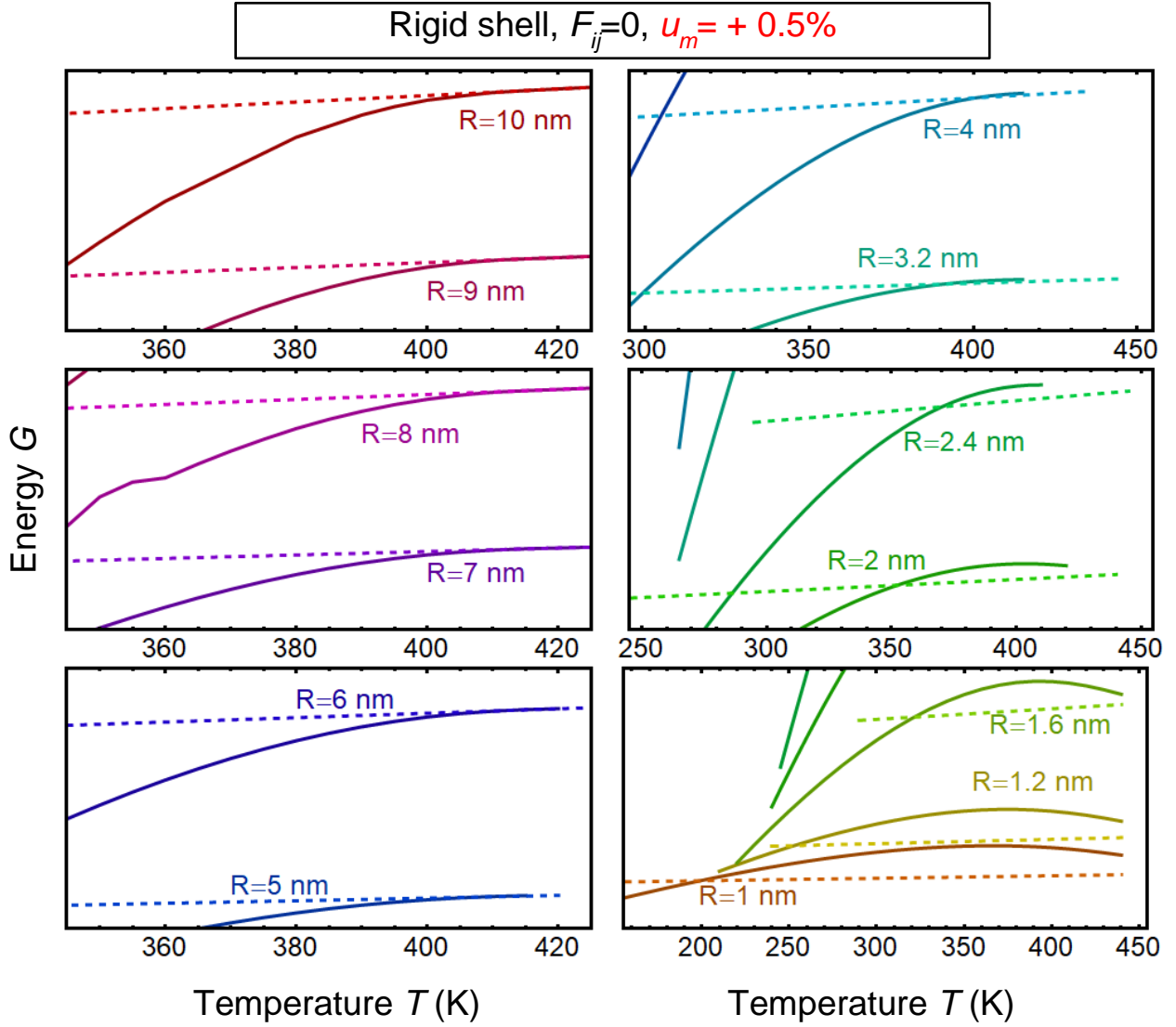

**Figure B3**. Minimal value of the total free energy as the function of temperature for the nanoparticles with different core radii $R =$ 1, 1.2, 1.6, 2, 2.4, 3.2, 4, 5, 6, 7, 8, 9, and 10 nm (specified near the curves). Solid and dashed curves correspond to the energy of ferroelectric and paraelectric phases, respectively. The case of a rigid shell imposing **tensile strain** on the core is considered here: the mismatch strain is $u_m = +0.5\%$; flexoelectric coefficients are set to zero. The dielectric and elastic properties of the SrTiO$_3$ shell and all other parameters are listed in **Table AI**.

## APPENDIX C. Calculations of the electrocaloric effect

Since the FEM results obtained in this work for core-shell nanoparticles with complex domain structure show that the transition temperature $T_{pt}(R)$ of a nanoparticle can be well fitted by an analytical expression (Eq. (1) in the main text), we can make analytical estimates for the EC temperature change $\Delta T_{EC}$ and EC coefficient $\Sigma$, and establish the role of a size effect. By definition, the values $\Delta T_{EC}$ and $\Sigma$ can be calculated from the expressions

$$\Delta T_{EC}(\vec{r}) = - \int_0^{E_{ext}} \frac{T}{\rho(\vec{r}) C_P(\vec{r})} \left(\frac{\partial P(\vec{r})}{\partial T}\right)_E dE, \qquad \text{(C.1a)}$$

$$\Sigma(\vec{r}) = \frac{d\Delta T_{EC}(\vec{r})}{dE_{ext}}, \tag{C.1b}$$

where $E_{ext}$ is an external field applied to the core-shell nanoparticle via effective media, $\rho$ is the mass density, $C_P$ is the heat capacity of the nanoparticle core or shell, and $P(\vec{r})$ is a scalar polarization magnitude, which depend on the point $\vec{r}$. Following Ref. [21], the spatially averaged values $\langle \overline{\Delta T_{EC}} \rangle$ and $\langle \overline{\Sigma} \rangle$ can be estimated as

$$\langle \overline{\Delta T_{EC}(E_{ext})} \rangle \approx \frac{T}{\eta \rho C_P} \left( \frac{\alpha_T}{2}[P^2(E_{ext}) - P^2(0)] + \frac{\beta_T}{4}[P^4(E_{ext}) - P^4(0)] + \frac{\gamma_T}{6}[P^6(E_{ext}) - P^6(0)] \right),$$
(C.2a)

$$\langle \overline{\Sigma(E_{ext},T)} \rangle = \langle \frac{T}{\rho C_P} \frac{\alpha_T P + \beta_T P^3 + \gamma_T P^5}{\alpha_T[T-T_{pt}(R)]+3\beta P^2 + 5\gamma P^4} \rangle. \tag{C.2b}$$

Here $P^2 = P_1^2 + P_2^2 + P_3^2 \equiv P_r^2$. When deriving these expressions (C.2), we used a polarization vector $\boldsymbol{P}$ averaged over core volume that is nearly zero at $E_{ext} \to 0$, and so $\overline{\boldsymbol{P}} = \overline{\boldsymbol{P^3}} = \overline{\boldsymbol{P^5}} \to 0$.

Using expression (1) for $T_{pt}(R)$ from the main text and Eqs.(C.2), we can make a prediction about the values of the $\langle \overline{\Delta T_{EC}} \rangle$ and $\langle \overline{\Sigma} \rangle$ for an ensemble of noninteracting ferroelectric core-shell nanoparticles in effective media approximation. Namely, the "effective" LGD free energy leads to the equation for the polarization magnitude

$$\alpha_T \left( T - T_{pt}(R) \right) P + \beta(T)P^3 + \gamma(T)P^5 = \eta E_{ext}, \tag{C.3a}$$

Equation (C.3a) allows the calculation of the field dependence of polarization $P(E_{ext})$ if the coefficients $\alpha_T$, $\beta(T)$, and $\gamma(T)$, and factor $\eta$ are known. Material parameters of BaTiO$_3$ in Eq.(C.3a) are listed in **Table CI**.

**Table CI.** LGD coefficients and other material parameters of a BaTiO$_3$ core

| $\varepsilon_{b,s,e}$ | $\varepsilon_b$=7 (core background),    $\varepsilon_e$=10 (surrounding), $\varepsilon_s(T) \approx 3000$ at $T$=50 K and $\varepsilon_s(T) \approx 300$ at T=298 K |
|---|---|
| $\alpha_T$ ($C^{-2}mJ/K$) | $6.68 \times 10^5$ |
| $T_C$ (K) | 381 |
| $\beta(C^{-4}m^5J)$ | $\beta_T(T-393) - 8.08 \times 10^8$; $\beta_T = 18.76 \times 10^6$ |
| $\gamma(C^{-6}m^9J)$ | $\gamma_T(T-393) + 16.56 \times 10^9$; $\gamma_T = -33.12 \times 10^{7*}$ |
| $g_{11}(m^3/F)$ | $5.1 \times 10^{-10}$ |
| $g_{44}(m^3/F)$ | $0.2 \times 10^{-10}$ |

*These parameters are valid until $\gamma > 0$, i.e. for $T < 445\ K$.

** $\rho = 6.02 \times 10^3\ kg/m^3$, $C_p = 4.6 \times 10^2\ J/(kg \cdot K)$ at room temperature.

In accordance with our estimates, listed below, the factor $\eta$ is given by the expression:

$$\eta = \frac{9(R+\Delta R)^3 \varepsilon_e \varepsilon_s}{2R^3(\varepsilon_e - \varepsilon_s)(\varepsilon_s - \varepsilon_b) + (R+\Delta R)^3(2\varepsilon_e + \varepsilon_s)(\varepsilon_b + 2\varepsilon_s)}. \tag{C.3b}$$

To estimate the factor $\eta$, let us consider a core-shell ferroelectric nanoparticle with a homogeneously polarized core of radius $R$. The nanoparticle is assumed to have a ferroelectric polarization averaged over the core volume, $\vec{P}_f$. The averaged polarization points along the z-axis, which coincides with the direction of a homogeneous external field $\vec{E}_{ext}$. Since the ferroelectric polarization is vortex-like at $\vec{E}_{ext} \to 0$ (see **Figs. 2-7** in the main text) the average polarization is absent.

The core, shell, and external effective medium are dielectrics, and so the electrostatic potential satisfies the Laplace equation in all of the regions:

$$\Delta\varphi_e = 0, \qquad \Delta\varphi_s = 0, \qquad \Delta\varphi_f = 0, \qquad (C.4)$$

where the subscripts "*f*", "*s*", and "*e*" denote the physical quantities related to the ferroelectric core, shell, and external media, respectively (see **Fig. 1a** in the main text). The electric field and displacement vectors are:

$$\mathbf{E}_{f,s,e} = -\nabla\varphi_{f,s,e}, \qquad \mathbf{D}_f = \varepsilon_0\varepsilon_b\mathbf{E}_f + \mathbf{e}_z P_f, \qquad \mathbf{D}_{s,e} = \varepsilon_0\varepsilon_{s,e}\mathbf{E}_{s,e}, \qquad (C.5)$$

here $\mathbf{e}_z$ is the unit vector along the axis z. The term $\varepsilon_0\varepsilon_b\mathbf{E}_f$ corresponds to the dielectric "background" reaction to the external field, and the term $\mathbf{e}_z P_f$ is the contribution of the core average ferroelectric polarization. The electric potential and radial displacement are continuous functions at all interfaces:

$$(\varphi_f - \varphi_s)\big|_{r=R} = 0, \qquad \mathbf{e}_r(\mathbf{D}_f - \mathbf{D}_s)\big|_{r=R} = 0, \qquad (C.6a)$$

$$(\varphi_s - \varphi_e)\big|_{r=R+\Delta R} = 0, \qquad \mathbf{e}_r(\mathbf{D}_s - \mathbf{D}_e)\big|_{r=R+\Delta R} = 0. \qquad (C.6b)$$

Let us switch coordinate systems from Cartesian to spherical coordinates with the origin {0,0,0} in the center of the core, and the polar axis along the z axis. It is natural to assume that the electrostatic field does not depend on the azimuthal coordinate $\psi$, so the general solution is:

$$\varphi_f = -E_f r\cos\theta, \quad \varphi_s = -E_s r\cos\theta + B_s\frac{\cos\theta}{r^2}, \quad \varphi_e = B_e\frac{\cos\theta}{r^2} - r\cos\theta E^{ext}. \qquad (C.7)$$

The potential $\varphi_e$ produces a homogeneous electric field $E^{ext}$ very far away from the particle. The radial components of the electric displacement are

$$\mathbf{e}_r\mathbf{D}_f = \varepsilon_0\varepsilon_b E_f \cos\theta + P_f\cos\theta, \quad \mathbf{e}_r\mathbf{D}_s = \varepsilon_0\varepsilon_s \cos\theta\left(E_s + \frac{2B_s}{r^3}\right), \qquad (C.8a)$$

$$\mathbf{e}_r\mathbf{D}_e = \varepsilon_0\varepsilon_e\left(2\frac{\cos\theta}{r^3}B_e + \cos\theta E^{ext}\right). \qquad (C.8b)$$

Substitution of the solution (C.7) - (C.8) into the boundary conditions (C.6) gives a system of linear equations for the unknown coefficients $E_{f,s}$ and $B_{s,e}$. Using their values yields the following expressions:

$$E_f = \frac{9(R+\Delta R)^3 \varepsilon_0\varepsilon_e\varepsilon_s E^{ext} - [2R^3(\varepsilon_s-\varepsilon_e)+(R+\Delta R)^3(2\varepsilon_e+\varepsilon_s)]P_f}{\varepsilon_0(2R^3(\varepsilon_e-\varepsilon_s)(\varepsilon_s-\varepsilon_b)+(R+\Delta R)^3(2\varepsilon_e+\varepsilon_s)(\varepsilon_b+2\varepsilon_s))}, \qquad (C.9a)$$

$$E_s = \frac{3(R+\Delta R)^3 \varepsilon_0\varepsilon_e(2\varepsilon_s+\varepsilon_f)E^{ext} - 2R^3(\varepsilon_s-\varepsilon_e)P_f}{\varepsilon_0(2R^3(\varepsilon_e-\varepsilon_s)(\varepsilon_s-\varepsilon_b)+(R+\Delta R)^3(2\varepsilon_e+\varepsilon_s)(\varepsilon_b+2\varepsilon_s))}, \qquad (C.9b)$$

$$B_s = \frac{R^3(R+\Delta R)^3[(2\varepsilon_e+\varepsilon_s)P_s+3\varepsilon_0(\varepsilon_b-\varepsilon_s)E^{ext}]}{\varepsilon_0(2R^3(\varepsilon_e-\varepsilon_s)(\varepsilon_s-\varepsilon_b)+(R+\Delta R)^3(2\varepsilon_e+\varepsilon_s)(\varepsilon_b+2\varepsilon_s))}, \qquad (C.9c)$$

$$B_e = \frac{3R^3(R+\Delta R)^3 P_S \varepsilon_s + (R+\Delta R)^3[R^3(\varepsilon_b-\varepsilon_s)(2\varepsilon_s+\varepsilon_e)-(\varepsilon_e-\varepsilon_s)(2\varepsilon_s+\varepsilon_b)(R+\Delta R)^3]\varepsilon_0 E^{ext}}{\varepsilon_0\big(2R^3(\varepsilon_e-\varepsilon_s)(\varepsilon_s-\varepsilon_b)+(R+\Delta R)^3(2\varepsilon_e+\varepsilon_s)(\varepsilon_b+2\varepsilon_s)\big)}. \quad \text{(C.9d)}$$

From Eqs.(C.7) and (C.9), the electric field inside the core is uniform and is equal to

$$\vec{E}_f = \frac{-[2R^3(\varepsilon_s-\varepsilon_e)+(R+\Delta R)^3(2\varepsilon_e+\varepsilon_s)]P_f \vec{e}_z}{\varepsilon_0\big(2R^3(\varepsilon_e-\varepsilon_s)(\varepsilon_s-\varepsilon_b)+(R+\Delta R)^3(2\varepsilon_e+\varepsilon_s)(\varepsilon_b+2\varepsilon_s)\big)} + \frac{9(R+\Delta R)^3 \varepsilon_e \varepsilon_s E^{ext} \vec{e}_z}{2R^3(\varepsilon_e-\varepsilon_s)(\varepsilon_s-\varepsilon_b)+(R+\Delta R)^3(2\varepsilon_e+\varepsilon_s)(\varepsilon_b+2\varepsilon_s)} \quad \text{(C.10a)}$$

The electric field is non-uniform in the shell and external media and is equal to:

$$\vec{E}_s = -\frac{3(R+\Delta R)^3 \varepsilon_0 \varepsilon_e (2\varepsilon_s+\varepsilon_b) E^{ext}\vec{e}_z - 2R^3(\varepsilon_s-\varepsilon_e)P_f \vec{e}_z}{\varepsilon_0\big(2R^3(\varepsilon_e-\varepsilon_s)(\varepsilon_s-\varepsilon_b)+(R+\Delta R)^3(2\varepsilon_e+\varepsilon_s)(\varepsilon_b+2\varepsilon_s)\big)} - \frac{R^3(R+\Delta R)^3[(2\varepsilon_e+\varepsilon_s)P_S + 3\varepsilon_0(\varepsilon_b-\varepsilon_s)E^{ext}]\nabla\left(\frac{z}{r^3}\right)}{\varepsilon_0\big(2R^3(\varepsilon_e-\varepsilon_s)(\varepsilon_s-\varepsilon_b)+(R+\Delta R)^3(2\varepsilon_e+\varepsilon_s)(\varepsilon_b+2\varepsilon_s)\big)},$$

(C.10b)

$$\vec{E}_e = -\nabla\left(\frac{z}{r^3}\right)\frac{3R^3(R+\Delta R)^3 P_f \varepsilon_s + (R+\Delta R)^3[R^3(\varepsilon_b-\varepsilon_s)(2\varepsilon_s+\varepsilon_e)-(\varepsilon_e-\varepsilon_s)(2\varepsilon_s+\varepsilon_b)(R+\Delta R)^3]\varepsilon_0 E^{ext}}{\varepsilon_0\big(2R^3(\varepsilon_e-\varepsilon_s)(\varepsilon_s-\varepsilon_b)+(R+\Delta R)^3(2\varepsilon_e+\varepsilon_s)(\varepsilon_b+2\varepsilon_s)\big)} - \vec{e}_z E^{ext}. \quad \text{(C.10c)}$$

The second term in Eq.(C.10a) gives Eq.(C.3b) for the factor $\eta = \frac{9(R+\Delta R)^3 \varepsilon_e \varepsilon_s}{2R^3(\varepsilon_e-\varepsilon_s)(\varepsilon_s-\varepsilon_b)+(R+\Delta R)^3(2\varepsilon_e+\varepsilon_s)(\varepsilon_b+2\varepsilon_s)}$.

# REFERENCES

1   A.K. Tagantsev and G. Gerra, Interface-induced phenomena in polarization response of ferroelectric thin films, J. Appl. Phys. **100**, 051607 (2006).

2   E. A. Eliseev, Y. M. Fomichov, S. V. Kalinin, Y. M. Vysochanskii, P. Maksymovich and A. N. Morozovska. Labyrinthine domains in ferroelectric nanoparticles: Manifestation of a gradient-induced morphological phase transition. Phys. Rev. B **98**, 054101 (2018).

3   J. J. Wang, E. A. Eliseev, X. Q. Ma, P. P. Wu, A. N. Morozovska, and Long-Qing Chen. Strain effect on phase transitions of $BaTiO_3$ nanowires. Acta Materialia **59**, 7189 (2011).

4   A. N. Morozovska, E. A. Eliseev, Y. A. Genenko, I. S. Vorotiahin, M. V. Silibin, Ye Cao, Y. Kim, M. D. Glinchuk, and S. V. Kalinin. Flexocoupling impact on the size effects of piezo- response and conductance in mixed-type ferroelectrics-semiconductors under applied pressure. Phys. Rev. B **94,** 174101 (2016).

5   M. S. Kilic, M. Z. Bazant, and A. Ajdari. Steric effects in the dynamics of electrolytes at large applied voltages. I. Double-layer charging. Phys. Rev. E **75**, 021502 (2007).

6   Y. T. Cheng, M. W. Verbrugge. Evolution of stress within a spherical insertion electrode particle under potentiostatic and galvanostatic operation. J. Power Sources **190**, 453 (2009).

7   X. Zhang, A. M. Sastry, W. Shyy. Intercalation-induced stress and heat generation within single lithium-ion battery cathode particles. J. Electrochem. Soc. **155**, A542 (2008).

8   D. A. Freedman, D. Roundy, and T. A. Arias. Elastic effects of vacancies in strontium titanate: Short-and long-range strain fields, elastic dipole tensors, and chemical strain. Phys. Rev. B **80**, 064108 (2009).

9   L. D. Landau, and I. M. Khalatnikov. On the anomalous absorption of sound near a second order phase transition point. In Dokl. Akad. Nauk SSSR, **96**, 469 (1954).

10   J. Hlinka, Mobility of Ferroelectric Domain Walls in Barium Titanate, Ferroelectrics **349,** 49 (2007).

11   J. Hlinka and P. Márton, Phenomenological model of 90-degree domain wall in BaTiO$_3$ type ferroelectrics.  Phys. Rev. B **74**, 104104 (2006).

12    N. A. Pertsev, A. K. Tagantsev, and N. Setter, Phase transitions and strain-induced ferroelectricity in SrTiO$_3$ epitaxial thin films. Phys. Rev. B **61**, R825 (2000).

13   M. Y. Gureev, A. K. Tagantsev, and N. Setter. Head-to-head and tail-to-tail 180 domain walls in an isolated ferroelectric. Phys. Rev. B **83,** 184104 (2011).

14   E. A. Eliseev, A. N. Morozovska, G. S. Svechnikov, P. Maksymovych, S. V. Kalinin. Domain wall conduction in multiaxial ferroelectrics: impact of the wall tilt, curvature, flexoelectric coupling, electrostriction, proximity and finite size effects. Phys. Rev. B **85**, 045312 (2012).

15   I. S. Vorotiahin, E. A. Eliseev, Q. Li, S. V. Kalinin, Y. A. Genenko and A. N. Morozovska. Tuning the Polar States of Ferroelectric Films via Surface Charges and Flexoelectricity. Acta Materialia **137** (15), 85–92 (2017).

16   E. A. Eliseev, A. N. Morozovska, C. T. Nelson, and S. V. Kalinin. Intrinsic structural instabilities of domain walls driven by gradient couplings: meandering anferrodistortive-ferroelectric domain walls in BiFeO$_3$. Phys. Rev. B **99**, 014112 (2019).

17   M. J. Han, E. A. Eliseev, A. N. Morozovska, Y. L. Zhu, Y. L. Tang, Y. J. Wang, X. W. Guo, X. L. Ma. Mapping gradient-driven morphological phase transition at the conductive domain walls of strained multiferroic films. Phys. Rev. B **100**, 104109 (2019).

18   E. A. Eliseev, I. S. Vorotiahin, Y. M. Fomichov, M. D. Glinchuk, S. V. Kalinin, Y. A. Genenko, and A. N. Morozovska. Defect driven flexo-chemical coupling in thin ferroelectric films. Phys. Rev. B **97,** 024102 (2018).

19   R. Hertel, and C. M. Schneider, Exchange Explosions: Magnetization Dynamics During Vortex-Antivortex Annihilation. Phys. Rev. Lett. **97,** 177202 (2006).

20   A. N. Morozovska, E. A. Eliseev, R. Hertel, Y. M. Fomichov, V. Tulaidan, V. Yu. Reshetnyak, and D. R. Evans. Electric Field Control of Three-Dimensional Vortex States in Core-Shell Ferroelectric Nanoparticles. Acta Materialia, **200**, 256-273 (2020).

21   A. N. Morozovska, E. A. Eliseev, M. D. Glinchuk, H. V. Shevliakova, G. S. Svechnikov, M. V. Silibin, A. V. Sysa, A. D. Yaremkevich, N. V. Morozovsky, and V. V. Shvartsman. Analytical description of the size effect on pyroelectric and electrocaloric properties of ferroelectric nanoparticles. Phys. Rev. Materials **3**, 104414 (2019).